\documentclass[twocolumn]{aastex62}
\usepackage{graphicx}	
\usepackage[caption=false]{subfig}
\usepackage{rotating}

\usepackage{amssymb}
\usepackage{color}
\usepackage{amsmath}
\usepackage{longtable}
\usepackage{threeparttablex}

\usepackage{natbib}
\newcommand{\ie}{\textit{i.e.,}}
\newcommand{\eg}{\textit{e.g.,}}

\newcommand{\Ks}{$K_s$}

\newcommand{\UVJ}{\textit{UVJ}}
\newcommand{\UVc}{\textit{(U-V)}}
\newcommand{\VJc}{\textit{(V-J)}}
\newcommand{\Dfour}{$D_n$(4000)}
\newcommand{\EWhd}{$EW_0$(H$\delta$)}

\newcommand{\logM}{$\log(M_*/\rm{M}_\odot)$}
\newcommand{\zphot}{$z_\textrm{phot}$}
\newcommand{\zspec}{$z_\textrm{spec}$}
\newcommand{\magazine}{MAGAZ3NE}

\newcommand{\OII}{\hbox{{\rm [O}\kern 0.1em{\sc ii}{\rm ]}}}
\newcommand{\OIIdub}{\hbox{{\rm [O}\kern 0.1em{\sc ii}{\rm ]$\lambda\lambda3726,3729$}}}
\newcommand{\OIII}{\hbox{{\rm [O}\kern 0.1em{\sc iii}{\rm ]}}}
\newcommand{\OIIIdub}{\hbox{{\rm [O}\kern 0.1em{\sc iii}{\rm ]$\lambda\lambda4959,5007$}}}
\newcommand{\OIIIfour}{\hbox{{\rm [O}\kern 0.1em{\sc iii}{\rm ]$\lambda4959$}}}
\newcommand{\OIIIfive}{\hbox{{\rm [O}\kern 0.1em{\sc iii}{\rm ]$\lambda5007$}}}

\newcommand{\Hbeta}{$\rm{H}\beta$}

\begin{document}

\title{\sc The Massive Ancient Galaxies At $z>3$ NEar-infrared (\magazine) Survey: Confirmation of Extremely Rapid Star-Formation and Quenching Timescales for Massive Galaxies in the Early Universe\footnote{The spectra presented herein were obtained at the W. M. Keck Observatory, which is operated as a scientific partnership among the California Institute of Technology, the University of California and the National Aeronautics and Space Administration. The Observatory was made possible by the generous financial support of the W. M. Keck Foundation.} }
\shorttitle{MAGAZ3NE: Rapid Star-Formation and Quenching Timescales}
\shortauthors{B. Forrest, et al.}

\correspondingauthor{Ben Forrest}
\email{benjamif@ucr.edu}

\author[0000-0001-6003-0541]{Ben Forrest}
	\affiliation{Department of Physics and Astronomy, University of California, Riverside, 900 University Avenue, Riverside, CA 92521, USA}
\author[0000-0002-7248-1566]{Z. Cemile Marsan}
	\affiliation{Department of Physics and Astronomy, York University, 4700,
Keele Street, Toronto, ON MJ3 1P3, Canada}
\author{Marianna Annunziatella}
	\affiliation{Department of Physics and Astronomy, Tufts University, 574 Boston
Avenue Suites 304, Medford, MA 02155, USA}
	\affiliation{Centro de Astrobiolog\'ia (CSIC-INTA), Ctra de Torrej\'on a Ajalvir, km 4, E-28850 Torrej\'on de Ardoz, Madrid, Spain}
\author[0000-0002-6572-7089]{Gillian Wilson}
	\affiliation{Department of Physics and Astronomy, University of California, Riverside, 900 University Avenue, Riverside, CA 92521, USA}
\author[0000-0002-9330-9108]{Adam Muzzin}
	\affiliation{Department of Physics and Astronomy, York University, 4700,
Keele Street, Toronto, ON MJ3 1P3, Canada}
\author[0000-0001-9002-3502]{Danilo Marchesini}
	\affiliation{Department of Physics and Astronomy, Tufts University, 574 Boston
Avenue Suites 304, Medford, MA 02155, USA}
\author[0000-0003-1371-6019]{M. C. Cooper}
	\affiliation{Center for Cosmology, Department of Physics and Astronomy, University of California, Irvine,  4129 Frederick Reines Hall, Irvine, CA, USA}
\author[0000-0001-6251-3125]{Jeffrey C. C. Chan}
	\affiliation{Department of Physics and Astronomy, University of California, Riverside, 900 University Avenue, Riverside, CA 92521, USA}
\author{Ian McConachie}
	\affiliation{Department of Physics and Astronomy, University of California, Riverside, 900 University Avenue, Riverside, CA 92521, USA}
\author{Percy Gomez}
	\affiliation{W.M. Keck Observatory, 65-1120 Mamalahoa Hwy., Kamuela, HI 96743, USA}

\author[0000-0002-0332-177X]{Erin Kado-Fong	}
	\affiliation{Department of Astrophysical Sciences, 4 Ivy Lane, Princeton University, Princeton, NJ 08544}
\author[0000-0003-1181-6841]{Francesco La Barbera}
	\affiliation{INAF - Osservatorio Astronomico di Capodimonte, sal. Moiariello 16, 80131 Napoli, Italy}
\author{Daniel Lange-Vagle}
	\affiliation{Department of Physics and Astronomy, Tufts University, 574 Boston Avenue Suites 304, Medford, MA 02155, USA}
\author[0000-0002-7356-0629]{Julie Nantais}
	\affiliation{Departamento de Ciencias F\'isicas, Universidad Andres Bello, Fernandez Concha 700, Las Condes 7591538, Santiago, Regi\'on Metropolitana, Chile}
\author[0000-0001-6342-9662]{Mario Nonino}
	\affiliation{INAF - Osservatorio Astronomico di Trieste, Via G. B. Tiepolo 11,
34143 Trieste, Italy}
\author[0000-0003-3959-2595]{Paolo Saracco}	
	\affiliation{INAF - Osservatorio Astronomico di Brera, via Brera 28, 20121 Milano, Italy}
\author[0000-0001-7768-5309]{Mauro Stefanon}
	\affiliation{Leiden Observatory, Leiden University, 2300 RA Leiden, The Netherlands}
\author[0000-0003-1535-2327]{Remco F. J. van der Burg}
	\affiliation{European Southern Observatory, Karl-Schwarzschild-Str. 2, 85748, Garching, Germany}


\begin{abstract}

We present near-infrared spectroscopic confirmations of a sample of 16 photometrically-selected galaxies with stellar masses \logM~$>11$ at redshift $z>3$ from the XMM-VIDEO and COSMOS-UltraVISTA fields using Keck/MOSFIRE as part of the \magazine\ survey.
Eight of the ultra-massive galaxies (UMGs) have specific star formation rates (sSFR)~$<0.03$~Gyr$^{-1}$, with negligible emission lines.
Another seven UMGs show emission lines consistent with active galactic nuclei and/or star formation, while only one UMG has sSFR~$>1$~Gyr$^{-1}$.
Model star formation histories of these galaxies describe systems that formed the majority of their stars in vigorous bursts of several hundred Myr duration around $4<z<6$ during which hundreds to thousands of solar masses were formed per year.
These formation ages of $<1$~Gyr prior to observation are consistent with ages derived from measurements of \Dfour\ and \EWhd.
Rapid quenching followed these bursty star-forming periods, generally occurring less than 350~Myr before observation, resulting in post-starburst SEDs and spectra for half the sample.
The rapid formation timescales are consistent with the extreme star formation rates observed in $4<z<7$ dusty starbursts observed with ALMA, suggesting that such dusty galaxies are progenitors of these UMGs.
While such formation histories have been suggested in previous studies, the large sample introduced here presents the most compelling evidence yet that vigorous star formation followed by rapid quenching is almost certainly the norm for high mass galaxies in the early universe.
The UMGs presented here were selected to be brighter than $K_s=21.7$ raising the intriguing possibility that even (fainter) older quiescent UMGs could exist at this epoch.  

\end{abstract}

\section{Introduction}

The observation of massive, quiescent galaxies has repeatedly forced astronomers to reconsider the paradigm of galaxy evolution and cosmology, particularly in the early Universe.
The spectrum of an old galaxy at $z=1.55$ \citep{Dunlop1996} brought serious challenges to the Einstein-de Sitter cosmology \citep{Einstein1932} popular at the time.
Further observations of similar galaxies at progressively higher redshifts \citep[\eg][]{Cimatti2004, Kriek2009, Gobat2012, vandeSande2013} led to the realization that galaxies could build up large stellar populations and quench star formation at early times, which required new suites of simulations to reproduce them \citep[\eg][]{Vogelsberger2014a, Vogelsberger2014b, Genel2014, Hopkins2014, Schaye2015, Crain2015, Henriques2015, Wellons2015, Dave2016, Feldmann2016a}.

It is now clear that most, if not all, ultra-massive galaxies (UMGs; \logM~$>11$) have assembled the majority of their mass by $z\sim1.5$ \citep[\eg][]{Nelan2005, Thomas2005, Treu2005, Gallazzi2005, Thomas2010, Estrada-Carpenter2020}.
Indeed, the number density of these galaxies evolves very little over the most recent 9 Gyr \citep{vanderWel2014, Gargiulo2016, Kawinwanichakij2020a}, and evidence suggests most of the mass assembly for these galaxies occurred in an early short-lived burst, while less massive quiescent galaxies have underwent longer, less intense periods of star formation \citep{Pacifici2016, Ciesla2017}.
Thus, determining the evolution of these UMGs at earlier times is of great interest as they hold clues to the processes required not only for intense, early star formation, but also rapid quenching.

Critical to the discovery of large numbers of quiescent UMGs at higher redshifts is the existence of wide and deep near-infrared (NIR) imaging combined with ancillary multi-wavelength observations.
The NIR wavelengths are where these galaxies are most easily detected and also constrain stellar masses, as the NIR corresponds to the rest-frame optical at $z\sim3$.
Additional photometry across a range of wavelengths is necessary to effectively characterize SEDs and rule out drastically different natures; in particular, IRAC photometry is critical to differentiating between dust-obscured and old populations at $z>2$.
A variety of recent surveys have made such imaging available, including CANDELS \citep{Koekemoer2011, Grogin2011}, NMBS \citep{VanDokkum2009, Whitaker2011}, UltraVISTA \citep{McCracken2012}, VIDEO \citep{Jarvis2013},  DES+VHS \citep{Banerji2015}, and ZFOURGE \citep{Straatman2016}.
This has resulted in an increasing number of photometric candidates that require follow-up to confirm their quiescent natures \citep{Marchesini2010, Straatman2014, Spitler2014, Guarnieri2019, Pampliega2019, Merlin2019, Shahidi2020}.

Selection of these objects is often done using color-color diagrams \citep[\eg][]{Labbe2005}, the most common of which is the restframe \UVc\ vs \VJc\ plane known as the \UVJ\ diagram \citep[\eg][]{Wuyts2007, Williams2009, Martis2016, Forrest2016}, where the red colors of quiescent objects make them stand out.
However, at high redshifts the clear color bimodality seen in \UVJ\ at low redshifts is not as well defined \citep{Whitaker2011, Muzzin2013a, Straatman2016}.
Additionally, the presence of strong emission lines can move galaxy colors significantly, thus affecting conclusions drawn from photometry alone \citep[\eg][Marsan et al. 2020, in prep]{Salmon2015, Forrest2018}.
Spectroscopic follow-up is thus required to 1) confirm the redshift, 2) quantify the strength of emission lines associated with ongoing star formation and 3) put limits on dust-obscured star formation.

NIR spectroscopy (corresponding to rest-frame optical wavelengths at $z\sim3$) can inform these areas by detecting either emission features such as \OIIdub, \Hbeta $\lambda4861$, and \OIIIdub\ or absorption features including the Balmer series and Calcium H$\lambda3970$ and K$\lambda3935$ lines \citep[\eg][]{Belli2014, Kriek2015, Belli2017, Newman2018}.
While the strongest constraints on dust obscured star formation can be provided by far-infrared data, a complete lack of emission lines in a spectrum is also highly constraining.
However, at $z>3$ candidate UMGs are faint and require significant amounts of exposure time, particularly for those which do not have emission lines.
Indeed while $<20$ candidate quiescent UMGs have been spectroscopically confirmed at $z>3$ \citep[\eg][Saracco et al. 2020, submitted]{Marsan2015, Marsan2017, Glazebrook2017, Schreiber2018b, Tanaka2019, Forrest2020, Valentino2020}, only a few have been confirmed to have no emission lines, and several have far-infrared detections indicating ongoing star formation \citep{Schreiber2018b, Schreiber2018a}.
All of these works have concluded that these galaxies formed their large stellar masses in short, intense bursts of star formation.

For this work we used the $H$- and $K$-band spectroscopic capabilities of Keck-MOSFIRE \citep{McLean2010, McLean2012} to follow up a set of massive candidates from XMM-VIDEO (Annunziatella et al., in prep) and COSMOS-UltraVISTA fields (Muzzin et al., in prep).
The survey aimed to characterize not only quiescent candidates, but also candidate UMGs in the blue star-forming region and red candidates consistent with large amounts of dust to inform the picture of galaxy formation in the early universe.
We present the photometric properties and selection method in Section \ref{Sec:TS}, the spectroscopic observations and data reduction in Section \ref{Sec:DR}, the redshift determination and galaxy property characterization in Section \ref{Sec:Analysis}, and then present the results in Section \ref{S:res} and the main conclusions in Section \ref{Sec:Conc}).
Throughout this work we assume a Chabrier IMF \citep{Chabrier2003} and a $\Lambda$CDM cosmology with $H_0=70$ km s$^{-1}$ Mpc$^{-1}$, $\Omega_M=0.3$, and $\Omega_\Lambda=0.7$.
We also adopt an AB magnitude system \citep{Oke1983}.

\section{Photometric Data and Target Selection} \label{Sec:TS}

\subsection{Photometric Catalogs}

\magazine\ targets were selected from parent photometric catalogs in the UltraVISTA DR1 \citep{Muzzin2013a}, UltraVISTA DR3 (Muzzin et al., in prep) and XMM-VIDEO (Annunziatella et al., in prep) fields.

The UltraVISTA survey \cite{McCracken2012} obtained deep near-infrared $Y$-, $J$-, $H$-, and $K$-band imaging over 1.62~deg$^2$ in the COSMOS field.
Combined with ancillary photometry from $0.15-24$~$\mu$m, this yielded 30 bandpasses allowing for accurate characterization of galaxy spectral energy distributions (SEDs) and precise photometric redshifts \citep[DR1 catalogs, 90\% completeness $K_s=23.4$ mag;][]{Muzzin2013a}.
Further deep imaging in the NIR yielded a set of ultra-deep strips over 0.84~deg$^2$ (DR3 catalogs; 90\% completeness $K_s=24.5$ mag; Marsan et al. 2020, in prep, Muzzin et al., in prep), which allowed for detection of massive, quiescent galaxies whose properties cannot be constrained from optical photometry alone.
Considerable further photometry was also added to obtain a total of 49 bandpasses in regions where there is overlap.
Critical to this work were Spitzer/IRAC mosaics using data from the S-COSMOS \citep{Sanders2007}, SPLASH \citep{Mehta2018} and SMUVS \citep{Ashby2018}, which resulted in IRAC imaging $\sim1.2$ mag deeper in the $3.6$ and $4.5~\mu$m bandpasses relative to DR1.

Similarly, the VISTA Deep Extragalactic Observations survey \citep[VIDEO;][]{Jarvis2013} acquired deep near-infrared $Z$-, $Y$-, $J$-, $H$-, and $K_s$-band imaging in three fields: the European Large Area Infrared Space Observatory S1 field, the extended Chandra Deep Field South, and the XMM-Newton Large Scale Structure (XMM) field.
Of the three VIDEO fields, we only present targets from the XMM field in this work, which also has substantial photometric observations in other optical and near-infrared bandpasses.
Catalogs were constructed using VIDEO DR4 data over 4.65~deg$^2$ with a $5\sigma$ depth of $K_s=23.8$ mag (Annunziatella, et al., in prep).
These include 22 photometric passbands ranging from u-band to IRAC $8.0~\mu$m including NIR observations from from the VIDEO survey \citep{Jarvis2013}, and deep IRAC data from the SERVS \citep{Mauduit2012} and DEEPDRILL (Lacy et al. 2020, submitted) surveys.

\subsection{Candidate UMG Targets}

We selected primary targets on the basis of photometric redshift ($3\leq$~\zphot~$\leq4$; roughly the window in which \Hbeta\ falls into the MOSFIRE $K$-band) and stellar mass (\logM~$>11.0$), as fit using EAzY \citep{Brammer2008} and FAST \citep{Kriek2009}
(see Table \ref{T:ps}).
We also made a $K$-band magnitude cut, $m_K<21.7$ AB.
According to the MOSFIRE ETC\footnote{https://www2.keck.hawaii.edu/inst/mosfire/etc.html} a 300 minute (18 ks) exposure would allow such an object to be detected with a continuum signal-to-noise ratio (SNR) of 5/pixel, sufficient to determine a redshift from a Balmer absorption feature such as \Hbeta.
In the $H$-band, the same detection threshold corresponds to $m_H<22.3$ AB.

In order to characterize a representative sample of UMGs at $3<z<4$, we aimed to observe galaxies in a variety of locations on the \UVJ\ diagram, potentially representing different stages of massive galaxy evolution.
The main goal of this survey was to obtain redshifts for UMGs, confirming that such a population exists at this epoch, and to subsequently characterize these galaxies using emission and absorption features.

Confirmation of these faint galaxies via detection of absorption lines is difficult under poor observing conditions.
As such, we selected a number of UMG candidates with blue SEDs, consistent with ongoing star formation, to observe when conditions were unfavorable.
These galaxies were more likely to have emission features which can more easily be detected.
In particular, we considered the amount of UV flux, strength of the Lyman and Balmer breaks, \UVJ\ colors, and photometrically-derived star-formation rate (SFR) for selection.
While the same mass and photometric redshift thresholds applied for selection of these targets, the magnitude cut was relaxed half a magnitude to $m_K<22.2$.

This selection method resulted in a sample of 33 candidate UMGs.
From this sample, 21 were observed and 16 were spectroscopically confirmed.
In this work we discuss these 16 objects, but a summary of the 5 remaining objects is presented in Appendix~\ref{A:Unconf}.
Note that none of these unconfirmed objects can be ruled out as a UMG at $3<z<4$ - three objects had insufficient signal-to-noise to confirm a redshift, one shows clear continuum but no defining features, and the fifth has a redshift $z=3.6$ but hosts a luminous quasar, making its mass uncertain.

\section{Observations and Data Reduction} \label{Sec:DR}

\subsection{Mask Construction \& Ancillary Targets}

MOSFIRE slits were arranged using the MOSFIRE Automatic GUI-based Mask Application (MAGMA) tool\footnote{http://www2.keck.hawaii.edu/inst/mosfire/magma.html}.
This software requires an input list of targets with priority weighting, and a range of allowable pointing positions and position angles (PA).
For each candidate UMG, \mbox{$K$-band} images were visually inspected to find nearby sources which could contaminate the spectra.
In several cases, PA was constrained and/or dither size altered to avoid spectral contamination.
We also constructed masks ahead of time for observing in poor conditions, targeting UMG candidates more likely to show emission features as discussed above.
Slit width for these masks was increased from $0.''7$ to $0.''9$.

All masks had one or two UMGs, 4-8 position alignment stars, and at least one star on a science slit, for use in monitoring the seeing and slit losses, as well as performing a telluric correction (see Section \ref{S:TC} and Appendix \ref{A:TC}).
Masks were designed to include as many bright, nearby galaxies with similar photometric redshifts as possible in order to characterize the local environment via emission line detection.
Ancillary targets were prioritized based on magnitude, SED shape, \zphot\ and redshift probability distribution from EAzY, and the stellar mass and SFR fit by FAST.

\subsection{Observing Plan}

Bright UMG candidates with red SEDs and \UVJ\ colors were prioritized for observation, as these were not expected to show strong emission features.
More time and better conditions were required to confirm redshifts for these objects through the detection of absorption features.
That said, classification as star-forming or quiescent via \UVJ\ colors is known to not be a pure indicator of the existence or lack of emission features, as spectroscopic determination of both redshift and emission line contamination can alter derived rest-frame colors significantly \citep[\eg][]{Schreiber2018b}.

We performed on-the-fly reductions throughout observing.
If a UMG showed clear emission lines in the 2D reduction, we moved on to the next target for the night.
However, if emission features from a UMG candidate were not clear after $2-3$ hours but stellar continuum was seen, we changed to a second mask on the same UMG.
This allowed us to spectroscopically confirm neighbor galaxies to the UMG and increase continuum SNR of the UMG for better characterization of absorption features.

Targets were initially observed with MOSFIRE \mbox{$K$-band} spectroscopy in order to detect any strong emission from \OIIIdub\ and \Hbeta$\lambda4861$ in either emission or absorption.
A single exception to this was COS-DR3-84674, which was observed in $H$-band, due to concerns that a small deviation from its low \zphot\ could cause lines to fall between the $H$- and \mbox{$K$-bands}.
As such, we targeted this object in \mbox{$H$-band} in an attempt to obtain an absorption line redshift or detection of \OII\ emission.

For half of the UMG targets, emission from \OIIIfive\ was detected in $K$-band with SNR~$>10$.
The remaining eight had continuum detections, but no clear emission features were seen.
While in three cases the redshift could be confirmed from absorption features, the remaining five UMGs without clear spectroscopic redshifts were followed up with $H$-band spectroscopy.
This allowed for detection of higher order Balmer absorption features, CaH and CaK absorption, and the 4000 \AA\ break, and also allowed for constraint of the spectral shape between the two bandpasses.

In total, 15 of the confirmed UMGs were observed for an average of 2 hours in $K$-band, and six confirmed UMGs were observed for an average of 5 hours in $H$-band.
All masks were observed using an ABBA dither pattern and \mbox{180 or 120 s} exposures for $K$- and \mbox{$H$-bands}, respectively.
A table of observed masks and UMG targets can be found in Appendix \ref{A:obs}.


	\begin{figure*}[t]
	\centerline{\includegraphics[width=0.9\textwidth,trim=0in 0in 0in 0in, clip=true]{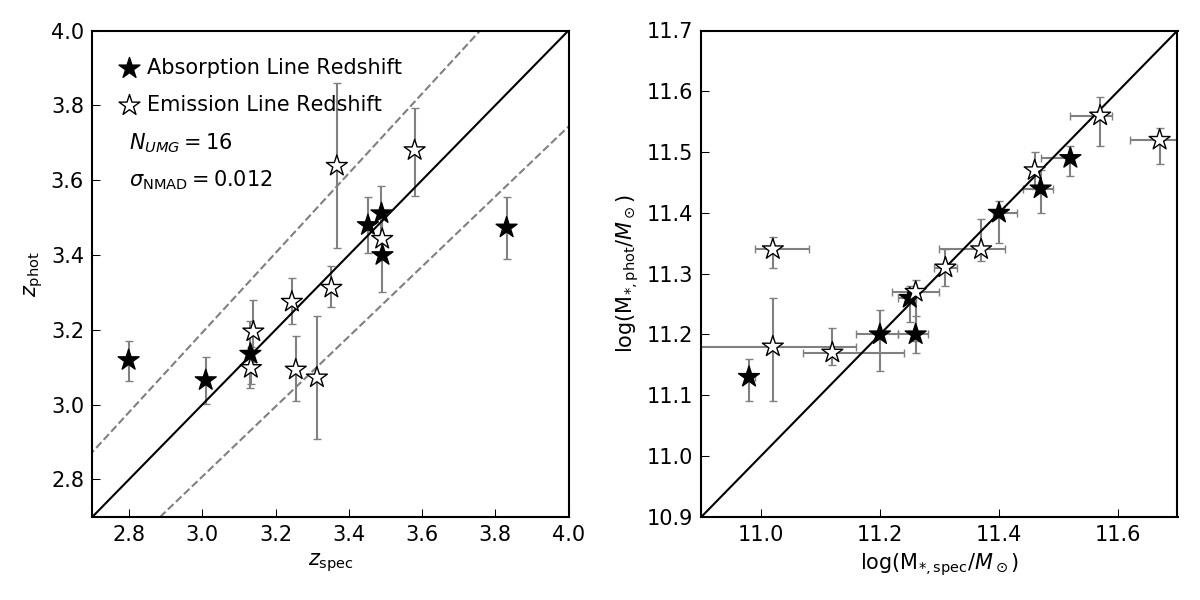}}
	\caption{Comparison of photometric and spectroscopic redshifts (left) and masses (right) for confirmed UMGs. Those with detected emission lines are shown as white stars, while UMGs confirmed via absorption features are in black. The black 1:1 line on the left is bounded by two dashed lines at 5$\sigma_{NMAD}$. Errors on \zspec\ are smaller than the marker size. See Table~\ref{T:ps} for numbers.}
	\label{fig:zpzs}
	\end{figure*}


\subsection{Spectral Reduction}

\subsubsection{Spectral Extraction}

Each mask was processed individually.
In cases where a galaxy was observed on multiple masks, spectra were only stacked after each one had been processed independently and a 1D spectrum extracted.
For these, the 1D spectra were coadded using inverse variance weighting based on the error spectra.

We began by running the MOSFIRE Data Reduction Pipeline (DRP)\footnote{https://github.com/Mosfire-DataReductionPipeline/MosfireDRP}, which constructs a pixel flat image, identifies slits, removes thermal contamination (\mbox{$K$-band}), performs wavelength calibration using sky lines, Neon arc lamps, and Argon arc lamps, removes sky background, and rectifies the spectra, yielding a reduced 2D spectrum for a given mask.

While the DRP also offers an option to extract a 1D spectrum, we wrote our own custom code to perform optimal extraction \citep{Horne1986}.
Briefly, this involves collapsing the 2D spectrum for each slit along the wavelength axis and fitting a Gaussian to the summed fluxes, with a prior on the predicted position of the object on each slit.
This Gaussian is then used as a weight when collapsing the spectrum along the spatial axis to obtain the 1D spectrum (and errors) for each object\footnote{The official MOSFIRE DRP appears to have not yet been updated to do this - https://github.com/Keck-DataReductionPipelines/MosfireDRP/issues/126}.

\subsubsection{Sky Line Identification and Telluric Corrections} \label{S:TC}

Strong sky lines are a serious contaminant in the NIR, particularly in the $H$-band, and while the DRP attempts to subtract off their flux, large flux variances remain at these locations even after subtraction.
For each wavelength in a reduced mask, we calculate the median variance per pixel on sky in the spatial direction, and derive thresholds above which wavelengths are determined to be contaminated by either weak or strong sky lines.
Strong sky lines are masked, while weak lines are interpolated over.
More details on this process are given in Appendix~\ref{A:SL}.
For plotting purposes, we filled masked wavelengths by using an inverse variance weighted average of nearby wavelengths which are not associated with sky lines.

A custom Python code was also written to perform telluric correction using stars observed in mask slits.
The method is briefly described here, while a more detailed description, as well as a comparison to standard star telluric correction is provided in Appendix~\ref{A:TC}.
Briefly, we used the spectra of stars observed on the slits within each mask to obtain a telluric correction by fitting PHOENIX stellar models \citep{Husser2013} to the NIR photometry of each star to obtain a model spectrum.
The ratio of the model to the observed stellar spectrum was taken to be the telluric correction and was applied to all slits on a given mask.
When multiple stars were observed on a mask, the resultant telluric corrections were coadded using inverse variance weighting of the spectral errors.
As the subsequent fitting procedure scales the spectra to match the photometry, we were here only concerned with the correction of the spectra in terms of shape.

\begin{table*}
  \caption{Properties of UMG candidates derived from fits to the combined spectroscopy and photometry (see Section~\ref{Sec:Analysis}), ordered by stellar mass. Errors on the photometric masses assume $z=$~\zphot. Star formation rates are taken from the best-fit SFH to the photometric and spectroscopic data, averaged over the previous 10 Myr. The formation and quenching times, $t_{50}$ and $t_{q}$, are given as lookback times from the spectroscopic redshift. Objects whose redshift or UMG nature were not confirmed are listed at the bottom (XMM-VID3-3941 is a quasar).}
  \resizebox{\textwidth}{!}{
    \begin{tabular}{ lcccccccc }
      UMG & \zphot & \zspec & {$\log(M_{*, \rm{phot}}/\rm{M}_\odot)$} & {$\log(M_{*, \rm{spec}}/\rm{M}_\odot)$} & SFR$_{10} (M_\odot/ \textrm{yr})$ & $t_{50}$ (Gyr) & $t_{q}$ (Gyr) & $A_\textrm{V}$\\
      \hline \hline
     COS-DR3-202019 & $3.10^{+0.06}_{-0.04}$    & $3.1326^{+0.0021}_{-0.0011}$   & $11.52^{+0.02}_{-0.04}$    & $11.67^{+0.04}_{-0.05}$    & $82.4^{+10.4}_{-58.2}$ 	& $0.76^{+0.20}_{-0.05}$  & $0.40^{+0.34}_{-0.40}$   & $0.6^{+0.0}_{-0.3}$  \\
      XMM-VID3-2293   & $3.07^{+0.17}_{-0.16}$   & $3.3132^{+0.0009}_{-0.0007}$    & $11.56^{+0.03}_{-0.05}$    & $11.57^{+0.02}_{-0.05}$    & $29.5^{+43.0}_{-22.3}$  & $1.12^{+0.40}_{-0.37}$  & $0.32^{+0.98}_{-0.32}$   & $0.8^{+0.2}_{-0.4}$  \\      
      XMM-VID1-2075   & $3.48^{+0.08}_{-0.07}$    & $3.4520^{+0.0014}_{-0.0017}$   & $11.49^{+0.02}_{-0.03}$    & $11.52^{+0.00}_{-0.05}$    & $0.0^{+1.0}_{-0.0}$	& $0.48^{+0.12}_{-0.07}$  & $0.42^{+0.12}_{-0.13}$   & $0.4^{+0.0}_{-0.2}$  \\
      XMM-VID3-1120   & $3.40^{+0.12}_{-0.10}$    & $3.4919^{+0.0018}_{-0.0029}$   & $11.44^{+0.03}_{-0.04}$    & $11.47^{+0.02}_{-0.03}$    & $0.0^{+0.1}_{-0.0}$	& $0.72^{+0.10}_{-0.07}$  & $0.51^{+0.14}_{-0.22}$   & $0.0^{+0.0}_{-0.0}$  \\
     COS-DR3-160748 & $3.35^{+0.02}_{-0.09}$    & $3.3524^{+0.0008}_{-0.0006}$   & $11.47^{+0.03}_{-0.03}$    & $11.46^{+0.01}_{-0.08}$    & $3.1^{+0.2}_{-0.2}$	& $0.37^{+0.10}_{-0.05}$  & $0.32^{+0.10}_{-0.05}$   & $0.0^{+0.0}_{-0.0}$  \\
     COS-DR3-201999 & $3.14^{+0.09}_{-0.09}$    & $3.1313^{+0.0014}_{-0.0012}$   & $11.40^{+0.02}_{-0.05}$    & $11.40^{+0.03}_{-0.01}$    & $1.3^{+0.3}_{-1.2}$	& $0.52^{+0.10}_{-0.08}$  & $0.32^{+0.10}_{-0.05}$   & $0.4^{+0.1}_{-0.0}$  \\
     COS-DR3-179370 & $3.14^{+0.72}_{-0.28}$    & $3.3673^{+0.0010}_{-0.0007}$   & $11.34^{+0.05}_{-0.02}$    & $11.37^{+0.04}_{-0.07}$    & $3.2^{+9.3}_{-2.4}$        & $1.20^{+0.10}_{-0.21}$  & $0.23^{+0.75}_{-0.14}$   & $0.8^{+0.2}_{-0.2}$  \\
     COS-DR3-195616 & $3.09^{+0.09}_{-0.08}$    & $3.2552^{+0.0012}_{-0.0009}$   & $11.31^{+0.03}_{-0.03}$    & $11.31^{+0.02}_{-0.02}$    & $21.7^{+2.8}_{-12.3}$    & $0.62^{+0.15}_{-0.05}$  & $0.32^{+0.21}_{-0.05}$   & $0.6^{+0.0}_{-0.2}$  \\
     COS-DR3-208070 & $3.44^{+0.06}_{-0.05}$    & $3.4912^{+0.0011}_{-0.0012}$   & $11.27^{+0.02}_{-0.10}$    & $11.26^{+0.04}_{-0.04}$    & $125^{+44.8}_{-60.5}$   & $0.30^{+0.13}_{-0.07}$  & $0.23^{+0.13}_{-0.23}$   & $1.0^{+0.1}_{-0.1}$  \\
      XMM-VID3-2457   & $3.51^{+0.07}_{-0.07}$    & $3.4892^{+0.0032}_{-0.0024}$   & $11.20^{+0.03}_{-0.01}$    & $11.26^{+0.02}_{-0.03}$    & $0.0^{+0.4}_{-0.0}$ 	& $0.40^{+0.13}_{-0.12}$  & $0.21^{+0.10}_{-0.09}$   & $0.5^{+0.2}_{-0.1}$  \\
      COS-DR3-84674  & $3.06^{+0.06}_{-0.06}$    & $3.0094^{+0.0015}_{-0.0011}$   & $11.26^{+0.02}_{-0.04}$    & $11.25^{+0.01}_{-0.02}$    & $1.1^{+0.6}_{-0.8}$ 	& $0.50^{+0.11}_{-0.06}$  & $0.32^{+0.14}_{-0.05}$   & $0.5^{+0.0}_{-0.1}$  \\
     COS-DR1-113684 & $3.47^{+0.08}_{-0.08}$    & $3.8309^{+0.0014}_{-0.0020}$   & $11.20^{+0.04}_{-0.06}$    & $11.20^{+0.03}_{-0.04}$    & $15.7^{+0.2}_{-1.1}$ 	& $0.45^{+0.10}_{-0.05}$  & $0.32^{+0.11}_{-0.05}$   & $0.0^{+0.0}_{-0.0}$  \\
     COS-DR3-131925 & $3.20^{+0.08}_{-0.08}$    & $3.1393^{+0.0008}_{-0.0013}$   & $11.17^{+0.04}_{-0.02}$    & $11.12^{+0.12}_{-0.05}$    & $83.9^{+152}_{-83.5}$   & $0.32^{+0.34}_{-0.15}$  & $0.05^{+0.29}_{-0.05}$   & $0.5^{+0.2}_{-0.3}$  \\
      COS-DR3-226441& $3.27^{+0.06}_{-0.06}$    & $3.2446^{+0.0014}_{-0.0012}$   & $11.34^{+0.02}_{-0.03}$    & $11.02^{+0.06}_{-0.03}$    & $10.9^{+8.8}_{-6.2}$ 	& $0.43^{+0.14}_{-0.06}$  & $0.32^{+0.14}_{-0.32}$   & $0.5^{+0.2}_{-0.1}$  \\
     XMM-VID1-2399    & $3.68^{+0.11}_{-0.12}$    & $3.5798^{+0.0010}_{-0.0009}$   & $11.18^{+0.08}_{-0.09}$    & $11.02^{+0.14}_{-0.13}$    & $504^{+20.9}_{-503}$    & $0.06^{+0.23}_{-0.05}$  & $0.01^{+0.17}_{-0.01}$   & $1.8^{+0.0}_{-0.7}$  \\
     COS-DR3-111740  & $3.12^{+0.05}_{-0.06}$   & $2.7988^{+0.0013}_{-0.0011}$    & $11.13^{+0.03}_{-0.04}$    & $10.98^{+0.01}_{-0.00}$    & $0.7^{+0.2}_{-0.6}$  	& $0.33^{+0.10}_{-0.05}$  & $0.26^{+0.10}_{-0.07}$   & $0.0^{+0.0}_{-0.0}$  \\
            \hline
     COS-DR1-79837 	  & $3.30^{+0.19}_{-0.20}$   & ---    & $11.92^{+0.06}_{-0.05}$    & ---    & ---  	  & ---  & ---    \\
     XMM-VID3-3941   & $3.04^{+0.12}_{-0.11}$   & $3.5901^{+0.0010}_{-0.0008}$    & $11.71^{+0.07}_{-0.08}$    & ---    & ---  	  & ---  & ---    \\
     XMM-VID1-2761   & $3.60^{+0.18}_{-0.17}$   & ---    & $11.52^{+0.02}_{-0.03}$    & ---    & ---  	  & ---  & ---    \\
     COS-DR1-258857 & $3.27^{+0.16}_{-0.17}$   & ---    & $11.31^{+0.04}_{-0.02}$    & ---    & ---  	  & ---  & ---    \\
     XMM-VID2-270     & $3.23^{+0.16}_{-0.11}$   & ---    & $11.28^{+0.07}_{-0.02}$    & ---    & ---  	  & ---  & ---    \\
         \hline
    \end{tabular}}
  \label{T:ps}
\end{table*}

\section{Analysis}\label{Sec:Analysis}

In order to determine spectroscopic redshifts we used a combination of the programs \texttt{slinefit}\footnote{https://github.com/cschreib/slinefit} and FAST++\footnote{https://github.com/cschreib/fastpp} \citep{Schreiber2018b}, a variation of the popular FAST program \citep{Kriek2011} which fits to both photometric and spectroscopic data.
FAST++ allows for different functional star formation histories (SFHs), scaling of spectra to match photometry, and fitting of spectra with different wavelength resolutions, among many other features.
The program \texttt{slinefit} fits spectra with emission (and absorption) features across a range of redshifts using gaussians of varying widths and amplitudes.
In this work, we are concerned with fitting \OIIdub\ (line ratio fixed to unity), \Hbeta, and \OIIIdub\ (line ratio fixed to 0.3) for the UMGs and galaxies at similar redshifts.
The program also allows for continuum fitting to enable more accurate characterization of line properties such as equivalent widths.


\begin{figure*}[tp]
	\centering{\includegraphics[width=\textwidth,trim=0in 0.25in 0in 0in, clip=true]{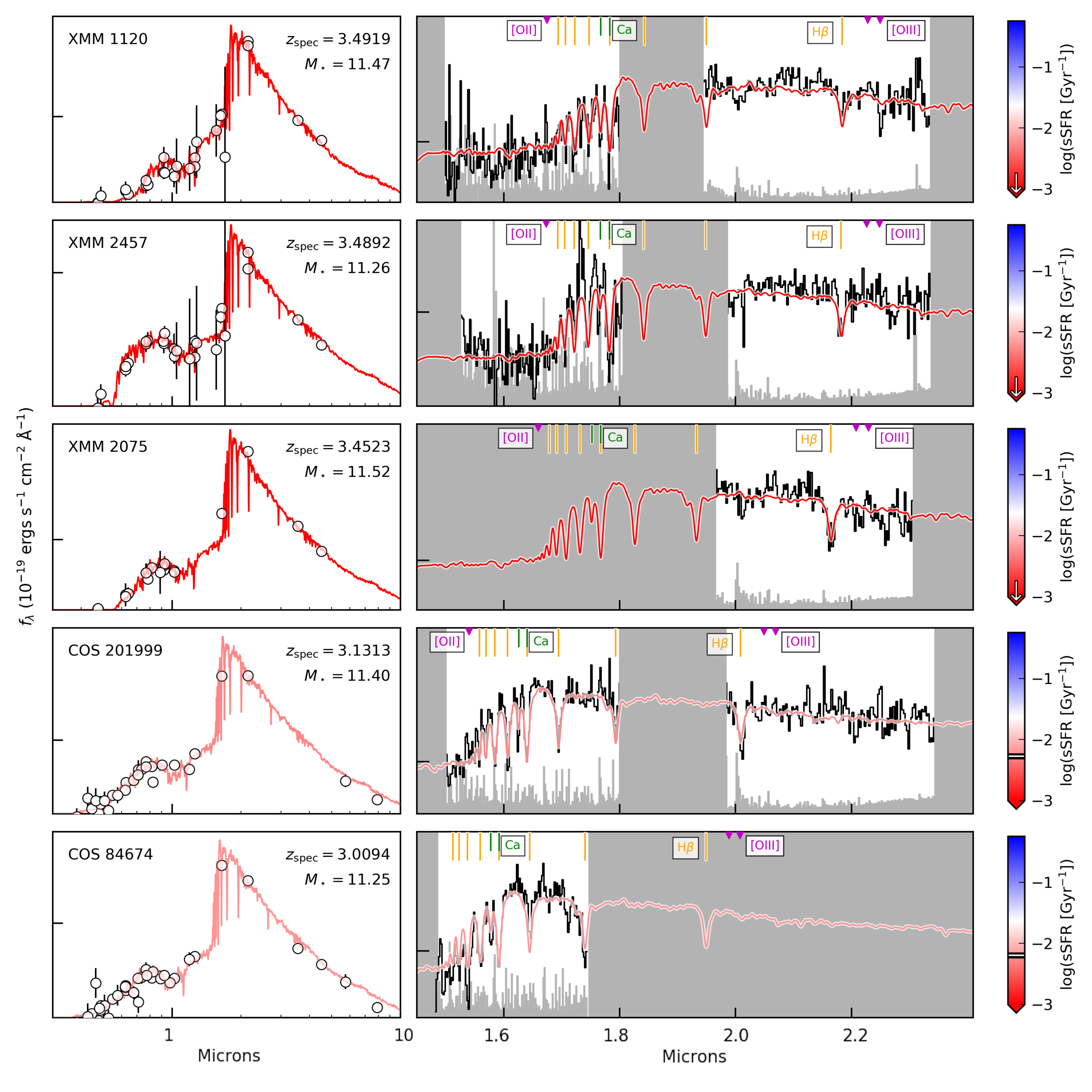}}
	\caption{The photometry (left) and spectroscopy (right) for each spectroscopically confirmed UMG. The best-fit template to the combined photometry and spectroscopy is shown colored by the template's sSFR averaged over the last 10 Myr, which is indicated by a black and white line/arrow on the colorbar to the right. Spectra are plotted in bins of 12 pixels, weighted by inverse variance, with errors shown in gray.  Wavelengths corresponding to prominent spectral features from hydrogen (orange), oxygen (magenta), and calcium (green) are also labeled. The dash on each y-axis corresponds to a flux density of \mbox{$5\times10^{-19}$~ergs s$^{-1}$ cm$^{-2}$ \AA$^{-1}$.}}
	\label{fig:uvj_spec_a}
\end{figure*}



\begin{figure*}[tp]
	\centering{\includegraphics[width=\textwidth,trim=0in 0in 0.25in 0in, clip=true]{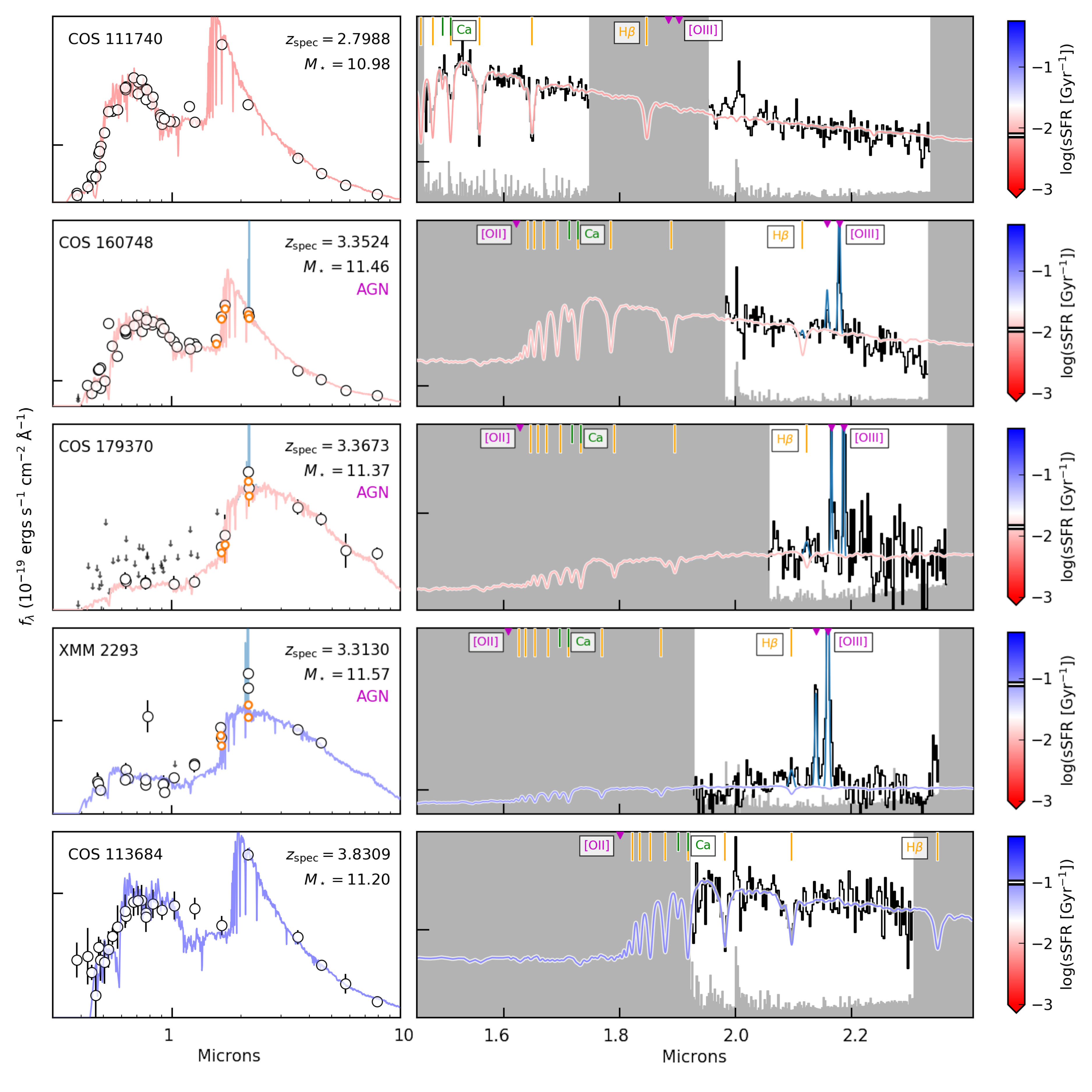}}
	\caption{Same as Figure \ref{fig:uvj_spec_a}. For UMGs with strongly detected emission features, the best-fit model to the observed emission-lines is overlaid in teal. Orange points indicate the photometry corrected for the flux of emission lines (see Section~\ref{S:ELG}), while photometric points with SNR~$<3$ are shown as downward facing arrows. }
	\label{fig:uvj_spec_b}
\end{figure*}



\begin{figure*}[tp]
	\centering{\includegraphics[width=\textwidth,trim=0in 1in 3.05in 0in, clip=true]{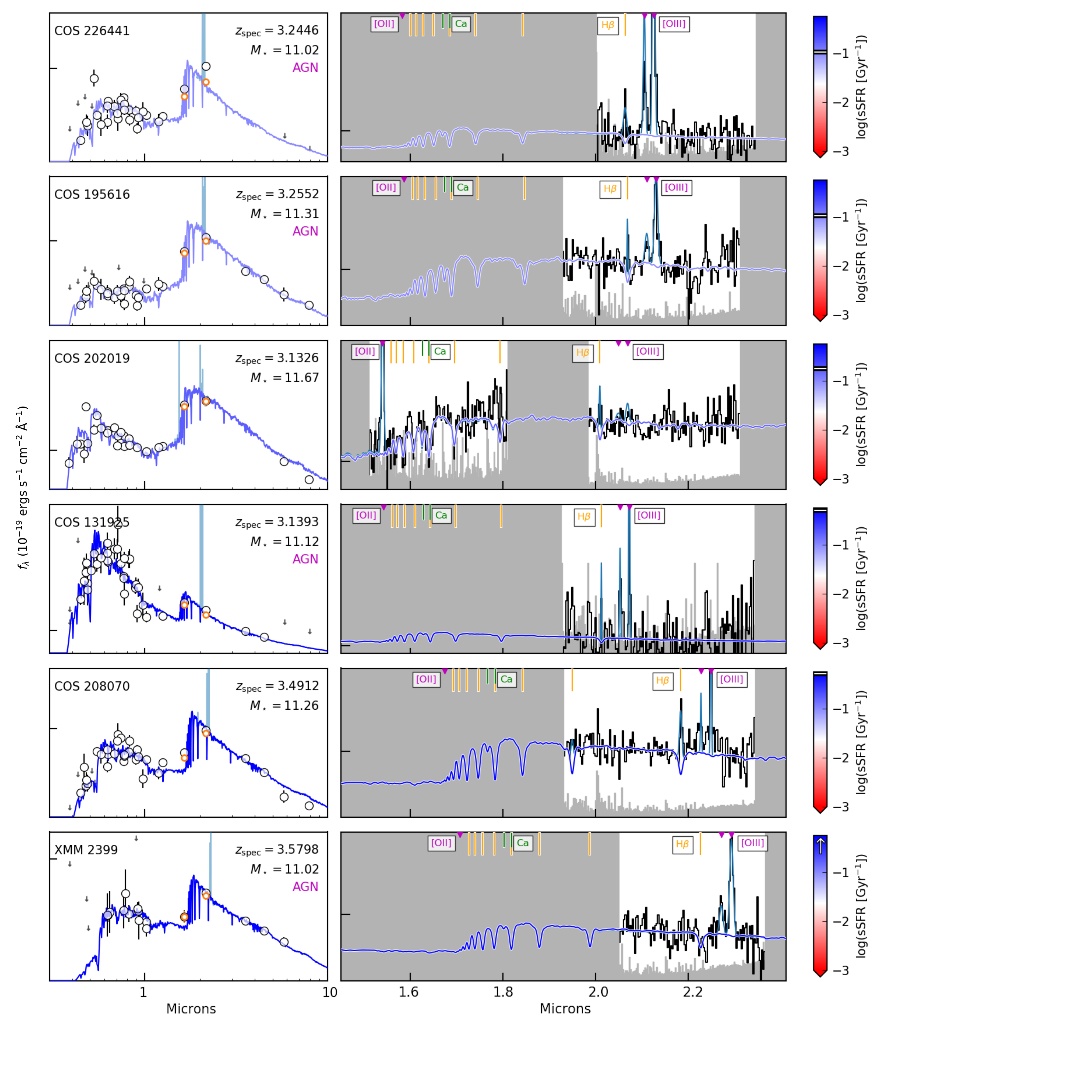}}
	\caption{Same as Figures \ref{fig:uvj_spec_a} and \ref{fig:uvj_spec_b}.}
	\label{fig:uvj_spec_c}
\end{figure*}


\subsection{Fitting Photometric and Spectroscopic Data with FAST++}\label{S:fpp}

In fitting with FAST++ we use stellar population models from \citet[][BC03]{Bruzual2003} with a starburst dust law \citep{Calzetti2000}.
We used the following grids for various input parameters:
\begin{eqnarray*}
&7<\log{(\tau/\textrm{yr})}<10, & \Delta \tau=0.1\\
&7<\log{(\textrm{age}/\textrm{yr})}<9.25, & \Delta \textrm{age} = 0.05\\
&0<A_V/\textrm{mag}<4, & \Delta A_V=0.1
\end{eqnarray*}

In general, metallicity was fixed to $Z=0.02=Z_\odot$, as several studies have shown this to be broadly correct for high-redshift massive galaxies \citep[\eg][Jafariyazani et al., in prep; but also see Saracco et al. 2020 for a spectroscopic measurement of super-solar stellar metallicity for one of our galaxies, COS-DR3-160748]{Belli2019, Kriek2019, Estrada-Carpenter2019}.
However, we did test metallicities of $0.4~Z_\odot$ and $2.5~Z_\odot$ as well, which yielded only slightly worse fits to the data, and not significantly different galaxy properties.

Velocity dispersion, which must be fixed for each run of FAST++, was set to 300 km/s. 
While velocity dispersions of similar galaxies at these redshifts and masses are rare, \citet{Tanaka2019} calculate $\sigma=268\pm59$ km/s for a galaxy at $z=4.01$ with \logM~11.
Their Figure 4 also presents published velocity dispersions for massive quiescent galaxies at $z>1.5$, which range from $\sim100$ to $\sim500$ km/s and show a slight positive correlation with stellar mass.
A more in-depth analysis of velocity dispersions is beyond the scope of this work.

Characterizing the SFHs of these galaxies addresses directly the potential identification of the population from which they descended.
Additionally, the maximum SFR and the time at which these galaxies quenched inform models of star formation in the young cosmos, and can provide important insights into the evolution of massive galaxies in the local universe.
Here we test both a delayed exponentially declining SFH\footnote{SFR($t$)~$\propto t e^{-t/\tau}$} (using the parameter grid above) 
as well as a double exponential SFH:
\begin{eqnarray}
\rm{SFR}_{\rm base}(t) \propto
\begin{cases}
  \rm{e}^{(t_{\rm burst}-t)/\tau_{\rm rise}},& \text{for } t>t_{\rm burst}\\
  \rm{e}^{(t-t_{\rm burst})/\tau_{\rm decl}},& \text{for } t\leq t_{\rm burst}
\end{cases}\\
\rm{SFR}(t) = \rm{SFR}_{\rm base}(t) \times
\begin{cases}
  1,& \text{for } t>t_{\rm free}\\
  R_{\rm SFR},& \text{for } t\leq t_{\rm free}
\end{cases}
\end{eqnarray}

This SFH has been used in other works studying massive galaxies at high redshift \citep{Schreiber2018b, Valentino2020, Forrest2020}, and we adopt the same parameter grid described therein:
\begin{eqnarray*}
& 7.0<\log(t_{\rm burst}/\rm{yr})<9.2, & \Delta \log(t_{\rm burst}/\rm{yr})=0.05\\
& 7.0<\log(\tau_{\rm rise}/\rm{yr})<9.5, & \Delta \log(\tau_{\rm rise}/\rm{yr})=0.1\\
& 7<\log(\tau_{\rm decl}/\rm{yr})<9.5, & \Delta \log(\tau_{\rm decl}/\rm{yr})=0.1\\
& 7<\log(t_{\rm free}/\rm{yr})<8.5, & \Delta \log(t_{\rm free}/\rm{yr})=0.5\\
& -2.0<\log(R_{\rm SFR})<5.0, & \Delta \log(R_{\rm SFR})=0.2
\end{eqnarray*}

We emphasize that while the exact shape of the best-fit SFH is model-dependent and can vary depending upon assumptions of the parametric form, general characteristics, such as when the majority of stars were formed and when the galaxy quenched, can be well constrained provided that the SFH inputs are sufficiently flexible and appropriate for the object in question.
In particular, \cite{Belli2019} show that various SFH parameterizations yield similar values of star forming timescale and sSFR.
We find the same for the \mbox{delayed-$\tau$} and double-exponential models described here, and will use values from the double-exponential fits in the remainder of the work.

\subsection{Redshift Determination and Template Fitting for Absorption Line UMGs} 

When stellar continuum is detected with no clear emission features, we use FAST++ to determine the redshift.
For galaxies with spectra in both the $H$- and $K$-bands, we fit the spectrum of each band combined with the photometry independently in order to remove differences in flux scaling between the spectra when matching the photometry.
The two spectra are then scaled relative to each other and refit in combination with the photometry.
This relative rescaling between bands can be up to 15\% depending on the accuracy of the flux calibration factor in the slit star telluric correction.

For redshift fitting, we fit to the combined photometry and spectroscopy of the galaxy in question.
We do three runs per galaxy, the first with the model redshifts ranging $2.5<z_\textrm{model}<4.0$ with $\Delta z=0.01$ to obtain an initial fit, \zspec$_{, 0}$, and a second run with model redshifts \zspec$_{, 0}-0.02<z_\textrm{model}<$\zspec$_{, 0}+0.02$ and $\Delta z=0.0001$ to obtain a final \zspec.
While these two runs are completed with a delayed-$\tau$ SFH, the third run uses a double-exponential SFH with the possibility of additional low levels of late-time star formation \citep[see Section \ref{S:fpp}, as well as][]{Schreiber2018a, Schreiber2018b}.
This SFH is more computationally intensive and thus having the redshift fixed to \zspec\ improves runtime significantly.
More details on the fitting input parameters are given in Section \ref{S:fpp}.

We subsequently run \texttt{slinefit} on the spectrum, with the best-fit model from FAST++ used as the stellar continuum template.
This allows for measurement or limits on the strength of weak emission features in these galaxies.
This does not affect the redshift determination at all, only the line measurements.
Final line fluxes and equivalent widths are presented in Table~\ref{T:spec}.

\subsection{Redshift Determination and Template Fitting for Emission Line UMGs} \label{S:ELG}

For galaxies where emission features are detected we use \texttt{slinefit} to determine the spectroscopic redshift, \zspec.
In this first fit we use a set of high resolution galaxy spectral templates based on those provided in EAzY \citep{Brammer2008} to generate a stellar continuum on top of which to fit lines.
The resultant \zspec\ is then fixed when fitting with FAST++.

Before fitting these galaxies with FAST++ however, we must correct the observed photometry for the effects of the emission lines, as the template libraries do not include them.
To do this, the observed magnitude is converted to a total flux by multiplying the average flux density by the width of the bandpass.
In the $K$-band, the photometric flux is reduced by the sum of the emission line fluxes,
\small
\begin{eqnarray}
F_{K, corr} = F_{K, obs} - \bigg{(}F_{\textrm{H}\beta} + F_{\textrm{[O III]}\lambda4959} + F_{\textrm{[OIII]}\lambda5007}\bigg{)}
\end{eqnarray}
\normalsize
where the line fluxes are taken from the model output of \texttt{slinefit}.

In most cases, we do not have $H$-band spectra, and so we must estimate the effects of photometric contamination by \OII.
We do this by making the following assumptions:\\
1) the \OII\ line flux is due to a combination of star formation and AGN activity, 
\begin{eqnarray}
F_{\textrm{[OII], total}} = F_{\textrm{[OII], SFR}} + F_{\textrm{[OII], AGN}} \label{E:fo2tot}
\end{eqnarray}
2) case B recombination can be used to obtain a SFR from \Hbeta\ \citep{Kennicutt1998b,Moustakas2006}, 
\begin{eqnarray}
\rm{SFR}_{\textrm{H}\alpha} &=& 4.65\times10^{-42}~L_{\textrm{H}\alpha}\\ 
L_{\textrm{H}\alpha} &=& 2.86~L_{\textrm{H}\beta}\\
\rm{SFR}_{\textrm{H}\beta} &=& 1.33 \times 10^{-41}~L_{\textrm{H}\beta},
\end{eqnarray}
where $L_i$ is in units of \mbox{erg s$^{-1}$} and SFRs are in units of \mbox{M$_\odot$ yr$^{-1}$}.\\
3) the SFR derived from \Hbeta\ is equal to that derived from \OII, 
\begin{eqnarray}
\rm{SFR}_{\textrm{H}\beta} &=& \rm{SFR}_\textrm{[OII]}\\
\rm{SFR}_\textrm{[OII]} &=& 4.14 \times 10^{-42}~L_{\textrm{[OII], SFR}}
\end{eqnarray}
4) the ratio of \OIIIfive\ to \OII\ due to AGN activity is given by typical values \citep{Silverman2009}
\begin{eqnarray}
L_\textrm{[OII], AGN} / L_{\textrm{[OIII]}\lambda5007\textrm{, AGN}} = 0.21
\end{eqnarray}
Tying all these pieces together leaves us with an \OII\ line flux estimation as a function of the \Hbeta\ and \OIIIfive\ line fluxes observed in the $K$-band spectra. Equation \ref{E:fo2tot} then becomes
\begin{eqnarray}
F_{\textrm{[OII], total}} = 3.21~F_{\textrm{H}\beta} + 0.21~F_{\textrm{[OIII]}\lambda5007}
\end{eqnarray} 
and we can correct the $H$-band photometry accordingly
\begin{eqnarray}
F_{H, corr} = F_{H, obs}  - F_{\textrm{[OII]}}.
\end{eqnarray}

We note that while the various assumptions made in these calculations may not be accurate in every case, the typical effect of the $H$-band photometric correction is only a few percent, and a factor of two in the correction does not change the resultant best-fit parameters for the equations significantly, particularly since the spectroscopic redshift is known.
Furthermore, any effects of dust would only decrease the strength of \OII\ relative to \Hbeta\ and \OIII\, thus any correction would be smaller than calculated here.

Next we use FAST++ to fit the galaxy with the redshift fixed to that obtained with \texttt{slinefit}.
For this process we input the corrected photometry, as well as spectra with the emission line regions masked, again due to the fact that the template libraries in use do not include emission features.
As above, galaxies with spectra in both the $H$- and $K$-bands are fit with each spectral band independently and then scaled to match the entire spectrum. 
Once a best-fit template is obtained from FAST++ we then repeat the entire process, \ie\ we rerun \texttt{slinefit} with the best fit FAST++ template input as the continuum model, use the resulting revised line strength estimates to recorrect the photometry, and refit with FAST++.
In all cases, the new line flux and equivalent width values are similar to the initial values, with differences of $<15\%$.

The spectroscopic and photometric redshifts agree quite well (see Figure~\ref{fig:zpzs} and Table~\ref{T:ps}), with $\sigma_{\rm NMAD}=0.012$ for the UMGs.
As defined in \citet{Brammer2008},
\begin{eqnarray*}
\sigma_{\rm NMAD}=1.48 \times {\rm median} \bigg{(} \bigg{\lvert} \frac{\Delta z - {\rm median}(\Delta z)}{1+z_{\rm spec}} \bigg{\rvert}\bigg{)}
\end{eqnarray*}

Similarly, the final stellar masses agree very well with those derived from the photometric catalogs alone. 
The largest discrepancies occur for galaxies with strong emission line contamination and one with a large redshift discrepancy.
Best-fit spectral models are shown plotted with the observed photometry and spectra in Figures~\ref{fig:uvj_spec_a},  \ref{fig:uvj_spec_b}, and \ref{fig:uvj_spec_c}.
These figures also show the line fits and effects of photometric corrections for line emission.

\begin{sidewaystable}
  \centering
  \caption{Properties of confirmed UMGs derived from fits to the spectra. Fluxes are in units of $10^{-18}$ erg/s/cm$^2$. Equivalent widths are in rest-frame \AA ngstroms and in the case of \Hbeta\ is calculated after taking stellar continuum absorption into account. Negative values indicate emission.}
  \resizebox{\textwidth}{!}{
      \begin{tabular}{ lcccccccccccc }
      UMG & $f_{\rm{[OII]}}$ & $EW_0$(\OII) & SFR$_{\rm{[OII]}}$ & \Dfour & \EWhd & $f_{H\beta}$ & $EW_0$(H$\beta$) & SFR$_{H\beta}$\footnote{This calculation assumes all flux from the line is due to star formation. This may be considered an upper limit for those galaxies which host AGN.}  &  $f_{\rm [OIII]\lambda4959}$ & $EW_0$(\OIIIfour) & $f_{\rm [OIII]\lambda5007}$ & $EW_0$(\OIIIfive) \\
      \hline \hline
     COS-DR3-202019 &	$162\pm7$ & $-30.1\pm3.4$ & $57.6\pm2.6$ & $1.15 \pm 0.10$ & $8.3 \pm 1.7$ & 	$32.0\pm11.6$ & $-6.5\pm2.4$ & $36.5\pm13.3$ & $9.9\pm2.4$ & $-2.0\pm0.5$ & $33.2\pm7.9$ & $-6.7\pm1.6$\\
     XMM-VID3-2293\footnote{$f_{\rm [OIII]\lambda5007} / f_{H\beta} > 6$, consistent with AGN at these large stellar masses.} & 	$-$ & 		$-$ & 		$-$ & 		$-$ & 			$-$ & 		$44.3\pm6.9$ & $-19.3\pm3.3$ & $57.9\pm9.1$ & $95.8\pm2.0$ & $-40.8\pm3.0$ & $320\pm7$ & $-136\pm10$\\
     XMM-VID1-2075 & 	$-$ & 		$-$ & 		$-$ & 		$-$ & 			$-$ & 		$6.3\pm1.5$ & $-1.4\pm0.3$ & $9.1\pm2.2$ & $1.2\pm1.5$ & $-0.3\pm0.3$ & $4.0\pm4.9$ & $-0.9\pm1.1$\\
     XMM-VID3-1120 & 	$6.0\pm5.0$ & 	$-1.4\pm1.2$ & $2.8\pm2.3$ & 	$-$ & 			$-$ & 		$2.8\pm3.3$ & $-1.5\pm0.9$ & $4.2\pm4.9$ & $-4.4\pm1.2$ & $1.1\pm0.3$ & $-14.6\pm4.1$ & $3.8\pm1.1$\\
     COS-DR3-160748$^b$ &	$-$ & 		$-$ & 		$-$ & 		$-$ & 			$-$ & 		$17.8\pm7.2$ & $-5.7\pm0.9$ & $23.9\pm9.7$ & $47.6\pm2.1$ & $-5.9\pm0.3$ & $158.7\pm7.1$ & $-19.8\pm1.0$\\
     COS-DR3-201999$^b$ & 	$17.8\pm3.5$ & $-4.2\pm0.9$ & $6.3\pm1.2$ & $1.21 \pm 0.05$ & $9.1 \pm 0.8$ & 	$1.3\pm2.2$ & $-0.3\pm2.6$ & $1.5\pm2.5$ & $3.4\pm1.0$ & $-0.8\pm0.2$ & $11.3\pm3.2$ & $-2.8\pm0.8$\\
     COS-DR3-179370 & 	$-$ & 		$-$ & 		$-$ & 		$-$ & 			$-$ & 		$10.8\pm3.5$ & $-8.8\pm2.9$ & $14.6\pm4.7$ & $19.2\pm0.7$ & $-15.5\pm1.1$ & $64.0\pm2.2$ & $-51.6\pm3.5$\\
     COS-DR3-195616$^b$ & 	$-$ & 		$-$ & 		$-$ & 		$-$ & 			$-$ & 		$12.4\pm2.2$ & $-5.5\pm1.0$ & $15.5\pm2.8$ & $23.1\pm1.3$ & $-10.0\pm0.7$ & $77.1\pm4.5$ & $-33.4\pm2.3$\\
     COS-DR3-208070 & 	$-$ & 		$-$ &		$-$ & 		$-$ & 			$-$ & 		$32.3\pm2.8$ & $-15.6\pm1.4$ & $47.8\pm4.1$ & $12.4\pm0.6$ & $-5.8\pm0.3$ & $41.5\pm2.1$ & $-19.4\pm1.2$\\      
     XMM-VID3-2457 & 	$5.3\pm4.1$ & 	$-2.0\pm1.6$ & $2.5\pm1.9$ & 	$-$ & 			$-$ & 		$1.5\pm2.6$ & $-0.6\pm1.1$ & $2.2\pm3.8$ & $3.1\pm1.0$ & $-1.3\pm0.4$ & $10.5\pm3.3$ & $-4.3\pm1.4$\\
      COS-DR3-84674 & 	$6.2\pm2.1$ & $-2.0\pm3.6$ & $2.0\pm0.7$ & 	$1.47 \pm 0.06$ & $7.2 \pm 0.7$ & 	$-$ & 		$-$ & 		$-$ & 		$-$ & 		$-$ & 		$-$ & 		$-$\\
     COS-DR1-113684 & 	$-$ & 		$-$ & 		$-$ & 		$-$ & 			$4.8 \pm 2.1$ & $-$ & 		$-$ & 		$-$ & 		$-$ & 		$-$ & 		$-$ & 		$-$\\
     COS-DR3-131925$^b$ & 	$-$ & 		$-$ & 		$-$ & 		$-$ & 			$-$ & 		$76.5\pm52.6$ & $-19.4\pm17.0$ & $87.8\pm60.3$ & $156\pm12$ & $-38.9\pm12.9$ & $518\pm40$ & $-130\pm43$\\
     COS-DR3-226441$^b$ & 	$-$ & 		$-$ & 		$-$ & 		$-$ & 			$-$ & 		$50.8\pm6.5$ & $-29.7\pm5.7$ & $63.1\pm8.0$ & $116\pm2$ & $-65.7\pm7.2$ & $386.8\pm7.8$ & $-219.1\pm24.1$\\
      XMM-VID1-2399$^b$ & 	$-$ & 		$-$ & 		$-$ & 		$-$ & 			$-$ & 		$0.4\pm4.4$ & $-0.3\pm2.9$ & $0.7\pm7.0$ & $21.5\pm2.0$ & $-13.7\pm1.6$ & $71.8\pm6.6$ & $-45.6\pm5.2$\\      
     COS-DR3-111740 & 	$-$ & 		$-$ & 		$-$ & 		$1.10 \pm 0.03$ & $11.4 \pm 0.5$ & 	$-$ & 		$-$ & 		$-$ & 		$-$ & 		$-$ & 		$-$ & 		$-$\\
            \hline
      \end{tabular}}
      \label{T:spec}
\end{sidewaystable}

\section{Results} \label{S:res}

\subsection{Absorption Line Galaxies}

Of the 16 confirmed UMGs in the sample, 7 show well-detected absorption features and lack emission features.
Five of these absorption-line galaxies are \UVJ-quiescent residing in the lower region of the quiescent wedge typically associated with young post-starburst galaxies.
Notably, XMM-VID3-1120 was published in \cite{Forrest2020} as XMM-2599 - the updated identifier is due to a new catalog version.

The remaining two absorption line galaxies (COS-DR3-111740, $z=2.80$, and COS-DR1-113684, $z=3.83$) are at redshifts where a combination of the wavelength coverage of MOSFIRE, and position of the object on the chip, and the bandpasses observed do not allow access to the \OII, \Hbeta, or \OIII\ emission features, and as such, emission features could be present.
This also results from the spectroscopic redshifts of these two objects being discrepant from their photometric redshifts - indeed, they are clear outliers on Figure \ref{fig:zpzs}.
Both have similar \UVJ\ colors and SEDs to COS-DR3-160748 (discussed below), consistent with galaxies that have very recently ceased forming new stars, or have small residual amounts of star formation.


\begin{figure}[tp]
	\centering{\includegraphics[width=0.5\textwidth,trim=0.25in 0in 0in 0in, clip=true]{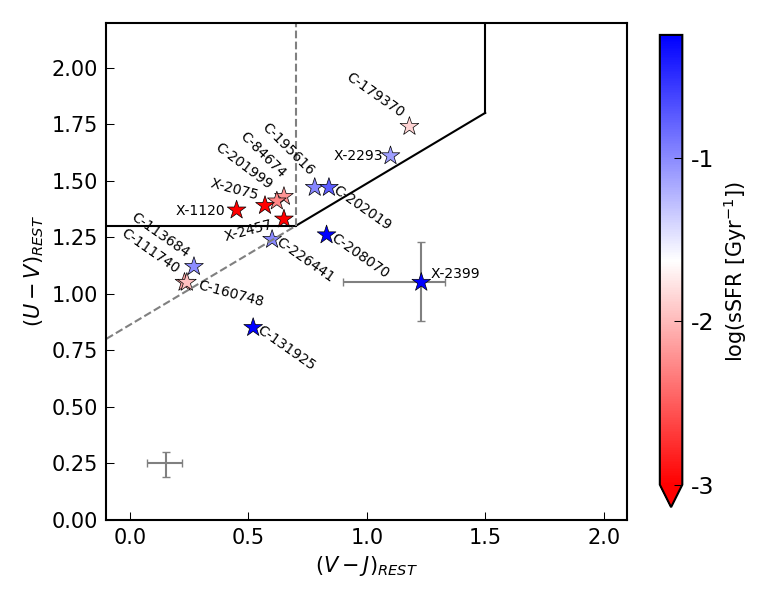}}
	\caption{The restframe \UVJ\ diagram for all confirmed UMGs, colored by sSFR averaged over the last 10 Myr based on the best-fit SFH. The median error bar is shown in the lower left, and the only UMG with an error bar twice this size has the error plotted on the point. }
	\label{fig:uvj}
\end{figure}


\subsection{Emission Line Galaxies}

The remaining 9 UMGs show clear emission features with SNR~$>10$ from \OII\ or \OIIIfive.
Of these, 7 have a line ratio $f_{\OIIIfive}/f_{H\beta}>6$ (labeled in Table~\ref{T:spec} and Figures~\ref{fig:uvj_spec_a}, \ref{fig:uvj_spec_b}, \ref{fig:uvj_spec_c}), which is typically associated with AGN activity, especially at high mass \citep[\eg][]{Baldwin1981,Juneau2011,Trump2013,Kewley2013,Shapley2015,Strom2016,Reddy2018}.
This agrees with numerous previous studies which suggest that AGN activity is more common at high redshifts \citep[\eg][]{Marsan2017}, though this is the largest spectroscopic sample of this mass and redshift to be analyzed.

One of these, COS-DR3-160748 was first spectroscopically confirmed in \citet{Marsan2015}, and a deep spectrum has recently been analyzed in Saracco et al. (2020, submitted).
Stellar masses derived in all three works are in rough agreement ($11.3<$~\logM~$<11.5$), as are the conclusions that an AGN exists in this galaxy and there is little ongoing star formation.


	\begin{figure}[tp]
	\centerline{\includegraphics[width=0.52\textwidth,trim=0in 0in 0in 0in, clip=true]{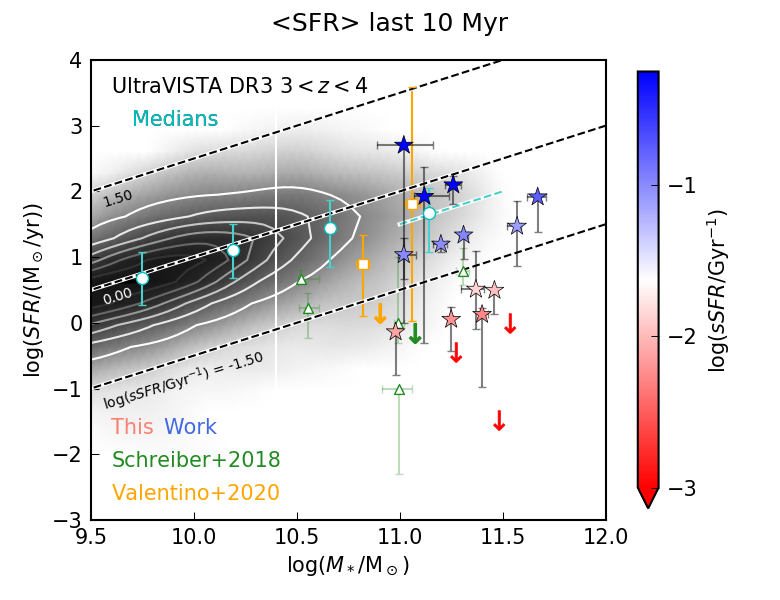}}
	\caption{The SFR-stellar mass plane. UMGs are shown as stars with the same color scheme as in previous figures. The population of galaxies at $3<z<4$ in the UltraVISTA DR3 catalog are shown in black, with the stellar mass completeness limit of the catalog as a white vertical line and medians in bins of 0.5 dex in stellar mass also shown in cyan. The highest mass bin average corresponds to $\log(sSFR/$Gyr$^{-1})$~$\sim-0.5$, which is represented by a dashed cyan line spanning the width of the bin. Massive $z>3$ galaxies from \citet{Schreiber2018b} and \citet{Valentino2020} are shown in green and orange, respectively.}
	\label{fig:sfms}
	\end{figure}


	\begin{figure*}[tp]
	\centerline{\includegraphics[width=\textwidth,trim=0in 0in 0in 0in, clip=true]{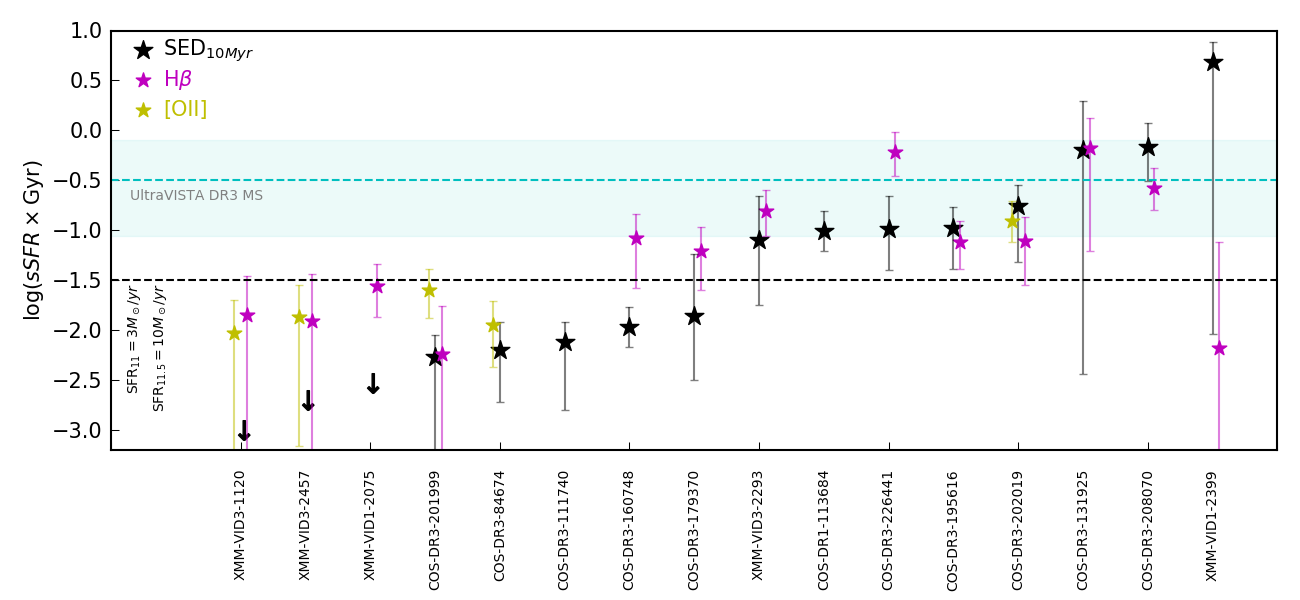}}
	\caption{Comparison of sSFR for the UMGs using various star formation indicators. Black stars show the SFH from the best-fit template averaged over 10~Myr, gold stars use SFR from the \OII\ line, and magenta stars use SFR from the \Hbeta\ line. Note that the \Hbeta\ value assumes all line flux is from star formation, which is not accurate for AGN hosts, and may therefore be an upper limit. The sSFR from the massive bin of the DR3 catalog plotted in Figure \ref{fig:sfms} is shown in cyan. A sSFR one dex below this value, corresponding to SFR~$=3$~M$_\odot$/yr for a \logM~$=11$ UMG, and SFR~$=10$~M$_\odot$/yr for a \logM$=11.5$ UMG, is shown as a black dashed line.}
	\label{fig:sfr_comp}
	\end{figure*}


\subsection{Rest-Frame Colors}

Figure \ref{fig:uvj} shows the spectroscopically confirmed galaxies on the restframe \UVJ\ color-color diagram, which has been well-used as a discriminator between star-forming and quiescent galaxies \citep[\eg][]{Labbe2006, Williams2009, Forrest2018, Schreiber2018b}.
However, at high redshifts, the clear bimodality in colors between these populations appears to erode, no longer providing a selection as pure as at lower redshifts \citep[\eg][]{Whitaker2011,Muzzin2013a, Straatman2016}.
Part of the reason for this is that galaxies can be passive for up to several hundred Myr (depending on SFH) before their colors reach the quiescent wedge on the \UVJ\ diagram \citep{Merlin2018}.
While this is a short time relative to the lifetime of local galaxies, it is non-negligible in the high-redshift universe, and thus one would expect a larger fraction of galaxies to be observed in such a phase.

\citet{Schreiber2018b} suggest simply extending the diagonal wedge line to bluer colors for samples of massive galaxies, though they then encounter some star-forming contaminants in the resulting selection.
We add this extension in Figure \ref{fig:uvj} and find that 3 additional UMGs are included in the resulting quiescent wedge. 
All have SEDs with small amounts of UV flux, consistent with low levels of star formation as would be expected in a galaxy that has recently undergone a starburst event and is now quenching its star formation, as well as very blue \VJc\ indicating a lack of an older, passive stellar population.

Three galaxies with strongly detected emission lines are also \UVJ-quiescent, even when accounting for the effects of their emission features. 
Two of these have large \OIIIfive /\Hbeta\ ratios, implying that their emission lines are due to AGN activity and not ongoing star formation, while the third has clear \OII\ emission and negligible \OIII\, suggestive of some ongoing star formation.
We thus conclude that the standard \UVJ\ diagram, while still doing a decent job of separating star-forming and quiescent galaxies, provides neither a complete nor pure selection of either population at this mass and redshift regime.

\subsection{Star Formation Rates}

We compare the UMGs on the SFR vs. stellar-mass plane to other $3<z<4$ galaxies in the UltraVISTA DR3 catalogs (Muzzin et al., in prep), for which the SFRs and stellar masses were calculated using an exponentially declining SFH (Figure~\ref{fig:sfms}).
All SFRs plotted here are averages over the last 10 Myr as calculated from each galaxy's best-fit SFH.
Using the instantaneous SFR or averaging over 100 Myr does not make any significant difference to our conclusions.
The absorption line UMGs fall at least one dex below the star-forming main sequence (SFMS), clearly consistent with galaxies having highly inhibited, if not completely quenched, star formation.
This is also the case for several other massive $z>3$ galaxies from \citet{Schreiber2018b}.
The confirmed emission line galaxies are generally consistent with the SFMS, though they average only \mbox{$\log$(sSFR [Gyr$^{-1}$]$)\sim-1$} , or SFR~$\sim15$~M$_\odot$/yr.

In addition to the SFRs derived from the emission-line corrected photometry and masked spectroscopy, we also derive SFRs from \Hbeta\ and \OII\ lines when observed in the spectra, providing other probes of star formation.
Comparing the results of these star formation indicators is important in assessing their accuracy, and we show this in Figure \ref{fig:sfr_comp}.
Though these probes do measure star formation over slightly different timescales, agreement between various measurements for individual galaxies is quite good.
In logarithmic space the differences are most significant for low sSFR, but differences in absolute terms here are on the order of $\Delta SFR \sim 1~\rm{M}_\odot/\rm{yr}$.


	\begin{figure}[tp]
	\centerline{\includegraphics[width=0.55\textwidth,trim=0in 0in 0in 0in, clip=true]{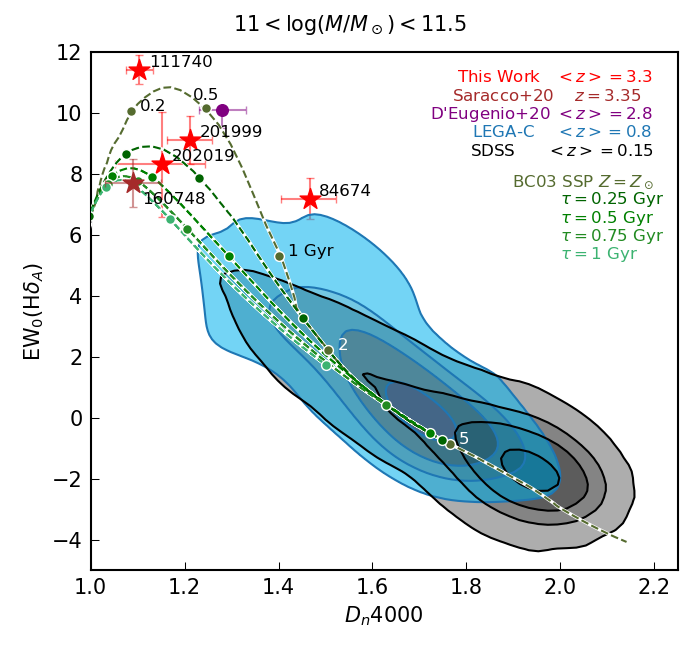}}
	\caption{The \EWhd-\Dfour\ plane, used to constrain age. Lower redshift samples from SDSS and LEGA-C, presented in \citep{Wu2018a}, are shown in black and blue, respectively. A stack of 9 galaxies from \citet{DEugenio2020} is shown in purple. The UMGs with spectroscopic coverage of both features are shown in red, while COS-160748, presented in 
	Saracco et al. (2020, submitted),
	 is shown in brown. Stellar evolution tracks for a simple stellar population (SSP) and exponentially-declining SFH with varying $\tau$ are shown in shades of green. Ages of 0.2, 0.5, 1, 2, and 5 Gyr are marked with a circle for each track.}
	\label{fig:d4hd}
	\end{figure}



	\begin{figure*}[tp]
	\centerline{\includegraphics[width=\textwidth,trim=0in 0in 0in 0in, clip=true]{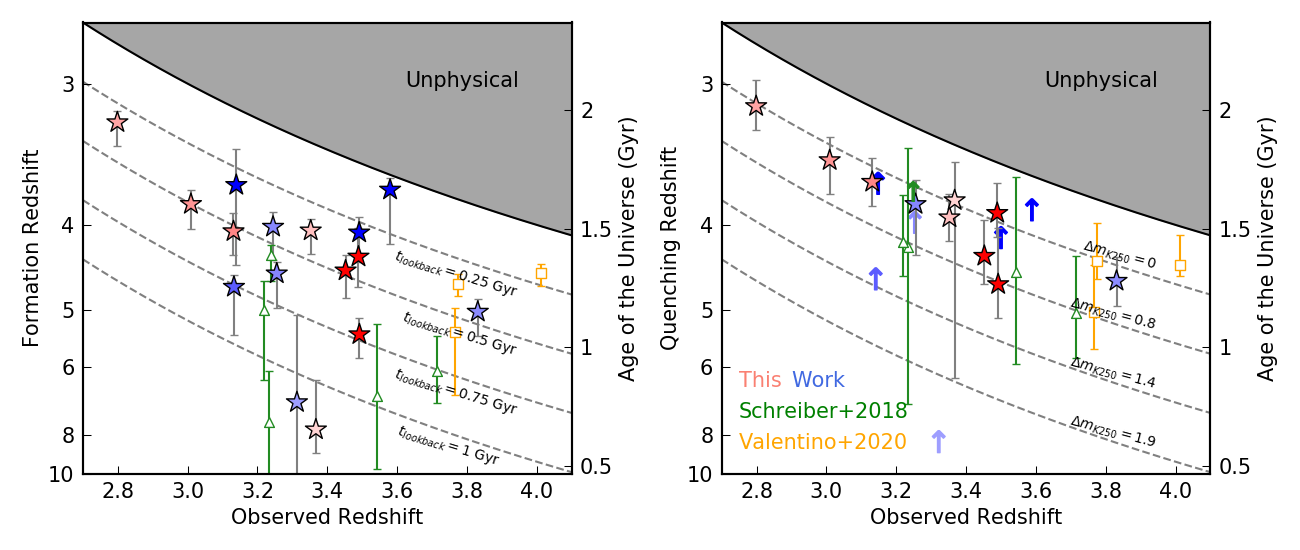}}
	\caption{The evolution of spectroscopically confirmed UMGs. As previously, the sample from this work is shown colored by sSFR, galaxies from \citet{Schreiber2018b} are green, and those from \citet{Valentino2020} are orange. \textbf{Left:} The redshift at which the galaxy had formed 50\% of its stellar mass, as determined from the best-fit SFH. The dashed lines indicate lookback time in intervals of 250 Myr. \textbf{Right:} The redshift at which the SFR fell below 10\% of its average during the main starburst period. Upward arrows indicate UMGs which are consistent with still forming stars above this threshold.
	The difference in $K$-band magnitude a galaxy would present due to quenching at a lookback time greater than 250 Myr ($\Delta m_{K250}$) is shown on each dashed line (see also Figure \ref{fig:DeltaM} and Appendix \ref{A:DeltaM}).
	  This assumes a SFH consisting of a period of constant star formation for 100 Myr, followed by an exponentially declining SFR with $\tau=100$ Myr, as well as no dust attenuation.}
	\label{fig:zfzq}
	\end{figure*}


\subsection{Stellar Ages} \label{S:ages}

The spectral indices \Dfour\ and \EWhd\ \citep{Balogh1999} combine to form an effective probe of stellar age which breaks a degeneracy with metallicity \citep[\eg][]{Kauffmann2003a}.
In the local universe, massive galaxies tend to show larger \Dfour\ and lower \EWhd, as they quenched long ago, while less massive galaxies, on average containing younger stellar populations have lower \Dfour\ and higher \EWhd.

We are able to measure both spectral features for four of the UMGs, and use published values for a fifth using spectra from LBT/LUCI in Saracco et al. (2020, submitted).
The spectral wavelength coverage of the other UMGs does not allow for one or both measurements due to different redshifts and bandpasses observed.
These are shown in Figure \ref{fig:d4hd}, along with large samples of massive galaxies at $z\sim0.1$ and $z\sim0.8$ published in \citet{Wu2018a}.
Also plotted is a stack of massive galaxies at $z\sim2.8$ from \citet{DEugenio2020}.
Taken together, these data paint a picture showing the most massive galaxies at high redshifts are younger than galaxies of similar mass at lower redshifts.
This makes sense due to the younger age of the universe, as well as numerous studies suggesting that high-mass galaxies appear to have formed much of their stellar mass at earlier times than lower mass galaxies at a given epoch \citep[\eg][]{Pacifici2016,Bellstedt2020}.
Indeed when compared to various evolutionary tracks, the UMGs are consistent with stellar ages of 200-800 Myr.

\subsection{Star Formation Histories} \label{S:sfh}

These ages can be compared to the ages determined from the best-fit SFHs of the modeling.
As a reminder, we use FAST++ to simultaneously fit the photometry and spectroscopy, and thereby derive the formation lookback time, by which point half of the observed stellar mass has been formed, and the quenching lookback time, at which time star formation in a galaxy has fallen below 10\% of the average during the main burst of star formation (Table~\ref{T:ps}).
While a double-exponential SFH parameterization is used, conclusions do not change if a delayed-exponential profile is used instead (see Section~\ref{S:fpp}).

Using this methodology, the resulting best-fit SFHs show short (median timescale of 120~Myr), intense (median of 1500~M$_\odot$/yr) bursts of star formation responsible for the vast majority of stellar mass growth.
These bursts occurred $320<t_{50}/\textrm{Myr}<740$ in the past, consistent with the ages surmised from spectral indices above.
Following this, star formation rates truncated quickly.
UMG quenching times are narrowly constrained to lookback times of $240<t_q/\textrm{Myr}<510$ for those UMGs no longer forming stars, and the median value is 310 Myr (Figure~\ref{fig:zfzq}).
This quenching happens rapidly with median timescale $t_{50}-t_q=180$~Myr.
The derived timescales are in good agreement with those from Saracco et al. (2020), which performed an independent analysis on deeper H+K spectra for COS-DR3-160748.

\subsubsection{Rapidly Star-forming Progenitors}

The confirmation of a number of highly dust-obscured massive star-forming galaxies at $z>5$ \citep[][]{Capak2011, Riechers2013, Ma2015, Riechers2017, Strandet2017, Marrone2018, Pavesi2018, Zavala2018, Jin2019} has raised the question of how these galaxies evolve and what their descendants look like.
Naturally, massive quiescent galaxies must be an eventual evolutionary stage, as these galaxies cannot continue forming stars at the observed rates for more than several hundred Myr (assuming no large inflows of cool gas to use as fuel).
The idea that such galaxies might be the progenitors of lower redshift massive quiescent galaxies has been suggested in previous works \citep[\eg][]{Toft2014, Belli2019, Forrest2020}.

\begin{figure}[tp]
	\centerline{\includegraphics[width=0.55\textwidth,trim=0in 0in 0in 0in, clip=true]{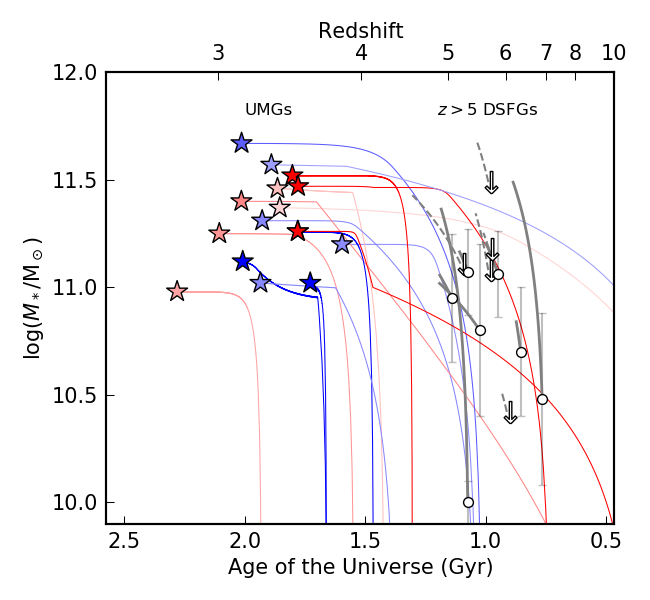}}
	\caption{The stellar mass evolution of spectroscopically confirmed UMGs according to their best-fit SFHs. A set of spectroscopically confirmed DSFGs at higher redshifts is shown in gray. The tracks on this population correspond to the growth assuming the observed SFR continuing for half of the depletion timescale, \ie\ consuming half the available gas to produce stars.}
	\label{fig:dsfg}
	\end{figure}

It now appears that significant numbers of these dust-obscured galaxies exist at $z>3$, many of which may not be photometrically-detected in the optical or even the $H$- and $K-$ bandpasses due to large amounts of dust \citep{Twang2019,Williams2019,Riechers2020a}.
The inferred number densities of these objects has also increased as a result of larger and deeper surveys with ALMA \citep[\eg][]{Riechers2020a}, and agree with number densities of UMGs (star-forming and quiescent combined) observed photometrically at $3<z<4$ which has been found to be about $N\sim1\times10^{-5}$ Mpc$^{-3}$ \citep[][Marsan et al. 2020, in prep]{Schreiber2018b}.
Of course since star-forming UMGs contribute to this number density, it is unreasonable to assume that all $z>5$ DSFGs are quenched by $z=3$ \citep[see][for an in depth discussion]{Valentino2020}, though it seems reasonable to assume that the progenitors of the quiescent UMGs were at one time starbursts very similar to these DSFGs.  
Best-fit SFHs suggest that this period of intense star formation occurred at $z\gtrsim4$ for the majority of the UMGs (Figure~\ref{fig:dsfg}), perhaps also suggesting the existence of fainter, quenched UMGs as descendants of the $z>5$ DSFG population.

Indeed, the evolution of the most massive early-type galaxies in the local Universe along a path involving rapid star-formation as a DSFG at $z>5$, quenching around $z>3$, and subsequent stellar mass growth via mergers is supported by several lines of evidence.
Studies of local galaxies suggest a scenario in which the most massive galaxies formed their stars in bursts at very early times \citep[\eg][]{Kauffmann2003a, Gallazzi2005, Thomas2010, Smith2012}, with subsequent mass growth dominated by mergers \citep{Hill2017a}.
The size growth of quiescent galaxies from compact at high redshift objects to more extended at lower redshifts is also consistent with being merger-driven \citep[\eg][]{Barro2013, Kubo2018, Estrada-Carpenter2020}.
The quenching timescales presented here for such a large sample of UMGs at $3<z<4$ convincingly support this evolutionary hypothesis of the most massive galaxies.


\subsubsection{On the Possibility of Older, Quiescent Galaxies}

The quenching timescales derived for this sample suggest that all the quiescent UMGs are younger post-starburst galaxies, and not old, long-dead passive populations.
Indeed at this epoch, no such old passive populations have been spectroscopically confirmed \citep{DEugenio2020}, which we also find when comparing the UMGs presented here to those published in \citet{Schreiber2018b} and \citet{Valentino2020}.
While the age of the universe at $z>3$ is quite young, there is still sufficient time for a UMG to have formed its stars, quenched, and evolved passively for perhaps as long as 1 Gyr by $z=3$, especially given the short timescales on which UMGs at this epoch appear to form their stars \citep[][Saracco et al. 2020]{Schreiber2018b,Valentino2020,Forrest2020}.

\begin{figure*}
	\centerline{\includegraphics[width=\textwidth,trim=0in 0in 0in 0in, clip=true]{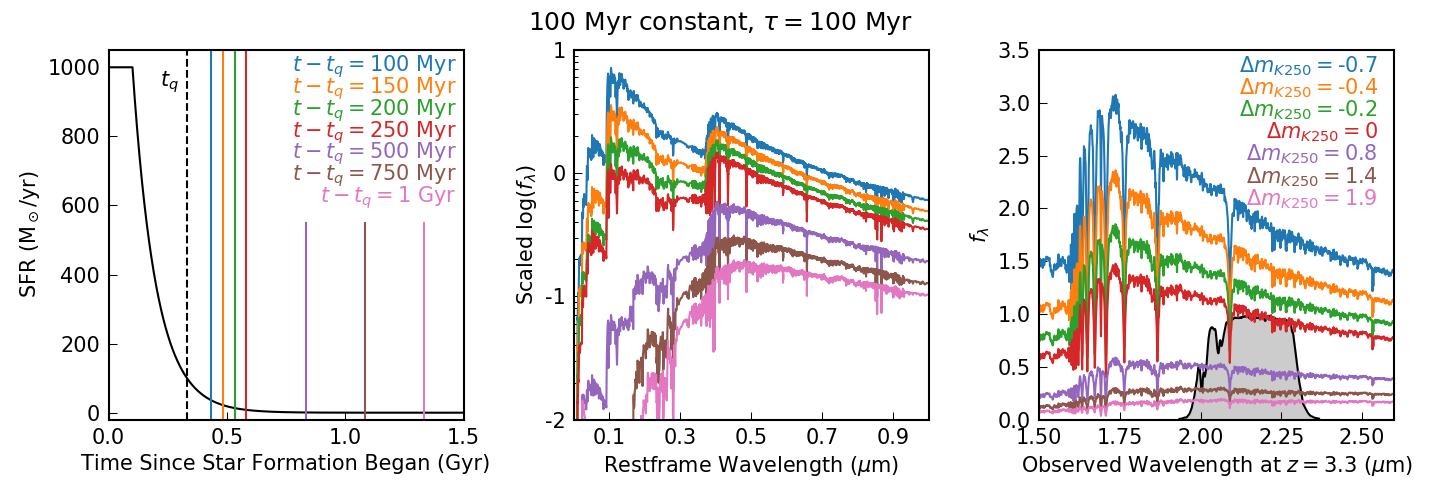}}
	\caption{The evolution of \Ks-band magnitude for an object with a 100 Myr of constant star formation followed by an exponential decline parameterised by $\tau=100$ Myr. \textbf{Left:} The SFH (black) used to construct the SEDs at various times (solid, colored lines). The vertical dashed line represents the nominal quenching time, $t_q$, taken to be when the SFR drops below 10\% that of the main starburst period. \textbf{Center:} The resultant SEDs at each timestep, corresponding to 100, 150, 200, 250, 500, 750, and 1000 Myr after the quenching time, and scaled relative to the $t-t_q=250$ Myr SED (red). \textbf{Right:} The SEDs redshifted to $z=3.3$. The gray curve corresponds to the \Ks-band response function, and magnitude differences are given relative to the $t-t_q=250$ Myr SED (red).}
	\label{fig:DeltaM}
	\end{figure*}

Such a galaxy would be considerably fainter than a post-starburst however, and selection for spectroscopic follow-up is necessarily biased towards brighter targets.
We quantify this magnitude difference by generating galaxy SEDs at various stages of evolution according to several SFHs using BC03 SPS models.
Solar metallicity models are used, as this metallicity has been shown to be roughly appropriate for these massive, early galaxies \citep[][Saracco et al. 2020, submitted]{Belli2019}.
We take the SED at several timesteps in this evolution, redshift it to $z=3.3$ (the median redshift of the sample), and calculate the magnitude in the \Ks-band (see Figure~\ref{fig:DeltaM}, as well as Appendix \ref{A:DeltaM} for details).
Compared to a baseline observation of 0.25 Gyr after the simulated galaxy quenches (roughly consistent with the quenched UMGs),
a galaxy observed 1 Gyr after quenching would be 1-2 magnitudes fainter in the \Ks-band depending upon the SFH, corresponding to $m_K\sim22-23$ and fainter than the magnitude cut used here.
For an object observed at $z=3.3$, this would correspond to $z_q=6.1$, and would require extreme star formation just prior to this time, consistent with the $z=6.34$ DSFG HFLS-3 \citep{Riechers2013, Cooray2014}.

Galaxies with the requisite \Ks-band magnitudes, photometric redshifts, stellar masses, as well as red \UVJ\ colors do exist in the UltraVISTA DR3 catalog.
There are 56 objects with photometric fits suggesting $3<z<4$, \logM$>11$, and SFR~$<10$ M$_\odot$/yr, 47 of which are fainter than $m_K=22$.
Of the nine brighter targets, we targeted and confirmed three - COS-DR3-84674, COS-DR3-111740, and COS-DR3-201999, two of which are post-starbursts, and one of which is at a lower redshift.
The faintest of the spectroscopically confirmed UMGs (COS-DR3-179370) has $m_K=22.14$, but we note that this was confirmed via detection of strong emission lines.

The fainter candidate objects generally have lower photometric signal-to-noise, leading to less well constrained photometric redshifts and template fits.
As a result, selecting reliable candidates at fainter magnitudes from the existing photometry for spectroscopic follow-up is more difficult.
Targeting these candidate old red galaxies with instruments such as Keck/MOSFIRE would require multiple nights to confirm their redshifts and natures through high signal-to-noise detection of stellar continuum features.
Still, the intriguing possibility of older, more passive galaxies at this early time does exist.
Proving it will require significant time investment either from the ground (several nights on 8-10 m class telescopes) or from space (\eg\ JWST).
Upcoming 30 m class telescopes could also play an important role in confirming these objects as well.


\section{Conclusions}\label{Sec:Conc}

We have presented the largest sample to date of spectroscopically confirmed ultra-massive galaxies with \mbox{\logM~$>11$} at $z>3$.
The sample is selected from the COSMOS-UltraVISTA and XMM-VIDEO fields which have a critical combination of impressive depth and area, allowing these catalogs to excel at finding such rare objects.
Galaxies with well-defined SEDs were observed as part of the \magazine\ survey with the \mbox{Keck/MOSFIRE} instrument in the $K$-band, where strong emission lines from \Hbeta\ and \OIII\ are observed in 9/16 UMGs.
The $H$-band was also used to confirm redshifts via detection of Balmer series absorption, which is clearly visible in all 6 UMGs targeted, and also to confirm any \OII\ emission.
The confirmation of this population is further evidence that such massive galaxies form the vast majority of their stars early in the universe.

While these UMGs show a wide range of SFRs, at least 5/16 have ceased star formation in the last several hundred Myr (post-starbursts), and 3 others only have residual amounts of new star formation ongoing ($\lesssim5$~M$_\odot$/yr), which puts them $>1$ dex below the star-forming main sequence for galaxies of this redshift and mass.
The remainder are forming at least 10~M$_\odot$/yr, and two UMGs are consistent with SFR~$\gtrsim100$~M$_\odot$/yr.

Of those with emission features, 7 show line ratios of \OIIIfive /\Hbeta~$>6$, typically observed in such massive galaxies only with AGN activity.
This is consistent with previous studies suggesting that massive galaxies at these early times are often AGN hosts, and that AGN feedback may play a significant role in quenching star formation in this population.

We compare the \Dfour\ and \EWhd\ values for five of the UMGs to high mass galaxies at lower redshifts as well as stellar evolution models.
The most massive galaxies at higher redshifts are younger, with the sample suggesting the bulk of stars were formed $<1$ Gyr before observation.

The build up of stellar mass from the best-fit SFHs for some of these UMGs are consistent with $z>5$ DSFGs, again suggesting that a period of intense star formation is a critical step in the evolution of this population before quenching star formation at $z>3$.
This is also consistent with the SFHs derived for massive local galaxies from stellar archaeology.

Fitting the full multi-wavelength SEDs (corrected for emission line contamination) and the near-IR spectroscopy yield best-fit SFHs which similarly indicate that most star formation occurred at $4<z<5$ in short-lived, explosive bursts of $500-3000$~M$_\odot$/yr for $100-400$~Myr before quenching rapidly.
While several of the UMGs have quenched before $z=4$ as in \citet{Forrest2020}, most UMGs which have quenched their star formation did so around $250-400$~Myr before observation, and we do not confirm any significantly evolved passive galaxies, implying that post-starbursts are the most evolved galaxies in the massive population at this epoch.
However, the possibility of older, more passive galaxies does exist, even at this remarkably early time.  Identifying compelling candidates will require deeper NIR photometry, and spectroscopically confirming them will require substantial observations with either large ground-based instruments or sensitive space-based telescopes.

\section{Acknowledgements}
The authors wish to recognize and acknowledge the very significant cultural role and reverence that the summit of Maunakea has always had within the indigenous Hawaiian community.
We are most fortunate to have the opportunity to conduct observations from this mountain.
Additionally, we thank the anonymous referee for suggestions that improved the manuscript.

This work is supported by the National Science Foundation through grants AST-1517863, AST-1518257, and AST-1815475, by {\it HST} program number GO-15294, and by grant numbers 80NSSC17K0019 and NNX16AN49G issued through the NASA Astrophysics Data Analysis Program (ADAP).
Support for program number GO-15294 was provided by NASA through a grant from the Space Telescope Science Institute, which is operated by the Association of Universities for Research in Astronomy, Incorporated, under NASA contract NAS5-26555.
Further support was provided by the Faculty Research Fund (FRF) of Tufts University and by Universidad Andr\'es Bello grant number DI-12-19/R.
M.N. acknowledges INAF 1.05.01.86.20 and PRIN MIUR F.OB. 1.05.01.83.08.
Data presented herein were obtained using the UCI Remote Observing Facility, made possible 
by a generous gift from John and Ruth Ann Evans.

This work has relied heavily upon code developed by other people, for which we are quite thankful.
\software{
Astropy \citep{Astropy2013,Astropy2018},
EAzY \citep{Brammer2008},
FAST \citep{Kriek2009},
FAST++ \citep{Schreiber2018a},
IPython \citep{Perez2007},
Matplotlib \citep{Hunter2007},
NumPy \citep{Oliphant2006},
slinefit \citep{Schreiber2018b}
}

\bibliography{library}

\begin{thebibliography}{}
\expandafter\ifx\csname natexlab\endcsname\relax\def\natexlab#1{#1}\fi
\providecommand{\url}[1]{\href{#1}{#1}}
\providecommand{\dodoi}[1]{doi:~\href{http://doi.org/#1}{\nolinkurl{#1}}}
\providecommand{\doeprint}[1]{\href{http://ascl.net/#1}{\nolinkurl{http://ascl.net/#1}}}
\providecommand{\doarXiv}[1]{\href{https://arxiv.org/abs/#1}{\nolinkurl{https://arxiv.org/abs/#1}}}

\bibitem[{{Alcalde Pampliega} {et~al.}(2019){Alcalde Pampliega},
  P{\'{e}}rez-Gonz{\'{a}}lez, Barro, S{\'{a}}nchez, Eliche-Moral, Cardiel,
  Hern{\'{a}}n-Caballero, Rodriguez-Mu{\~{n}}oz, Bl{\'{a}}zquez, \&
  Esquej}]{Pampliega2019}
{Alcalde Pampliega}, B., P{\'{e}}rez-Gonz{\'{a}}lez, P.~G., Barro, G., {et~al.}
  2019, The Astrophysical Journal, 876, 135, \dodoi{10.3847/1538-4357/ab14f2}

\bibitem[{Ashby {et~al.}(2018)Ashby, Caputi, Cowley, Deshmukh, Dunlop,
  Milvang-Jensen, Fynbo, Muzzin, McCracken, F{\`{e}}vre, Huang, \&
  Zhang}]{Ashby2018}
Ashby, M. L.~N., Caputi, K.~I., Cowley, W., {et~al.} 2018, The Astrophysical
  Journal Supplement Series, 237, 39, \dodoi{10.3847/1538-4365/aad4fb}

\bibitem[{Baldwin {et~al.}(1981)Baldwin, Phillips, \& Terlevich}]{Baldwin1981}
Baldwin, J.~A., Phillips, M.~M., \& Terlevich, R. 1981, Publications of the
  Astronomical Society of the Pacific, 93, 5, \dodoi{10.1086/130766}

\bibitem[{Balogh {et~al.}(1999)Balogh, Morris, Yee, Carlberg, \&
  Ellingson}]{Balogh1999}
Balogh, M.~L., Morris, S.~L., Yee, H. K.~C., Carlberg, R.~G., \& Ellingson, E.
  1999, The Astrophysical Journal, 527, 54, \dodoi{10.1086/308056}

\bibitem[{Banerji {et~al.}(2015)Banerji, Jouvel, Lin, McMahon, Lahav,
  Castander, Abdalla, Bertin, Bosman, Carnero, {Carrasco Kind}, {Da Costa},
  Gerdes, Gschwend, Lima, Maia, Merson, Miller, Ogando, Pellegrini, Reed,
  Saglia, S{\'{a}}nchez, Allam, Annis, Bernstein, Bernstein, Bernstein,
  Capozzi, Childress, Cunha, Davis, DePoy, Desai, Diehl, Doel, Findlay, Finley,
  Flaugher, Frieman, Gaztanaga, Glazebrook, Gonz{\'{a}}lez-Fern{\'{a}}ndez,
  Gonzalez-Solares, Honscheid, Irwin, Jarvis, Kim, Koposov, Kuehn,
  Kupcu-Yoldas, Lagattuta, Lewis, Lidman, Makler, Marriner, Marshall, Miquel,
  Mohr, Neilsen, Peoples, Sako, Sanchez, Scarpine, Schindler, Schubnell,
  Sevilla, Sharp, Soares-Santos, Swanson, Tarle, Thaler, Tucker, Uddin,
  Wechsler, Wester, Yuan, \& Zuntz}]{Banerji2015}
Banerji, M., Jouvel, S., Lin, H., {et~al.} 2015, Monthly Notices of the Royal
  Astronomical Society, 446, 2523, \dodoi{10.1093/mnras/stu2261}

\bibitem[{Barro {et~al.}(2013)Barro, Faber, P{\'{e}}rez-Gonz{\'{a}}lez, Koo,
  Williams, Kocevski, Trump, Mozena, McGrath, van~der Wel, Wuyts, Bell, Croton,
  Ceverino, Dekel, Ashby, Cheung, Ferguson, Fontana, Fang, Giavalisco, Grogin,
  Guo, Hathi, Hopkins, Huang, Koekemoer, Kartaltepe, Lee, Newman, Porter,
  Primack, Ryan, Rosario, Somerville, Salvato, \& Hsu}]{Barro2013}
Barro, G., Faber, S.~M., P{\'{e}}rez-Gonz{\'{a}}lez, P.~G., {et~al.} 2013, The
  Astrophysical Journal, 765, 104, \dodoi{10.1088/0004-637X/765/2/104}

\bibitem[{Belli {et~al.}(2017)Belli, Newman, \& Ellis}]{Belli2017}
Belli, S., Newman, A.~B., \& Ellis, R.~S. 2017, The Astrophysical Journal, 834,
  18, \dodoi{10.3847/1538-4357/834/1/18}

\bibitem[{Belli {et~al.}(2019)Belli, Newman, \& Ellis}]{Belli2019}
---. 2019, The Astrophysical Journal, 874, 17, \dodoi{10.3847/1538-4357/ab07af}

\bibitem[{Belli {et~al.}(2014)Belli, Newman, Ellis, \& Konidaris}]{Belli2014}
Belli, S., Newman, A.~B., Ellis, R.~S., \& Konidaris, N.~P. 2014, Astrophysical
  Journal Letters, 788, \dodoi{10.1088/2041-8205/788/2/L29}

\bibitem[{Bellstedt {et~al.}(2020)Bellstedt, Robotham, Driver, Thorne, Davies,
  Lagos, Stevens, Taylor, Baldry, Moffett, Hopkins, \&
  Phillipps}]{Bellstedt2020}
Bellstedt, S., Robotham, A. S.~G., Driver, S.~P., {et~al.} 2020.
\newblock \doarXiv{2005.11917}

\bibitem[{Brammer {et~al.}(2008)Brammer, van Dokkum, \& Coppi}]{Brammer2008}
Brammer, G.~B., van Dokkum, P.~G., \& Coppi, P. 2008, The Astrophysical
  Journal, 686, 1503, \dodoi{10.1086/591786}

\bibitem[{Bruzual \& Charlot(2003)}]{Bruzual2003}
Bruzual, G., \& Charlot, S. 2003, Monthly Notices of the Royal Astronomical
  Society, 344, 1000, \dodoi{10.1046/j.1365-8711.2003.06897.x}

\bibitem[{Calzetti {et~al.}(2000)Calzetti, Armus, Bohlin, Kinney, Koornneef, \&
  Storchi-Bergmann}]{Calzetti2000}
Calzetti, D., Armus, L., Bohlin, R.~C., {et~al.} 2000, The Astrophysical
  Journal, 533, 682, \dodoi{10.1086/308692}

\bibitem[{Capak {et~al.}(2011)Capak, Riechers, Scoville, Carilli, Cox, Neri,
  Robertson, Salvato, Schinnerer, Yan, Wilson, Yun, Civano, Elvis, Karim,
  Mobasher, \& Staguhn}]{Capak2011}
Capak, P.~L., Riechers, D., Scoville, N.~Z., {et~al.} 2011, Nature, 470, 233,
  \dodoi{10.1038/nature09681}

\bibitem[{Chabrier(2003)}]{Chabrier2003}
Chabrier, G. 2003, Publications of the Astronomical Society of the Pacific,
  115, 763, \dodoi{10.1086/376392}

\bibitem[{Ciesla {et~al.}(2017)Ciesla, Elbaz, \& Fensch}]{Ciesla2017}
Ciesla, L., Elbaz, D., \& Fensch, J. 2017, Astronomy {\&} Astrophysics, 608,
  A41, \dodoi{10.1051/0004-6361/201731036}

\bibitem[{Cimatti {et~al.}(2004)Cimatti, Daddi, Renzini, Cassata, Vanzella,
  Pozzetti, Cristiani, Fontana, Rodighiero, Mignoll, \& Zamorani}]{Cimatti2004}
Cimatti, A., Daddi, E., Renzini, A., {et~al.} 2004, Nature, 430, 184,
  \dodoi{10.1038/nature02668}

\bibitem[{Cooray {et~al.}(2014)Cooray, Calanog, Wardlow, Bock, Bridge,
  Burgarella, Bussmann, Casey, Clements, Conley, Farrah, Fu, Gavazzi, Ivison,
  {La Porte}, {Lo Faro}, Ma, Magdis, Oliver, Osage, P{\'{e}}rez-Fournon,
  Riechers, Rigopoulou, Scott, Viero, \& Watson}]{Cooray2014}
Cooray, A., Calanog, J., Wardlow, J.~L., {et~al.} 2014, The Astrophysical
  Journal, 790, 40, \dodoi{10.1088/0004-637X/790/1/40}

\bibitem[{Crain {et~al.}(2015)Crain, Schaye, Bower, Furlong, Schaller, Theuns,
  Vecchia, Frenk, McCarthy, Helly, Jenkins, Rosas-Guevara, White, \&
  Trayford}]{Crain2015}
Crain, R.~A., Schaye, J., Bower, R.~G., {et~al.} 2015, Monthly Notices of the
  Royal Astronomical Society, 450, 1937, \dodoi{10.1093/mnras/stv725}

\bibitem[{Dav{\'{e}} {et~al.}(2016)Dav{\'{e}}, Thompson, \& Hopkins}]{Dave2016}
Dav{\'{e}}, R., Thompson, R., \& Hopkins, P.~F. 2016, Monthly Notices of the
  Royal Astronomical Society, 462, 3265, \dodoi{10.1093/mnras/stw1862}

\bibitem[{D'Eugenio {et~al.}(2020)D'Eugenio, Daddi, Gobat, Strazzullo, Lustig,
  Delvecchio, Jin, Puglisi, Calabr{\'{o}}, Mancini, Dickinson, Cimatti, \&
  Onodera}]{DEugenio2020}
D'Eugenio, C., Daddi, E., Gobat, R., {et~al.} 2020, The Astrophysical Journal,
  892, L2, \dodoi{10.3847/2041-8213/ab7a96}

\bibitem[{Dunlop {et~al.}(1996)Dunlop, Peacock, Spinrad, Dey, Jimenez, Stern,
  \& Windhorst}]{Dunlop1996}
Dunlop, J., Peacock, J., Spinrad, H., {et~al.} 1996, Nature, 381, 581,
  \dodoi{10.1038/381581a0}

\bibitem[{Einstein \& de~Sitter(1932)}]{Einstein1932}
Einstein, A., \& de~Sitter, W. 1932, Proceedings of the National Academy of
  Sciences, 18, 213, \dodoi{10.1073/pnas.18.3.213}

\bibitem[{Estrada-Carpenter {et~al.}(2019)Estrada-Carpenter, Papovich,
  Momcheva, Brammer, Long, Quadri, Bridge, Dickinson, Ferguson, Finkelstein,
  Giavalisco, Gosmeyer, Lotz, Salmon, Skelton, Trump, \&
  Weiner}]{Estrada-Carpenter2019}
Estrada-Carpenter, V., Papovich, C., Momcheva, I., {et~al.} 2019, The
  Astrophysical Journal, 870, 133, \dodoi{10.3847/1538-4357/aaf22e}

\bibitem[{Estrada-Carpenter {et~al.}(2020)Estrada-Carpenter, Papovich,
  Momcheva, Brammer, Simons, Bridge, Cleri, Ferguson, Finkelstein, Giavalisco,
  Jung, Matharu, Trump, \& Weiner}]{Estrada-Carpenter2020}
---. 2020.
\newblock \doarXiv{2005.12289}

\bibitem[{Feldmann {et~al.}(2016)Feldmann, Hopkins, Quataert,
  Faucher-Gigu{\`{e}}re, \& Ker{\v{e}}s}]{Feldmann2016a}
Feldmann, R., Hopkins, P.~F., Quataert, E., Faucher-Gigu{\`{e}}re, C.~A., \&
  Ker{\v{e}}s, D. 2016, Monthly Notices of the Royal Astronomical Society:
  Letters, 458, L14, \dodoi{10.1093/mnrasl/slw014}

\bibitem[{Forrest {et~al.}(2016)Forrest, Tran, Tomczak, Broussard, Labb{\'{e}},
  Papovich, Kriek, Allen, Cowley, Dickinson, Glazebrook, van Houdt, Inami,
  Kacprzak, Kawinwanichakij, Kelson, McCarthy, Monson, Morrison, Nanayakkara,
  Persson, Quadri, Spitler, Straatman, \& Tilvi}]{Forrest2016}
Forrest, B., Tran, K.-V.~H., Tomczak, A.~R., {et~al.} 2016, The Astrophysical
  Journal, 818, L26, \dodoi{10.3847/2041-8205/818/2/L26}

\bibitem[{Forrest {et~al.}(2018)Forrest, Tran, Broussard, Cohn, Kennicutt,
  Papovich, Allen, Cowley, Glazebrook, Kacprzak, Kawinwanichakij, Nanayakkara,
  Salmon, Spitler, \& Straatman}]{Forrest2018}
Forrest, B., Tran, K.-V.~H., Broussard, A., {et~al.} 2018, The Astrophysical
  Journal, 863, 131, \dodoi{10.3847/1538-4357/aad232}

\bibitem[{Forrest {et~al.}(2020)Forrest, Annunziatella, Wilson, Marchesini,
  Muzzin, Cooper, Marsan, McConachie, Chan, Gomez, Kado-Fong, Barbera,
  Labb{\'{e}}, Lange-Vagle, Nantais, Nonino, Pe{\~{n}}a, Saracco, Stefanon, \&
  van~der Burg}]{Forrest2020}
Forrest, B., Annunziatella, M., Wilson, G., {et~al.} 2020, The Astrophysical
  Journal, 890, L1, \dodoi{10.3847/2041-8213/ab5b9f}

\bibitem[{Gallazzi {et~al.}(2005)Gallazzi, Charlot, Brinchmann, White, \&
  Tremonti}]{Gallazzi2005}
Gallazzi, A., Charlot, S., Brinchmann, J., White, S.~D., \& Tremonti, C.~A.
  2005, Monthly Notices of the Royal Astronomical Society, 362, 41,
  \dodoi{10.1111/j.1365-2966.2005.09321.x}

\bibitem[{Gargiulo {et~al.}(2016)Gargiulo, Saracco, Tamburri, Lonoce, \&
  Ciocca}]{Gargiulo2016}
Gargiulo, A., Saracco, P., Tamburri, S., Lonoce, I., \& Ciocca, F. 2016,
  Astronomy and Astrophysics, 592, 1, \dodoi{10.1051/0004-6361/201526563}

\bibitem[{Genel {et~al.}(2014)Genel, Vogelsberger, Springel, Sijacki, Nelson,
  Snyder, Rodriguez-Gomez, Torrey, \& Hernquist}]{Genel2014}
Genel, S., Vogelsberger, M., Springel, V., {et~al.} 2014, Monthly Notices of
  the Royal Astronomical Society, 445, 175, \dodoi{10.1093/mnras/stu1654}

\bibitem[{Glazebrook {et~al.}(2017)Glazebrook, Schreiber, Labb{\'{e}},
  Nanayakkara, Kacprzak, Oesch, Papovich, Spitler, Straatman, Tran, \&
  Yuan}]{Glazebrook2017}
Glazebrook, K., Schreiber, C., Labb{\'{e}}, I., {et~al.} 2017, Nature
  Publishing Group, 544, 71, \dodoi{10.1038/nature21680}

\bibitem[{Gobat {et~al.}(2012)Gobat, Strazzullo, Daddi, Onodera, Renzini,
  B{\'{e}}thermin, Dickinson, Carollo, \& Cimatti}]{Gobat2012}
Gobat, R., Strazzullo, V., Daddi, E., {et~al.} 2012, The Astrophysical Journal,
  759, L44, \dodoi{10.1088/2041-8205/759/2/L44}

\bibitem[{Grogin {et~al.}(2011)Grogin, Kocevski, Faber, Ferguson, Koekemoer,
  Riess, Acquaviva, Alexander, Almaini, Ashby, Barden, Bell, Bournaud, Brown,
  Caputi, Casertano, Cassata, Castellano, Challis, Chary, Cheung, Cirasuolo,
  Conselice, Cooray, Croton, Daddi, Dahlen, Dav{\'{e}}, de~Mello, Dekel,
  Dickinson, Dolch, Donley, Dunlop, Dutton, Elbaz, Fazio, Filippenko,
  Finkelstein, Fontana, Gardner, Garnavich, Gawiser, Giavalisco, Grazian, Guo,
  Hathi, H{\"{a}}ussler, Hopkins, Huang, Huang, Jha, Kartaltepe, Kirshner, Koo,
  Lai, Lee, Li, Lotz, Lucas, Madau, McCarthy, McGrath, McIntosh, McLure,
  Mobasher, Moustakas, Mozena, Nandra, Newman, Niemi, Noeske, Papovich,
  Pentericci, Pope, Primack, Rajan, Ravindranath, Reddy, Renzini, Rix, Robaina,
  Rodney, Rosario, Rosati, Salimbeni, Scarlata, Siana, Simard, Smidt,
  Somerville, Spinrad, Straughn, Strolger, Telford, Teplitz, Trump, van~der
  Wel, Villforth, Wechsler, Weiner, Wiklind, Wild, Wilson, Wuyts, Yan, \&
  Yun}]{Grogin2011}
Grogin, N.~A., Kocevski, D.~D., Faber, S.~M., {et~al.} 2011, The Astrophysical
  Journal Supplement Series, 197, 35, \dodoi{10.1088/0067-0049/197/2/35}

\bibitem[{Guarnieri {et~al.}(2019)Guarnieri, Maraston, Thomas, Pforr,
  Gonzalez-Perez, Etherington, Carlsen, Morice-Atkinson, Conselice, Gschwend,
  {Carrasco Kind}, Abbott, Allam, Brooks, Burke, {Carnero Rosell}, Carretero,
  Cunha, D'Andrea, da~Costa, {De Vincente}, DePoy, {Thomas Diehl}, Doel,
  Frieman, Garcia-Bellido, Gruen, Gutierrez, Hanley, Hollowood, Honscheid,
  James, Jeltema, Kuehn, Lima, Maia, Marshall, Martini, Melchior, Menanteau,
  Miquel, {Plazas Malagon}, Richardson, Romer, Sanchez, Scarpine, Schindler,
  Sevilla, Smith, Soares-Santos, Sobreira, Suchyta, Tarle, Walker, \&
  Wester}]{Guarnieri2019}
Guarnieri, P., Maraston, C., Thomas, D., {et~al.} 2019, Monthly Notices of the
  Royal Astronomical Society, 483, 3060, \dodoi{10.1093/mnras/sty3305}

\bibitem[{Henriques {et~al.}(2015)Henriques, White, Thomas, Angulo, Guo,
  Lemson, Springel, \& Overzier}]{Henriques2015}
Henriques, B. M.~B., White, S. D.~M., Thomas, P.~A., {et~al.} 2015, Monthly
  Notices of the Royal Astronomical Society, 451, 2663,
  \dodoi{10.1093/mnras/stv705}

\bibitem[{Hill {et~al.}(2017)Hill, Muzzin, Franx, Clauwens, Schreiber,
  Marchesini, Stefanon, Labbe, Brammer, Caputi, Fynbo, Milvang-Jensen, Skelton,
  van Dokkum, \& Whitaker}]{Hill2017a}
Hill, A.~R., Muzzin, A., Franx, M., {et~al.} 2017, The Astrophysical Journal,
  837, 147, \dodoi{10.3847/1538-4357/aa61fe}

\bibitem[{Hopkins {et~al.}(2014)Hopkins, Kere{\v{s}}, O{\~{n}}orbe,
  Faucher-Gigu{\`{e}}re, Quataert, Murray, \& Bullock}]{Hopkins2014}
Hopkins, P.~F., Kere{\v{s}}, D., O{\~{n}}orbe, J., {et~al.} 2014, Monthly
  Notices of the Royal Astronomical Society, 445, 581,
  \dodoi{10.1093/mnras/stu1738}

\bibitem[{Horne(1986)}]{Horne1986}
Horne, K. 1986, Publications of the Astronomical Society of the Pacific, 98,
  609, \dodoi{10.1086/131801}

\bibitem[{Hunter(2007)}]{Hunter2007}
Hunter, J.~D. 2007, Computing in Science and Engineering,
  \dodoi{10.1109/MCSE.2007.55}

\bibitem[{Husser {et~al.}(2013)Husser, {Wende-von Berg}, Dreizler, Homeier,
  Reiners, Barman, \& Hauschildt}]{Husser2013}
Husser, T.-O., {Wende-von Berg}, S., Dreizler, S., {et~al.} 2013, Astronomy
  {\&} Astrophysics, 553, A6, \dodoi{10.1051/0004-6361/201219058}

\bibitem[{Jarvis {et~al.}(2013)Jarvis, Bonfield, Bruce, Geach, McAlpine,
  McLure, Gonz{\'{a}}lez-Solares, Irwin, Lewis, Yoldas, Andreon, Cross,
  Emerson, Dalton, Dunlop, Hodgkin, Le, Karouzos, Meisenheimer, Oliver,
  Rawlings, Simpson, Smail, Smith, Sullivan, Sutherland, White, \&
  Zwart}]{Jarvis2013}
Jarvis, M.~J., Bonfield, D.~G., Bruce, V.~A., {et~al.} 2013, Monthly Notices of
  the Royal Astronomical Society, 428, 1281, \dodoi{10.1093/mnras/sts118}

\bibitem[{Jin {et~al.}(2019)Jin, Daddi, Magdis, Liu, Schinnerer, Papadopoulos,
  Gu, Gao, \& Calabr{\`{o}}}]{Jin2019}
Jin, S., Daddi, E., Magdis, G.~E., {et~al.} 2019, The Astrophysical Journal,
  887, 144, \dodoi{10.3847/1538-4357/ab55d6}

\bibitem[{Juneau {et~al.}(2011)Juneau, Dickinson, Alexander, \&
  Salim}]{Juneau2011}
Juneau, S., Dickinson, M., Alexander, D.~M., \& Salim, S. 2011, The
  Astrophysical Journal, 736, 104, \dodoi{10.1088/0004-637X/736/2/104}

\bibitem[{Kauffmann {et~al.}(2003)Kauffmann, Heckman, White, Charlot, Tremonti,
  Peng, Seibert, Brinkmann, Nichol, SubbaRao, \& York}]{Kauffmann2003a}
Kauffmann, G., Heckman, T.~M., White, S. D.~M., {et~al.} 2003, Monthly Notices
  of the Royal Astronomical Society, 341, 54,
  \dodoi{10.1046/j.1365-8711.2003.06292.x}

\bibitem[{Kawinwanichakij {et~al.}(2020)Kawinwanichakij, Papovich, Ciardullo,
  Finkelstein, Stevans, Wold, Jogee, Sherman, Florez, \&
  Gronwall}]{Kawinwanichakij2020a}
Kawinwanichakij, L., Papovich, C., Ciardullo, R., {et~al.} 2020, The
  Astrophysical Journal, 892, 7, \dodoi{10.3847/1538-4357/ab75c4}

\bibitem[{Kennicutt(1998)}]{Kennicutt1998b}
Kennicutt, R.~C. 1998, Annual Review of Astronomy and Astrophysics, 36, 189,
  \dodoi{10.1146/annurev.astro.36.1.189}

\bibitem[{Kewley {et~al.}(2013)Kewley, Dopita, Leitherer, Dav{\'{e}}, Yuan,
  Allen, Groves, \& Sutherland}]{Kewley2013}
Kewley, L.~J., Dopita, M.~A., Leitherer, C., {et~al.} 2013, The Astrophysical
  Journal, 774, 100, \dodoi{10.1088/0004-637X/774/2/100}

\bibitem[{Koekemoer {et~al.}(2011)Koekemoer, Faber, Ferguson, Grogin, Kocevski,
  Koo, Lai, Lotz, Lucas, McGrath, Ogaz, Rajan, Riess, Rodney, Strolger,
  Casertano, Castellano, Dahlen, Dickinson, Dolch, Fontana, Giavalisco,
  Grazian, Guo, Hathi, Huang, van~der Wel, Yan, Acquaviva, Alexander, Almaini,
  Ashby, Barden, Bell, Bournaud, Brown, Caputi, Cassata, Challis, Chary,
  Cheung, Cirasuolo, Conselice, Cooray, Croton, Daddi, Dav{\'{e}}, de~Mello,
  de~Ravel, Dekel, Donley, Dunlop, Dutton, Elbaz, Fazio, Filippenko,
  Finkelstein, Frazer, Gardner, Garnavich, Gawiser, Gruetzbauch, Hartley,
  H{\"{a}}ussler, Herrington, Hopkins, Huang, Jha, Johnson, Kartaltepe,
  Khostovan, Kirshner, Lani, Lee, Li, Madau, McCarthy, McIntosh, McLure,
  McPartland, Mobasher, Moreira, Mortlock, Moustakas, Mozena, Nandra, Newman,
  Nielsen, Niemi, Noeske, Papovich, Pentericci, Pope, Primack, Ravindranath,
  Reddy, Renzini, Rix, Robaina, Rosario, Rosati, Salimbeni, Scarlata, Siana,
  Simard, Smidt, Snyder, Somerville, Spinrad, Straughn, Telford, Teplitz,
  Trump, Vargas, Villforth, Wagner, Wandro, Wechsler, Weiner, Wiklind, Wild,
  Wilson, Wuyts, \& Yun}]{Koekemoer2011}
Koekemoer, A.~M., Faber, S.~M., Ferguson, H.~C., {et~al.} 2011, The
  Astrophysical Journal Supplement Series, 197, 36,
  \dodoi{10.1088/0067-0049/197/2/36}

\bibitem[{Kriek {et~al.}(2009)Kriek, van Dokkum, Labb{\'{e}}, Franx,
  Illingworth, Marchesini, \& Quadri}]{Kriek2009}
Kriek, M., van Dokkum, P.~G., Labb{\'{e}}, I., {et~al.} 2009, The Astrophysical
  Journal, 700, 221, \dodoi{10.1088/0004-637X/700/1/221}

\bibitem[{Kriek {et~al.}(2011)Kriek, van Dokkum, Whitaker, Labb{\'{e}}, Franx,
  \& Brammer}]{Kriek2011}
Kriek, M., van Dokkum, P.~G., Whitaker, K.~E., {et~al.} 2011, The Astrophysical
  Journal, 743, 168, \dodoi{10.1088/0004-637X/743/2/168}

\bibitem[{Kriek {et~al.}(2015)Kriek, Shapley, Reddy, Siana, Coil, Mobasher,
  Freeman, de~Groot, Price, Sanders, Shivaei, Brammer, Momcheva, Skelton, van
  Dokkum, Whitaker, Aird, Azadi, Kassis, Bullock, Conroy, Dav{\'{e}}, Keres, \&
  Krumholz}]{Kriek2015}
Kriek, M., Shapley, A.~E., Reddy, N.~A., {et~al.} 2015, The Astrophysical
  Journal Supplement Series, 218, 1, \dodoi{10.1088/0067-0049/218/2/15}

\bibitem[{Kriek {et~al.}(2019)Kriek, Price, Conroy, Suess, Mowla, Pasha,
  Bezanson, van Dokkum, \& Barro}]{Kriek2019}
Kriek, M., Price, S.~H., Conroy, C., {et~al.} 2019, The Astrophysical Journal,
  880, L31, \dodoi{10.3847/2041-8213/ab2e75}

\bibitem[{Kubo {et~al.}(2018)Kubo, Tanaka, Yabe, Toft, Stockmann, \&
  G{\'{o}}mez-Guijarro}]{Kubo2018}
Kubo, M., Tanaka, M., Yabe, K., {et~al.} 2018, The Astrophysical Journal, 867,
  1, \dodoi{10.3847/1538-4357/aae3e8}

\bibitem[{Labb{\'{e}} {et~al.}(2006)Labb{\'{e}}, Bouwens, Illingworth, \&
  Franx}]{Labbe2006}
Labb{\'{e}}, I., Bouwens, R.~J., Illingworth, G.~D., \& Franx, M. 2006, The
  Astrophysical Journal, 649, L67, \dodoi{10.1086/508512}

\bibitem[{Labb{\'{e}} {et~al.}(2005)Labb{\'{e}}, Huang, Franx, Rudnick, Barmby,
  Daddi, van Dokkum, Fazio, Schreiber, Moorwood, Rix, R{\"{o}}ttgering,
  Trujillo, \& van~der Werf}]{Labbe2005}
Labb{\'{e}}, I., Huang, J., Franx, M., {et~al.} 2005, The Astrophysical
  Journal, 624, L81, \dodoi{10.1086/430700}

\bibitem[{Ma {et~al.}(2015)Ma, Gonzalez, Spilker, Strandet, Ashby, Aravena,
  B{\'{e}}thermin, Bothwell, Breuck, Brodwin, Chapman, Fassnacht, Greve,
  Gullberg, Hezaveh, Malkan, Marrone, Saliwanchik, Vieira, Weiss, \&
  Welikala}]{Ma2015}
Ma, J., Gonzalez, A.~H., Spilker, J.~S., {et~al.} 2015, The Astrophysical
  Journal, 812, 88, \dodoi{10.1088/0004-637X/812/1/88}

\bibitem[{Marchesini {et~al.}(2010)Marchesini, Whitaker, Brammer, van Dokkum,
  Labb{\'{e}}, Muzzin, Quadri, Kriek, Lee, Rudnick, Franx, Illingworth, \&
  Wake}]{Marchesini2010}
Marchesini, D., Whitaker, K.~E., Brammer, G., {et~al.} 2010, The Astrophysical
  Journal, 725, 1277, \dodoi{10.1088/0004-637X/725/1/1277}

\bibitem[{Marrone {et~al.}(2018)Marrone, Spilker, Hayward, Vieira, Aravena,
  Ashby, Bayliss, B{\'{e}}thermin, Brodwin, Bothwell, Carlstrom, Chapman, Chen,
  Crawford, Cunningham, {De Breuck}, Fassnacht, Gonzalez, Greve, Hezaveh,
  Lacaille, Litke, Lower, Ma, Malkan, Miller, Morningstar, Murphy, Narayanan,
  Phadke, Rotermund, Sreevani, Stalder, Stark, Strandet, Tang, \&
  Wei{\ss}}]{Marrone2018}
Marrone, D.~P., Spilker, J.~S., Hayward, C.~C., {et~al.} 2018, Nature, 553, 51,
  \dodoi{10.1038/nature24629}

\bibitem[{Marsan {et~al.}(2017)Marsan, Marchesini, Brammer, Geier, Kado-Fong,
  Labb{\'{e}}, Muzzin, \& Stefanon}]{Marsan2017}
Marsan, Z.~C., Marchesini, D., Brammer, G.~B., {et~al.} 2017, The Astrophysical
  Journal, 842, 21, \dodoi{10.3847/1538-4357/aa7206}

\bibitem[{Marsan {et~al.}(2015)Marsan, Marchesini, Brammer, Stefanon, Muzzin,
  Fern{\'{a}}ndez-soto, Geier, Hainline, Intema, Karim, \&
  Labb{\'{e}}}]{Marsan2015}
---. 2015, The Astrophysical Journal, 801, 133,
  \dodoi{10.1088/0004-637X/801/2/133}

\bibitem[{Martis {et~al.}(2016)Martis, Marchesini, Brammer, Muzzin,
  Labb{\'{e}}, Momcheva, Skelton, Stefanon, van Dokkum, \&
  Whitaker}]{Martis2016}
Martis, N.~S., Marchesini, D., Brammer, G.~B., {et~al.} 2016, The Astrophysical
  Journal, 827, L25, \dodoi{10.3847/2041-8205/827/2/L25}

\bibitem[{Mauduit {et~al.}(2012)Mauduit, Lacy, Farrah, Surace, Jarvis, Oliver,
  Maraston, Vaccari, Marchetti, Zeimann, Gonz{\'{a}}les-Solares, Pforr, Petric,
  Henriques, Thomas, Afonso, Rettura, Wilson, Falder, Geach, Huynh, Norris,
  Seymour, Richards, Stanford, Alexander, Becker, Best, Bizzocchi, Bonfield,
  Castro, Cava, Chapman, Christopher, Clements, Covone, Dubois, Dunlop, Dyke,
  Edge, Ferguson, Foucaud, Franceschini, Gal, Grant, Grossi, Hatziminaoglou,
  Hickey, Hodge, Huang, Ivison, Kim, LeFevre, Lehnert, Lonsdale, Lubin, McLure,
  Messias, Mart{\'{i}}nez-Sansigre, Mortier, Nielsen, Ouchi, Parish,
  Perez-Fournon, Pierre, Rawlings, Readhead, Ridgway, Rigopoulou, Romer,
  Rosebloom, Rottgering, Rowan-Robinson, Sajina, Simpson, Smail, Squires,
  Stevens, Taylor, Trichas, Urrutia, van Kampen, Verma, \& Xu}]{Mauduit2012}
Mauduit, J.-C., Lacy, M., Farrah, D., {et~al.} 2012, Publications of the
  Astronomical Society of the Pacific, 124, 1135, \dodoi{10.1086/668290}

\bibitem[{McCracken {et~al.}(2012)McCracken, Milvang-Jensen, Dunlop, Franx,
  Fynbo, {Le F{\`{e}}vre}, Holt, Caputi, Goranova, Buitrago, Emerson,
  Freudling, Hudelot, L{\'{o}}pez-Sanjuan, Magnard, Mellier, M{\o}ller,
  Nilsson, Sutherland, Tasca, \& Zabl}]{McCracken2012}
McCracken, H.~J., Milvang-Jensen, B., Dunlop, J., {et~al.} 2012, Astronomy {\&}
  Astrophysics, 544, A156, \dodoi{10.1051/0004-6361/201219507}

\bibitem[{McLean {et~al.}(2010)McLean, Steidel, Epps, Matthews, Adkins,
  Konidaris, Weber, Aliado, Brims, Canfield, Cromer, Fucik, Kulas, Mace,
  Magnone, Rodriguez, Wang, \& Weiss}]{McLean2010}
McLean, I.~S., Steidel, C.~C., Epps, H., {et~al.} 2010in , 77351E.
\newblock
  \url{http://proceedings.spiedigitallibrary.org/proceeding.aspx?doi=10.1117/12.856715}

\bibitem[{McLean {et~al.}(2012)McLean, Steidel, Epps, Konidaris, Matthews,
  Adkins, Aliado, Brims, Canfield, Cromer, Fucik, Kulas, Mace, Magnone,
  Rodriguez, Rudie, Trainor, Wang, Weber, \& Weiss}]{McLean2012}
McLean, I.~S., Steidel, C.~C., Epps, H.~W., {et~al.} 2012in .
\newblock
  \url{http://proceedings.spiedigitallibrary.org/proceeding.aspx?doi=10.1117/12.924794}

\bibitem[{Mehta {et~al.}(2018)Mehta, Scarlata, Capak, Davidzon, Faisst, Hsieh,
  Ilbert, Jarvis, Laigle, Phillips, Silverman, Strauss, Tanaka, Bowler, Coupon,
  Foucaud, Hemmati, Masters, McCracken, Mobasher, Ouchi, Shibuya, \&
  Wang}]{Mehta2018}
Mehta, V., Scarlata, C., Capak, P., {et~al.} 2018, The Astrophysical Journal
  Supplement Series, 235, 36, \dodoi{10.3847/1538-4365/aab60c}

\bibitem[{Merlin {et~al.}(2018)Merlin, Fontana, Castellano, Santini, Torelli,
  Boutsia, Wang, Grazian, Pentericci, Schreiber, Ciesla, McLure, Derriere,
  Dunlop, \& Elbaz}]{Merlin2018}
Merlin, E., Fontana, A., Castellano, M., {et~al.} 2018, Monthly Notices of the
  Royal Astronomical Society, 473, 2098, \dodoi{10.1093/mnras/stx2385}

\bibitem[{Merlin {et~al.}(2019)Merlin, Fortuni, Torelli, Santini, Castellano,
  Fontana, Grazian, Pentericci, Pilo, \& Schmidt}]{Merlin2019}
Merlin, E., Fortuni, F., Torelli, M., {et~al.} 2019, Monthly Notices of the
  Royal Astronomical Society, 490, 3309, \dodoi{10.1093/mnras/stz2615}

\bibitem[{Moustakas {et~al.}(2006)Moustakas, {Kennicutt, Jr.}, \&
  Tremonti}]{Moustakas2006}
Moustakas, J., {Kennicutt, Jr.}, R.~C., \& Tremonti, C.~A. 2006, The
  Astrophysical Journal, 642, 775, \dodoi{10.1086/500964}

\bibitem[{Muzzin {et~al.}(2013)Muzzin, Marchesini, Stefanon, Franx,
  Milvang-Jensen, Dunlop, Fynbo, Brammer, Labb{\'{e}}, \& van
  Dokkum}]{Muzzin2013a}
Muzzin, A., Marchesini, D., Stefanon, M., {et~al.} 2013, The Astrophysical
  Journal Supplement Series, 206, 8, \dodoi{10.1088/0067-0049/206/1/8}

\bibitem[{Nelan {et~al.}(2005)Nelan, Smith, Hudson, Wegner, Lucey, Moore,
  Quinney, \& Suntzeff}]{Nelan2005}
Nelan, J.~E., Smith, R.~J., Hudson, M.~J., {et~al.} 2005, The Astrophysical
  Journal, 632, 137, \dodoi{10.1086/431962}

\bibitem[{Newman {et~al.}(2018)Newman, Belli, Ellis, \& Patel}]{Newman2018}
Newman, A.~B., Belli, S., Ellis, R.~S., \& Patel, S.~G. 2018, The Astrophysical
  Journal, 862, 125, \dodoi{10.3847/1538-4357/aacd4d}

\bibitem[{Oke \& Gunn(1983)}]{Oke1983}
Oke, J.~B., \& Gunn, J.~E. 1983, The Astrophysical Journal, 266, 713,
  \dodoi{10.1086/160817}

\bibitem[{Oliphant \& Millma(2006)}]{Oliphant2006}
Oliphant, T., \& Millma, J.~k. 2006, {A guide to NumPy},
  \dodoi{DOI:10.1109/MCSE.2007.58}

\bibitem[{Pacifici {et~al.}(2016)Pacifici, Kassin, Weiner, Holden, Gardner,
  Faber, Ferguson, Koo, Primack, Bell, Dekel, Gawiser, Giavalisco, Rafelski,
  Simons, Barro, Croton, Dav{\'{e}}, Fontana, Grogin, Koekemoer, Lee, Salmon,
  Somerville, \& Behroozi}]{Pacifici2016}
Pacifici, C., Kassin, S.~A., Weiner, B.~J., {et~al.} 2016, The Astrophysical
  Journal, 832, 79, \dodoi{10.3847/0004-637X/832/1/79}

\bibitem[{Pavesi {et~al.}(2018)Pavesi, Riechers, Sharon, Smol{\v{c}}i{\'{c}},
  Faisst, Schinnerer, Carilli, Capak, Scoville, \& Stacey}]{Pavesi2018}
Pavesi, R., Riechers, D.~A., Sharon, C.~E., {et~al.} 2018, The Astrophysical
  Journal, 861, 43, \dodoi{10.3847/1538-4357/aac6b6}

\bibitem[{P{\'{e}}rez \& Granger(2007)}]{Perez2007}
P{\'{e}}rez, F., \& Granger, B.~E. 2007, Computing in Science and Engineering,
  \dodoi{10.1109/MCSE.2007.53}

\bibitem[{Price-Whelan {et~al.}(2018)Price-Whelan, Sipőcz, G{\"{u}}nther, Lim,
  Crawford, Conseil, Shupe, Craig, Dencheva, Ginsburg, VanderPlas, Bradley,
  P{\'{e}}rez-Su{\'{a}}rez, de~Val-Borro, Aldcroft, Cruz, Robitaille, Tollerud,
  Ardelean, Babej, Bach, Bachetti, Bakanov, Bamford, Barentsen, Barmby,
  Baumbach, Berry, Biscani, Boquien, Bostroem, Bouma, Brammer, Bray,
  Breytenbach, Buddelmeijer, Burke, Calderone, Rodr{\'{i}}guez, Cara, Cardoso,
  Cheedella, Copin, Corrales, Crichton, D'Avella, Deil, Depagne, Dietrich,
  Donath, Droettboom, Earl, Erben, Fabbro, Ferreira, Finethy, Fox, Garrison,
  Gibbons, Goldstein, Gommers, Greco, Greenfield, Groener, Grollier, Hagen,
  Hirst, Homeier, Horton, Hosseinzadeh, Hu, Hunkeler, Ivezi{\'{c}}, Jain,
  Jenness, Kanarek, Kendrew, Kern, Kerzendorf, Khvalko, King, Kirkby, Kulkarni,
  Kumar, Lee, Lenz, Littlefair, Ma, Macleod, Mastropietro, McCully, Montagnac,
  Morris, Mueller, Mumford, Muna, Murphy, Nelson, Nguyen, Ninan, N{\"{o}}the,
  Ogaz, Oh, Parejko, Parley, Pascual, Patil, Patil, Plunkett, Prochaska,
  Rastogi, Janga, Sabater, Sakurikar, Seifert, Sherbert, Sherwood-Taylor, Shih,
  Sick, Silbiger, Singanamalla, Singer, Sladen, Sooley, Sornarajah, Streicher,
  Teuben, Thomas, Tremblay, Turner, Terr{\'{o}}n, van Kerkwijk, de~la Vega,
  Watkins, Weaver, Whitmore, Woillez, \& Zabalza}]{Astropy2018}
Price-Whelan, A.~M., Sipőcz, B.~M., G{\"{u}}nther, H.~M., {et~al.} 2018, The
  Astronomical Journal, 156, 123, \dodoi{10.3847/1538-3881/aabc4f}

\bibitem[{Reddy {et~al.}(2018)Reddy, Shapley, Sanders, Kriek, Coil, Shivaei,
  Freeman, Mobasher, Siana, Azadi, Fetherolf, Fornasini, Leung, Price, Zick, \&
  Barro}]{Reddy2018}
Reddy, N.~A., Shapley, A.~E., Sanders, R.~L., {et~al.} 2018, The Astrophysical
  Journal, 869, 92, \dodoi{10.3847/1538-4357/aaed1e}

\bibitem[{Riechers {et~al.}(2013)Riechers, Bradford, Clements, Dowell,
  P{\'{e}}rez-Fournon, Ivison, Bridge, Conley, Fu, Vieira, Wardlow, Calanog,
  Cooray, Hurley, Neri, Kamenetzky, Aguirre, Altieri, Arumugam, Benford,
  B{\'{e}}thermin, Bock, Burgarella, Cabrera-Lavers, Chapman, Cox, Dunlop,
  Earle, Farrah, Ferrero, Franceschini, Gavazzi, Glenn, Solares, Gurwell,
  Halpern, Hatziminaoglou, Hyde, Ibar, Kov{\'{a}}cs, Krips, Lupu, Maloney,
  Martinez-Navajas, Matsuhara, Murphy, Naylor, Nguyen, Oliver, Omont, Page,
  Petitpas, Rangwala, Roseboom, Scott, Smith, Staguhn, Streblyanska, Thomson,
  Valtchanov, Viero, Wang, Zemcov, \& Zmuidzinas}]{Riechers2013}
Riechers, D.~A., Bradford, C.~M., Clements, D.~L., {et~al.} 2013, Nature, 496,
  329, \dodoi{10.1038/nature12050}

\bibitem[{Riechers {et~al.}(2017)Riechers, Leung, Ivison, P{\'{e}}rez-Fournon,
  Lewis, Marques-Chaves, Oteo, Clements, Cooray, Greenslade,
  Mart{\'{i}}nez-Navajas, Oliver, Rigopoulou, Scott, \& Weiss}]{Riechers2017}
Riechers, D.~A., Leung, T. K.~D., Ivison, R.~J., {et~al.} 2017, The
  Astrophysical Journal, 850, 1, \dodoi{10.3847/1538-4357/aa8ccf}

\bibitem[{Riechers {et~al.}(2020)Riechers, Hodge, Pavesi, Daddi, Decarli,
  Ivison, Sharon, Smail, Walter, Aravena, Capak, Carilli, Cox, da~Cunha,
  Dannerbauer, Dickinson, Neri, \& Wagg}]{Riechers2020a}
Riechers, D.~A., Hodge, J.~A., Pavesi, R., {et~al.} 2020, The Astrophysical
  Journal, 895, 81, \dodoi{10.3847/1538-4357/ab8c48}

\bibitem[{Robitaille {et~al.}(2013)Robitaille, Tollerud, Greenfield,
  Droettboom, Bray, Aldcroft, Davis, Ginsburg, Price-Whelan, Kerzendorf,
  Conley, Crighton, Barbary, Muna, Ferguson, Grollier, Parikh, Nair,
  G{\"{u}}nther, Deil, Woillez, Conseil, Kramer, Turner, Singer, Fox, Weaver,
  Zabalza, Edwards, {Azalee Bostroem}, Burke, Casey, Crawford, Dencheva, Ely,
  Jenness, Labrie, Lim, Pierfederici, Pontzen, Ptak, Refsdal, Servillat, \&
  Streicher}]{Astropy2013}
Robitaille, T.~P., Tollerud, E.~J., Greenfield, P., {et~al.} 2013, Astronomy
  {\&} Astrophysics, 558, A33, \dodoi{10.1051/0004-6361/201322068}

\bibitem[{Salmon {et~al.}(2015)Salmon, Papovich, Finkelstein, Tilvi, Finlator,
  Behroozi, Dahlen, Dav{\'{e}}, Dekel, Dickinson, Ferguson, Giavalisco, Long,
  Lu, Mobasher, Reddy, Somerville, \& Wechsler}]{Salmon2015}
Salmon, B., Papovich, C., Finkelstein, S.~L., {et~al.} 2015, The Astrophysical
  Journal, 799, 183, \dodoi{10.1088/0004-637X/799/2/183}

\bibitem[{Sanders {et~al.}(2007)Sanders, Salvato, Aussel, Ilbert, Scoville,
  Surace, Frayer, Sheth, Helou, Brooke, Bhattacharya, Yan, Kartaltepe, Barnes,
  Blain, Calzetti, Capak, Carilli, Carollo, Comastri, Daddi, Ellis, Elvis,
  Fall, Franceschini, Giavalisco, Hasinger, Impey, Koekemoer, {Le Fevre},
  Lilly, Liu, McCracken, Mobasher, Renzini, Rich, Schinnerer, Shopbell,
  Taniguchi, Thompson, Urry, \& Williams}]{Sanders2007}
Sanders, D.~B., Salvato, M., Aussel, H., {et~al.} 2007, The Astrophysical
  Journal Supplement Series, 172, 86, \dodoi{10.1086/517885}

\bibitem[{Schaye {et~al.}(2015)Schaye, Crain, Bower, Furlong, Schaller, Theuns,
  {Dalla Vecchia}, Frenk, Mccarthy, Helly, Jenkins, Rosas-Guevara, White, Baes,
  Booth, Camps, Navarro, Qu, Rahmati, Sawala, Thomas, \& Trayford}]{Schaye2015}
Schaye, J., Crain, R.~A., Bower, R.~G., {et~al.} 2015, Monthly Notices of the
  Royal Astronomical Society, 446, 521, \dodoi{10.1093/mnras/stu2058}

\bibitem[{Schreiber {et~al.}(2018{\natexlab{a}})Schreiber, Glazebrook,
  Nanayakkara, Kacprzak, Labb{\'{e}}, Oesch, Yuan, Tran, Papovich, Spitler, \&
  Straatman}]{Schreiber2018b}
Schreiber, C., Glazebrook, K., Nanayakkara, T., {et~al.} 2018{\natexlab{a}},
  Astronomy {\&} Astrophysics, 618, A85, \dodoi{10.1051/0004-6361/201833070}

\bibitem[{Schreiber {et~al.}(2018{\natexlab{b}})Schreiber, Labb{\'{e}},
  Glazebrook, Bekiaris, Papovich, Costa, Elbaz, Kacprzak, Nanayakkara, Oesch,
  Pannella, Spitler, Straatman, Tran, \& Wang}]{Schreiber2018a}
Schreiber, C., Labb{\'{e}}, I., Glazebrook, K., {et~al.} 2018{\natexlab{b}},
  Astronomy {\&} Astrophysics, 611, A22, \dodoi{10.1051/0004-6361/201731917}

\bibitem[{Shahidi {et~al.}(2020)Shahidi, Mobasher, Nayyeri, Hemmati, Wiklind,
  Chartab, Dickinson, Finkelstein, Pacifici, Papovich, Ferguson, Fontana,
  Giavalisco, Koekemoer, Newman, Sattari, \& Somerville}]{Shahidi2020}
Shahidi, A., Mobasher, B., Nayyeri, H., {et~al.} 2020, The Astrophysical
  Journal, 897, 44, \dodoi{10.3847/1538-4357/ab96c5}

\bibitem[{Shapley {et~al.}(2015)Shapley, Reddy, Kriek, Freeman, Sanders, Siana,
  Coil, Mobasher, Shivaei, Price, \& Groot}]{Shapley2015}
Shapley, A.~E., Reddy, N.~A., Kriek, M., {et~al.} 2015, Astrophysical Journal,
  801, 88, \dodoi{10.1088/0004-637X/801/2/88}

\bibitem[{Silverman {et~al.}(2009)Silverman, Lamareille, Maier, Lilly,
  Mainieri, Brusa, Cappelluti, Hasinger, Zamorani, Scodeggio, Bolzonella,
  Contini, Carollo, Jahnke, Kneib, {Le F{\`{e}}vre}, Merloni, Bardelli,
  Bongiorno, Brunner, Caputi, Civano, Comastri, Coppa, Cucciati, {De La Torre},
  {De Ravel}, Elvis, Finoguenov, Fiore, Franzetti, Garilli, Gilli, Iovino,
  Kampczyk, Knobel, Kova{\v{c}}, {Le Borgne}, {Le Brun}, Mignoli, Pello, Peng,
  {Perez Montero}, Ricciardelli, Tanaka, Tasca, Tresse, Vergani, Vignali,
  Zucca, Bottini, Cappi, Cassata, Fumana, Griffiths, Kartaltepe, Koekemoer,
  Marinoni, Mccracken, Memeo, Meneux, Oesch, Porciani, \&
  Salvato}]{Silverman2009}
Silverman, J.~D., Lamareille, F., Maier, C., {et~al.} 2009, Astrophysical
  Journal, 696, 396, \dodoi{10.1088/0004-637X/696/1/396}

\bibitem[{Smith {et~al.}(2012)Smith, Lucey, Price, Hudson, \&
  Phillipps}]{Smith2012}
Smith, R.~J., Lucey, J.~R., Price, J., Hudson, M.~J., \& Phillipps, S. 2012,
  Monthly Notices of the Royal Astronomical Society, 419, 3167,
  \dodoi{10.1111/j.1365-2966.2011.19956.x}

\bibitem[{Spitler {et~al.}(2014)Spitler, Straatman, Labb{\'{e}}, Glazebrook,
  Tran, Kacprzak, Quadri, Papovich, Persson, van Dokkum, Allen,
  Kawinwanichakij, Kelson, McCarthy, Mehrtens, {J. Monson}, Nanayakkara, Rees,
  Tilvi, \& Tomczak}]{Spitler2014}
Spitler, L.~R., Straatman, C. M.~S., Labb{\'{e}}, I., {et~al.} 2014, The
  Astrophysical Journal, 787, L36, \dodoi{10.1088/2041-8205/787/2/L36}

\bibitem[{Straatman {et~al.}(2014)Straatman, Labb{\'{e}}, Spitler, Allen,
  Altieri, Brammer, Dickinson, van Dokkum, Inami, Glazebrook, Kacprzak,
  Kawinwanichakij, Kelson, McCarthy, Mehrtens, Monson, Murphy, Papovich,
  Persson, Quadri, Rees, Tomczak, Tran, \& Tilvi}]{Straatman2014}
Straatman, C. M.~S., Labb{\'{e}}, I., Spitler, L.~R., {et~al.} 2014, The
  Astrophysical Journal, 783, L14, \dodoi{10.1088/2041-8205/783/1/L14}

\bibitem[{Straatman {et~al.}(2016)Straatman, Spitler, Quadri, Labb{\'{e}},
  Glazebrook, Persson, Papovich, Tran, Brammer, Cowley, Tomczak, Nanayakkara,
  Alcorn, Allen, Broussard, van Dokkum, Forrest, van Houdt, Kacprzak,
  Kawinwanichakij, Kelson, Lee, McCarthy, Mehrtens, Monson, Murphy, Rees,
  Tilvi, \& Whitaker}]{Straatman2016}
Straatman, C. M.~S., Spitler, L.~R., Quadri, R.~F., {et~al.} 2016, The
  Astrophysical Journal, 830, 51, \dodoi{10.3847/0004-637X/830/1/51}

\bibitem[{Strandet {et~al.}(2017)Strandet, Weiss, Breuck, Marrone, Vieira,
  Aravena, Ashby, B{\'{e}}thermin, Bothwell, Bradford, Carlstrom, Chapman,
  Cunningham, Chen, Fassnacht, Gonzalez, Greve, Gullberg, Hayward, Hezaveh,
  Litke, Ma, Malkan, Menten, Miller, Murphy, Narayanan, Phadke, Rotermund,
  Spilker, \& Sreevani}]{Strandet2017}
Strandet, M.~L., Weiss, A., Breuck, C.~D., {et~al.} 2017, The Astrophysical
  Journal, 842, L15, \dodoi{10.3847/2041-8213/aa74b0}

\bibitem[{Strom {et~al.}(2017)Strom, Steidel, Rudie, Trainor, Pettini, \&
  Reddy}]{Strom2016}
Strom, A.~L., Steidel, C.~C., Rudie, G.~C., {et~al.} 2017, The Astrophysical
  Journal, 836, 164, \dodoi{10.3847/1538-4357/836/2/164}

\bibitem[{Tanaka {et~al.}(2019)Tanaka, Valentino, Toft, Onodera, Shimakawa,
  Ceverino, Faisst, Gallazzi, G{\'{o}}mez-Guijarro, Kubo, Magdis, Steinhardt,
  Stockmann, Yabe, \& Zabl}]{Tanaka2019}
Tanaka, M., Valentino, F., Toft, S., {et~al.} 2019, The Astrophysical Journal,
  885, L34, \dodoi{10.3847/2041-8213/ab4ff3}

\bibitem[{Thomas {et~al.}(2005)Thomas, Maraston, Bender, \&
  de~Oliveira}]{Thomas2005}
Thomas, D., Maraston, C., Bender, R., \& de~Oliveira, C.~M. 2005, The
  Astrophysical Journal, 621, 673, \dodoi{10.1086/426932}

\bibitem[{Thomas {et~al.}(2010)Thomas, Maraston, Schawinski, Sarzi, \&
  Silk}]{Thomas2010}
Thomas, D., Maraston, C., Schawinski, K., Sarzi, M., \& Silk, J. 2010, Monthly
  Notices of the Royal Astronomical Society, 404, 1775,
  \dodoi{10.1111/j.1365-2966.2010.16427.x}

\bibitem[{Toft {et~al.}(2014)Toft, Smol{\v{c}}i{\'{c}}, Magnelli, Karim, Zirm,
  Michalowski, Capak, Sheth, Schawinski, Krogager, Wuyts, Sanders, Man, Lutz,
  Staguhn, Berta, Mccracken, Krpan, \& Riechers}]{Toft2014}
Toft, S., Smol{\v{c}}i{\'{c}}, V., Magnelli, B., {et~al.} 2014, The
  Astrophysical Journal, 782, 68, \dodoi{10.1088/0004-637X/782/2/68}

\bibitem[{Treu {et~al.}(2005)Treu, Ellis, Liao, \& van Dokkum}]{Treu2005}
Treu, T., Ellis, R.~S., Liao, T.~X., \& van Dokkum, P.~G. 2005, The
  Astrophysical Journal, 622, L5, \dodoi{10.1086/429374}

\bibitem[{Trump {et~al.}(2013)Trump, Konidaris, Barro, Koo, Kocevski, Juneau,
  Weiner, Faber, McLean, Yan, P{\'{e}}rez-Gonz{\'{a}}lez, \&
  Villar}]{Trump2013}
Trump, J.~R., Konidaris, N.~P., Barro, G., {et~al.} 2013, Astrophysical Journal
  Letters, 763, 1, \dodoi{10.1088/2041-8205/763/1/L6}

\bibitem[{Valentino {et~al.}(2020)Valentino, Tanaka, Davidzon, Toft,
  G{\'{o}}mez-Guijarro, Stockmann, Onodera, Brammer, Ceverino, Faisst,
  Gallazzi, Hayward, Ilbert, Kubo, Magdis, Selsing, Shimakawa, Sparre,
  Steinhardt, Yabe, \& Zabl}]{Valentino2020}
Valentino, F., Tanaka, M., Davidzon, I., {et~al.} 2020, The Astrophysical
  Journal, 889, 93, \dodoi{10.3847/1538-4357/ab64dc}

\bibitem[{van~de Sande {et~al.}(2013)van~de Sande, Kriek, Franx, van Dokkum,
  Bezanson, Bouwens, Quadri, Rix, \& Skelton}]{vandeSande2013}
van~de Sande, J., Kriek, M., Franx, M., {et~al.} 2013, The Astrophysical
  Journal, 771, 85, \dodoi{10.1088/0004-637X/771/2/85}

\bibitem[{van~der Wel {et~al.}(2014)van~der Wel, Franx, van Dokkum, Skelton,
  Momcheva, Whitaker, Brammer, Bell, Rix, Wuyts, Ferguson, Holden, Barro,
  Koekemoer, Chang, McGrath, H{\"{a}}ussler, Dekel, Behroozi, Fumagalli, Leja,
  Lundgren, Maseda, Nelson, Wake, Patel, Labb{\'{e}}, Faber, Grogin, \&
  Kocevski}]{vanderWel2014}
van~der Wel, A., Franx, M., van Dokkum, P.~G., {et~al.} 2014, The Astrophysical
  Journal, 788, 28, \dodoi{10.1088/0004-637X/788/1/28}

\bibitem[{van Dokkum {et~al.}(2009)van Dokkum, Labb{\'{e}}, Marchesini, Quadri,
  Brammer, Whitaker, Kriek, Franx, Rudnick, Illingworth, Lee, \&
  Muzzin}]{VanDokkum2009}
van Dokkum, P.~G., Labb{\'{e}}, I., Marchesini, D., {et~al.} 2009, Publications
  of the Astronomical Society of the Pacific, 121, 2, \dodoi{10.1086/597138}

\bibitem[{{Vanden Berk} {et~al.}(2001){Vanden Berk}, Richards, Bauer, Strauss,
  Schneider, Heckman, York, Hall, Fan, Knapp, Anderson, Annis, Bahcall,
  Bernardi, Briggs, Brinkmann, Brunner, Burles, Carey, Castander, Connolly,
  Crocker, Csabai, Doi, Finkbeiner, Friedman, Frieman, Fukugita, Gunn,
  Hennessy, Ivezi{\'{c}}, Kent, Kunszt, Lamb, Leger, Long, Loveday, Lupton,
  Meiksin, Merelli, Munn, Newberg, Newcomb, Nichol, Owen, Pier, Pope, Rockosi,
  Schlegel, Siegmund, Smee, Snir, Stoughton, Stubbs, SubbaRao, Szalay, Szokoly,
  Tremonti, Uomoto, Waddell, Yanny, \& Zheng}]{VandenBerk2001}
{Vanden Berk}, D.~E., Richards, G.~T., Bauer, A., {et~al.} 2001, The
  Astronomical Journal, 122, 549, \dodoi{10.1086/321167}

\bibitem[{Vogelsberger {et~al.}(2014{\natexlab{a}})Vogelsberger, Genel,
  Springel, Torrey, Sijacki, Xu, Snyder, Bird, Nelson, \&
  Hernquist}]{Vogelsberger2014a}
Vogelsberger, M., Genel, S., Springel, V., {et~al.} 2014{\natexlab{a}}, Nature,
  509, 177, \dodoi{10.1038/nature13316}

\bibitem[{Vogelsberger {et~al.}(2014{\natexlab{b}})Vogelsberger, Genel,
  Springel, Torrey, Sijacki, Xu, Snyder, Nelson, \&
  Hernquist}]{Vogelsberger2014b}
---. 2014{\natexlab{b}}, Monthly Notices of the Royal Astronomical Society,
  444, 1518, \dodoi{10.1093/mnras/stu1536}

\bibitem[{Wang {et~al.}(2019)Wang, Schreiber, Elbaz, Yoshimura, Kohno, Shu,
  Yamaguchi, Pannella, Franco, Huang, Lim, \& Wang}]{Twang2019}
Wang, T., Schreiber, C., Elbaz, D., {et~al.} 2019, Nature, 572, 211,
  \dodoi{10.1038/s41586-019-1452-4}

\bibitem[{Wellons {et~al.}(2015)Wellons, Torrey, Ma, Rodriguez-Gomez,
  Vogelsberger, Kriek, van Dokkum, Nelson, Genel, Pillepich, Springel, Sijacki,
  Snyder, Nelson, Sales, \& Hernquist}]{Wellons2015}
Wellons, S., Torrey, P., Ma, C.-P., {et~al.} 2015, Monthly Notices of the Royal
  Astronomical Society, 449, 361, \dodoi{10.1093/mnras/stv303}

\bibitem[{Whitaker {et~al.}(2011)Whitaker, Labb{\'{e}}, van Dokkum, Brammer,
  Kriek, Marchesini, Quadri, Franx, Muzzin, Williams, Bezanson, Illingworth,
  Lee, Lundgren, Nelson, Rudnick, Tal, \& Wake}]{Whitaker2011}
Whitaker, K.~E., Labb{\'{e}}, I., van Dokkum, P.~G., {et~al.} 2011, The
  Astrophysical Journal, 735, 86, \dodoi{10.1088/0004-637X/735/2/86}

\bibitem[{Williams {et~al.}(2019)Williams, Labbe, Spilker, Stefanon, Leja,
  Whitaker, Bezanson, Narayanan, Oesch, \& Weiner}]{Williams2019}
Williams, C.~C., Labbe, I., Spilker, J., {et~al.} 2019, The Astrophysical
  Journal, 884, 154, \dodoi{10.3847/1538-4357/ab44aa}

\bibitem[{Williams {et~al.}(2009)Williams, Quadri, Franx, van Dokkum, \&
  Labb{\'{e}}}]{Williams2009}
Williams, R.~J., Quadri, R.~F., Franx, M., van Dokkum, P., \& Labb{\'{e}}, I.
  2009, The Astrophysical Journal, 691, 1879,
  \dodoi{10.1088/0004-637X/691/2/1879}

\bibitem[{Wu {et~al.}(2018)Wu, van~der Wel, Gallazzi, Bezanson, Pacifici,
  Straatman, Franx, Bari{\v{s}}i{\'{c}}, Bell, Brammer, Calhau, Chauke, van
  Houdt, Maseda, Muzzin, Rix, Sobral, Spilker, van~de Sande, van Dokkum, \&
  Wild}]{Wu2018a}
Wu, P.-F., van~der Wel, A., Gallazzi, A., {et~al.} 2018, The Astrophysical
  Journal, 855, 85, \dodoi{10.3847/1538-4357/aab0a6}

\bibitem[{Wuyts {et~al.}(2007)Wuyts, Labbe, Franx, Rudnick, van Dokkum, Fazio,
  {Forster Schreiber}, Huang, Moorwood, Rix, Rottgering, \& van~der
  Werf}]{Wuyts2007}
Wuyts, S., Labbe, I., Franx, M., {et~al.} 2007, The Astrophysical Journal, 655,
  51, \dodoi{10.1086/509708}

\bibitem[{Zavala {et~al.}(2018)Zavala, Monta{\~{n}}a, Hughes, Yun, Ivison,
  Valiante, Wilner, Spilker, Aretxaga, Eales, Avila-Reese, Ch{\'{a}}vez,
  Cooray, Dannerbauer, Dunlop, Dunne, G{\'{o}}mez-Ruiz, Micha{\l}owski,
  Narayanan, Nayyeri, Oteo, {Rosa Gonz{\'{a}}lez},
  S{\'{a}}nchez-Arg{\"{u}}elles, Schloerb, Serjeant, Smith, Terlevich, Vega,
  Villalba, {Van Der Werf}, Wilson, \& Zeballos}]{Zavala2018}
Zavala, J.~A., Monta{\~{n}}a, A., Hughes, D.~H., {et~al.} 2018, Nature
  Astronomy, 2, 56, \dodoi{10.1038/s41550-017-0297-8}

\end{thebibliography}

\clearpage
\appendix
\renewcommand\thefigure{\thesection.\arabic{figure}}    
\renewcommand\thetable{\thesection.\arabic{table}}

\section{Summary of Observations} \label{A:obs}
\setcounter{figure}{0}
\setcounter{table}{0}

Spectra were taken on Keck/MOSFIRE under the program ``Ultramassive Galaxies and their Environments at $3 < z < 4$" (PI: Wilson), which was awarded a total of 11.5 nights over 4 semesters (2018B-2020A).
Of these, two nights were lost in their entirety due to weather and a total of about one additional night was lost due to various technical issues and shorter periods of inclement weather.
Data from a previous program ``Formation of Massive Quiescent Galaxies'' (PI: Wilson), which consisted of two half-nights in 2017B were also used in this analysis.
We present the observations by individual mask in Table \ref{T:Obs} and by UMG in Table \ref{T:Obs2}.

\begin{ThreePartTable}
\begin{TableNotes}
  \footnotesize
  \item [a] All individual $K-$band exposures were 180s, while $H-$band exposures were 120s.
  \item [b] The first seeing value is the estimated seeing from MIRA focusing, while the second is an average of the seeing derived from Gaussian fits to 1D spectra of slit stars on the given mask.
\end{TableNotes}
  \begin{longtable*}{lllcccc}
  \caption{Summary of UMG candidate masks observed by date.}\\
  Field & Observing & Mask Name & Filter & Total Integration & Average & Galaxies \\ 
  & Night & & & Time (ks)\tnote{a} & Seeing ($''$)\tnote{b}	& Targeted \\
  \hline \hline \endfirsthead
  Field & Observing & Mask Name & Filter & Total Integration & Average & Galaxies \\ 
  & Night & & & Time (ks)\footnote{Testing} & Seeing ($''$)	& Targeted \\
   \hline \hline \endhead
  COSMOS-UltraVISTA & 21Nov2017 & COS2017$_3$	& K	& 8.8 &	0.6 / 0.78	& 15 \\
  & 	22Nov2017 & 	COS2017$_5$	& 	K &	7.8 &		0.8 / 0.74	& 13  \\
  &	27Nov2018 &	COS-201999-K1 &	K &	7.2 &		1.1 / 0.63	& 21  \\
  &	14Dec2018 &	COS-201999-H1 &	H &	5.8 &		1.2 / 1.49	& 12  \\
  &			 &	COS-201999-H2 &	H &	4.8 &		1.2 / 1.40	& 10  \\
  &	15Dec2018 &	COS-84674-H1 &	H &	5.3 &		1.2 / 0.95	& 15  \\
  &			 &	COS-84674-H2 &	H &	5.3 &		1.2 / 0.94	& 13  \\	
  &	28Feb2019 &	COS-195616-K1 &	K &	7.9 &		1.3 / 0.89	& 18  \\
  &			 &	COS-226441-K1 &	K &	5.0 &		1.0 / 0.99	& 22  \\
  &	17Mar2019 &	COS-160748-K1 &	K &	3.6 &		0.5 / 0.66	& 18  \\
  &			 &	COS-201999-H3 &	H &	4.8 &		0.7 / 0.67	& 20  \\
  &			 &	COS-201999-H4 &	H &	5.8 &		0.6 / 0.61	& 18  \\
  &	14Nov2019 &	COS-84674-H3 &	H &	6.2 &		0.5 / 0.65	& 16  \\
  &	08Dec2019 &	COS-79837-K1 &	K &	2.9 &		0.5 / 0.74	& 9  \\
  &			 &	COS-113684-K1 &	K &	3.6 &		0.6 / 0.56	& 6  \\
  &			 &	COS-258857-K1 &	K &	3.6 &		0.6 / 0.80	& 8  \\
  &	02Feb2020 &	COS-131925-K1w &	K &	4.3 &		1.4 / 0.77	& 25  \\
  &			 &	COS-111740-K1w &	K &	9.4 &		1.4 / 1.40	& 21  \\
  &	03Feb2020 &	COS-258857-K12 &	K &	3.6 &		1.0 / 0.65	& 8  \\
  &			 &	COS-79837-H1 &	K &	12.5 &	0.8 / 0.87	& 9  \\
  &			 &	COS-208070-K1 &	K &	3.6 &		0.8 / 0.99	& 14  \\
  &	23Feb2020 &	COS-111740-H1 &	H &	4.3 &		1.0 / 0.66	& 21  \\
  &			 &	COS-208070-K2 &	K &	4.3 &		1.0 / 0.80	& 16  \\
  &			 &	COS-113684-K2 &	K &	8.6 &		0.7 / 0.68	& 6  \\
  \hline
  XMM-VIDEO & 21Nov2017 &	XMM2017$_2$ & K & 18.5 & 0.8 / 0.89 & 14 \\ 
  & 	22Nov2017 &	XMM2017$_4$ & 	K & 18.5 & 	0.8 / 0.88	& 15 \\
  &	27Nov2018 &	XMM-2599-K1 &	K &	10.8 &	0.6 / 0.69	& 21 \\
  &	14Dec2018 &	XMM-2599-H1 &	H &	8.6 &		0.6 / 0.94	& 14 \\
  &			  &	XMM-2599-H2f &	H &	4.8 &		1.0 / 1.46	& 13 \\
  &	15Dec2018 &	XMM-2599-H1f &	H &	9.6 &		0.7 / 1.13	& 15 \\
  &			  &	XMM-2962-K1 &	K &	5.0 &		0.6 / 0.70	& 20 \\
  &	14Nov2019 &	XMM-3941-H1 &	H &	5.3 &		0.5 / 1.21	& 17 \\
  &	08Dec2019 &	XMM-3941-K1 &	K &	5.8 &		0.8 / 1.20	& 17 \\
  &			 &	XMM-2293-K1 &	K &	2.9 &		0.8 / 0.88	& 24 \\
  &	03Feb2020 &	XMM-3941-K12 &	K &	5.0 &		1.0 / 1.25	& 17 \\
  &	23Feb2020 &	XMM-270-K1w	&	K &	0.7 &		1.0 / 1.18	& 10 \\
  \hline
  \insertTableNotes
  \label{T:Obs}
  \end{longtable*}
\end{ThreePartTable}

\begin{longtable*}{ll p{1cm} p{1.2cm} p{1cm} p{1.2cm} p{1.2cm} p{1.5cm} p{1cm} p{1cm}}
\caption{Summary of observations by UMG candidates ordered by $m_K$ within a field. Masks of the same object begin with a UMG indicator, then are appended with a number and band indicator (here represented by an asterisk). The number of masks observed in each bandpass is also given. }\\
UMG & masks & \multicolumn{2}{l}{Magnitude} & \multicolumn{2}{l}{Int. Time (ks)} & \multicolumn{2}{l}{Avg. Seeing (")} & \multicolumn{2}{l}{Avg. SNR per Pixel} \\ 
& ($n_K$, $n_H$) 	& $K$ & $H$ 		& $K$ & $H$		&  $K$ & $H$ 			& $K$ & $H$\\
\hline \hline \endfirsthead
UMG & masks & Magnitude &Int. Time & Avg. Seeing & Average SNR / pixel \\ 
& ($n_K$, $n_H$) & ($K$ / $H$ )& (ks; $K$ / $H$) &  ("; $K$ / $H$) & ($K$ / $H$)\\
 \hline \hline \endhead
COS-DR3-160748	& COS-160748* (1,0) & 20.25 & 21.09	& 3.6	 & ---		& 0.7	 & ---		&  3.44 & ---  \\
COS-DR3-202019	& COS-201999* (1,4) & 20.79 & 21.36	& 5.0	 & 21.1	& 0.7 & 0.6-1.5	&  2.04 & 1.92 \\
COS-DR3-131925	& COS-131925* (1,0) & 20.96 & 21.30	& 4.3	 & ---		& 0.8 & ---		&  0.25 & ---   \\
COS-DR3-201999	& COS-201999* (1,4) & 21.00 & 21.51	& 7.2	 & 21.1	& 0.6	 & 0.6-1.5	&  3.01 & 3.73  \\
COS-DR1-79837	& COS-79837* (1,1) 	 & 21.10 & 22.25	& 2.9	 & 12.5	& 0.7	 & 0.9		&  2.42 & 4.12   \\ 
COS-DR3-113684	& COS-113684* (2,0) & 21.10 & 22.57	& 12.2 & --- 	& 0.6-0.7 & ---	&  2.64 & ---   \\
COS-DR3-111740	& COS-111740* (1,1) & 21.10 & 21.09	& 9.4	 & 4.3		& 1.4 & 0.7		&  4.96 & 3.03  \\
COS-DR3-84674	& COS-84674* (0,3)	 & 21.28 & 21.67	& ---	& 16.8	& --- & 1.2		&  --- & 4.36  \\
COS-DR3-195616	& COS-195616* (1,0) & 21.64 & 22.34	& 8.1	 & ---		& 1.3	 & ---		&  1.63 & ---   \\
COS-DR3-226441	& COS-226441* (1,0) & 21.65 & 22.46	& 5.0	 & ---		& 1.0	 & ---		&  0.96 & ---   \\
COS-DR1-258857	& COS-258857* (2,0) & 21.66 & 22.46	& 7.2	 & ---		& 0.76-0.8 & ---	&  1.30 & ---   \\ 
COS-DR3-208070 	& COS-208070* (2,0) & 21.70 & 22.54	& 7.9	 & ---		& 0.8-1.0 & ---	&  2.18 & ---   \\
COS-DR3-179370	& COS2017* (2,0)	 & 22.14 & 23.51	& 16.6 & ---	& 0.7-0.8 & ---	&  1.19 & ---   \\
XMM-VID3-2075	& XMM-2692* (1,0)	 & 20.79 & 21.94	& 5.0 & ---		& 0.7	 & ---		&  3.51 & ---   \\
XMM-VID1-1120	& XMM-2599* (1,3)	 & 20.97 & 22.44	& 10.8 & 23.0 	& 0.7	 & 1.0-1.4	&  4.00 & 1.35  \\
XMM-VID1-2399	& XMM2017$_4$ (1,0) & 21.10 & 21.09	& 18.4 & --- 	& 0.9	 & ---		&  1.21 & ---   \\
XMM-VID3-3941	& XMM-3941* (2,1) 	 & 21.33 & 22.47	& 10.8 & 5.3	& 1.02 & 0.5	&  2.23 & 1.59  \\
XMM-VID3-2293	& XMM-2293* (1,0) 	 & 21.35 & 22.47	& 2.9	 & ---		& 0.9	 & ---		&  1.00 & ---   \\
XMM-VID2-270		& XMM-270* (1,0) 	 & 21.47 & 22.51	& 0.7	 & ---		& 1.0	 & ---		&  0.32 & ---   \\
XMM-VID3-2457	& XMM-2599* (1,3)	 & 21.52 & 22.61	& 10.8 & 23.0 	& 0.7	 & 1.0-1.4	&  3.40 & 1.30  \\
XMM-VID1-2761	& XMM2017$_2$ (1,0) & 21.74 & 23.22	& 18.5 & --- 	& 0.9	 & ---		&  1.68 & ---   \\
\label{T:Obs2}
\end{longtable*}

\clearpage
\renewcommand\thefigure{\thesection.\arabic{figure}}    

\section{Observations of Unconfirmed Objects} \label{A:Unconf}
\setcounter{figure}{0}

\begin{figure*}[tp]
	\centering{\includegraphics[width=0.9\textwidth,trim=0in 0in 0in 0in, clip=true]{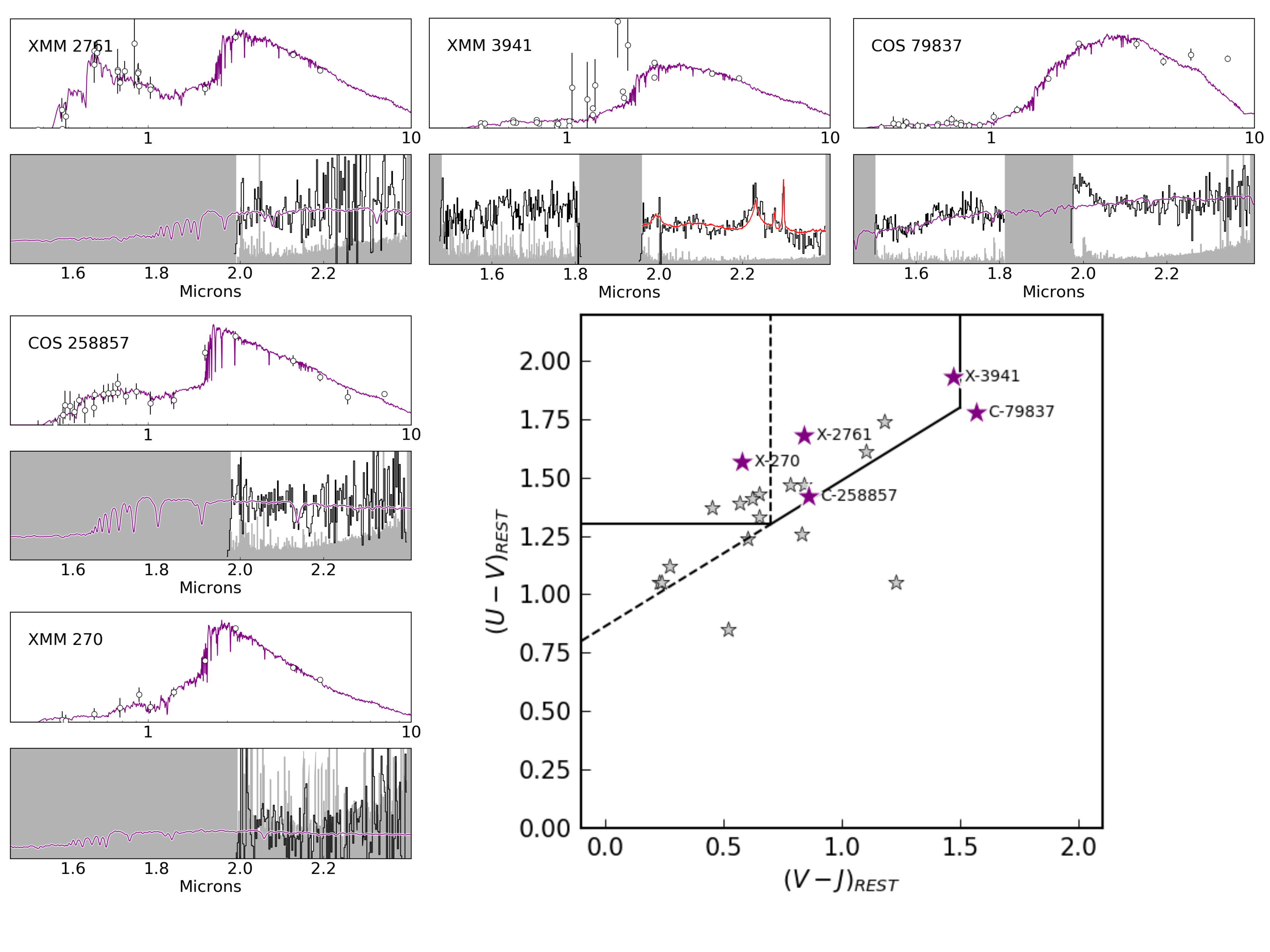}}
	\caption{\textbf{Central panel:} The restframe \UVJ-diagram showing the location of the five unconfirmed objects (purple) relative to the 16 confirmed UMGs (gray). \textbf{Surrounding panels:} The photometry (top) and spectroscopy (bottom) for each object, shown with the best-fit template in purple. For XMM-VID3-3941, the SDSS composite quasar template \citep{VandenBerk2001} redshifted to $z=3.59$ is shown in red.}
	\label{fig:uvj_unconf}
\end{figure*}

In this work we have presented spectra of 16 UMGs.
However, five other UMG candidates were also spectroscopically observed, and these objects are presented here.
The associated masks and observations for each UMG are included in Table~\ref{T:Obs} and Table~\ref{T:Obs2}.
SEDs and spectra are presented in Figure~\ref{fig:uvj_unconf}.

XMM-VID3-3941 has a very red SED, suggesting significant dust obscuration, and its photometry is well fit with a massive, dusty template at \zphot~$=3.04$.
However, it is spectroscopically confirmed to be a quasar at \zspec~$=3.68$.
Comparison of the $K$-band spectrum with the SDSS composite quasar spectrum shows good agreement with broad \Hbeta\ and H$\gamma$ emission, as well as narrow \OIIIdub\ emission (Figure~\ref{fig:uvj_unconf}).
Given that the quasar potentially contributes a substantial amount to the flux of this galaxy, we refrain from estimating other properties such as stellar mass.

A photometric redshift analysis indicates that COS-DR1-79837 is also a massive, dusty, star-forming galaxy at $z=3.3$.
Spectra do not reveal clear features to confirm a redshift, though a combined fit to spectra and photometry yields a best-fit $z=2.65$.
The upturn in flux in the IRAC 3.6 $\mu$m and 4.5 $\mu$m channels is also suggestive of AGN activity.

Both of these candidates lie in the upper right corner of the \UVJ\ diagram, nominally the location of dusty star-forming galaxies, and have large photometric stellar mass estimates of \logM~$>11.7$.
Galaxies in this region are difficult to spectroscopically confirm as many observable line features are expected to be obscured by dust.
Further probing of this region is warranted, and will be explored in future work.

COS-DR1-258857 lies on the wedge separating the star-forming and quiescent regions of the \UVJ\ diagram, with \mbox{\zphot~$=3.26$}.
The spectrum shows two significant dips around 2.14 and 2.17~$\mu$m, though the spectrum is noisy.
Assuming one of these is \Hbeta\ absorption, the redshift is then either $z=3.40$ or $z=3.46$.

The spectrum of XMM1-2761 does not have sufficient signal-to-noise to determine a redshift, though we note this target is very faint (\zphot~$=3.59$, $m_K=21.74$).
Photometric detections in the rest-frame UV suggest some ongoing star-formation.
The \UVJ\ colors may incorrectly place this galaxy in the quiescent wedge due to photometric contamination by strong emission lines.

XMM-VID2-270 is photometrically a post-starburst similar to a number of others confirmed in this work.
Integration time on this object was exceedingly short, taken in a short period after high winds had subsided and before the XMM field set, and no features are apparent in the resulting spectrum.

\clearpage
\section{Identification of Sky Line Contamination} \label{A:SL}
\setcounter{figure}{0}

\begin{figure*}
	\centering{\includegraphics[width=\textwidth,trim=0in 0in 0in 0in, clip=true]{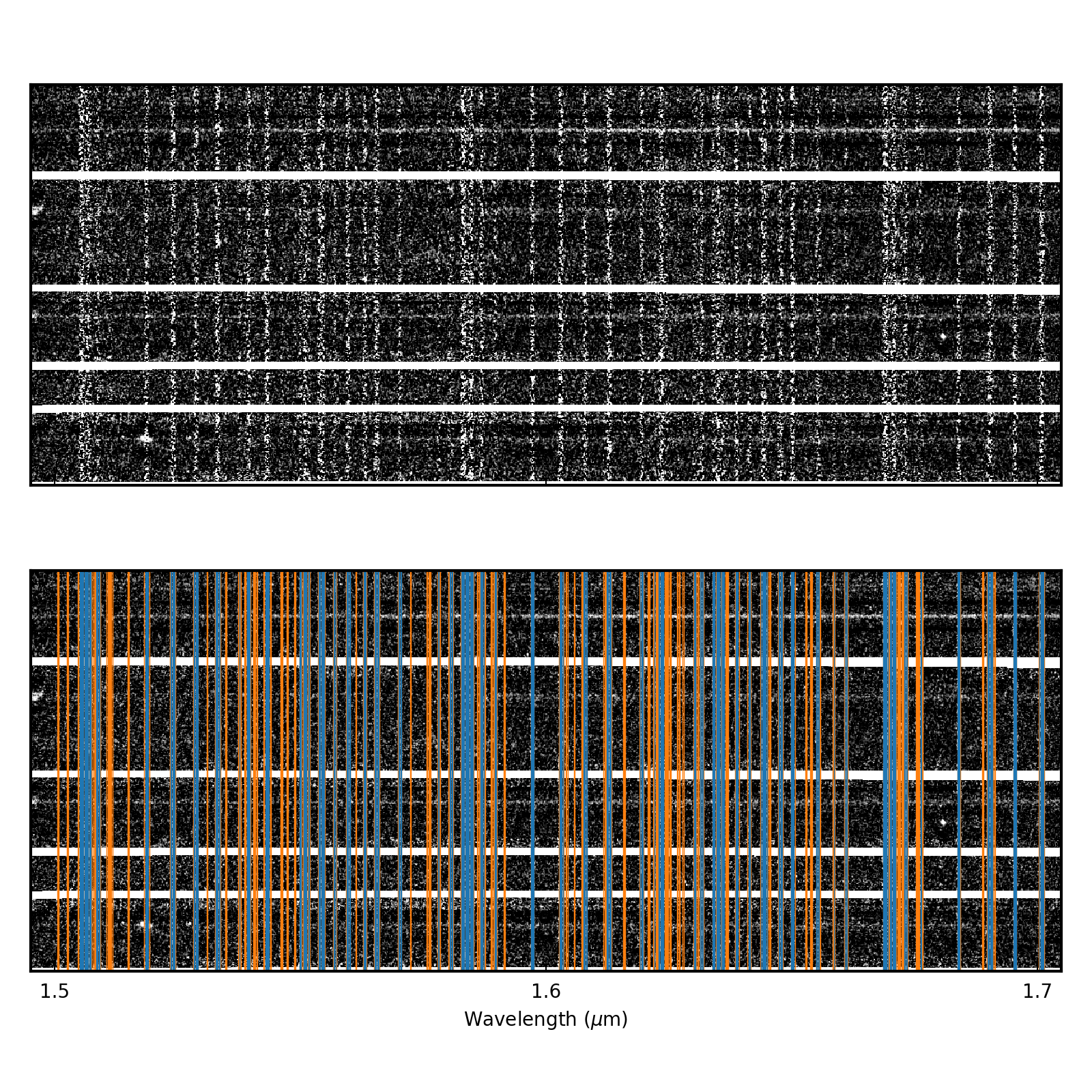}}
	\caption{\textbf{Top:} A portion of an example $H-$band mask reduced with the MOSFIRE DRP, with obvious sky lines. \textbf{Bottom:} The results of the sky line identification - orange are weak sky lines and blue are strong sky lines.}
	\label{fig:sl_h}
\end{figure*}

We use the uncertainties on pixel fluxes determined by the DRP, which we refer to below as pixel variance, to identify sky lines.
This is done by by comparing the average pixel variance at a given wavelength to the distribution of pixel variances across the entire mask.
For each wavelength $\lambda_i$ in a reduced $H$-band mask, we calculate the median variance per pixel on sky in the spatial direction, median($\sigma^2$($\lambda_i$)).
This is then compared to the distribution of variances across all sky pixels, $\sigma^2$($\lambda$, s).
If median($\sigma^2$($\lambda_i$))$>1.5\times P_{25}$($\sigma^2$($\lambda$, s)), where $P_n$ corresponds to the $n^{\rm th}$ percentile, then $\lambda_i$ is considered to have sky line contamination.
Additionally, if median($\sigma^2$($\lambda_i$))$>2.5\times P_{25}$($\sigma^2$($\lambda$, s)), then $\lambda_i$ is considered to have contamination from a strong sky line, while between the two thresholds is termed contamination from a weak sky line.
Regions with strong lines are masked when doing all fits.
We interpolate fluxes over regions with weak lines, but do not change the variance, which is generally higher at these wavelengths.
An example from a portion of an $H$-band mask is shown in Figure \ref{fig:sl_h}.

\begin{figure}
	\centering{\includegraphics[width=0.7\textwidth,trim=0in 0in 0in 0in, clip=true]{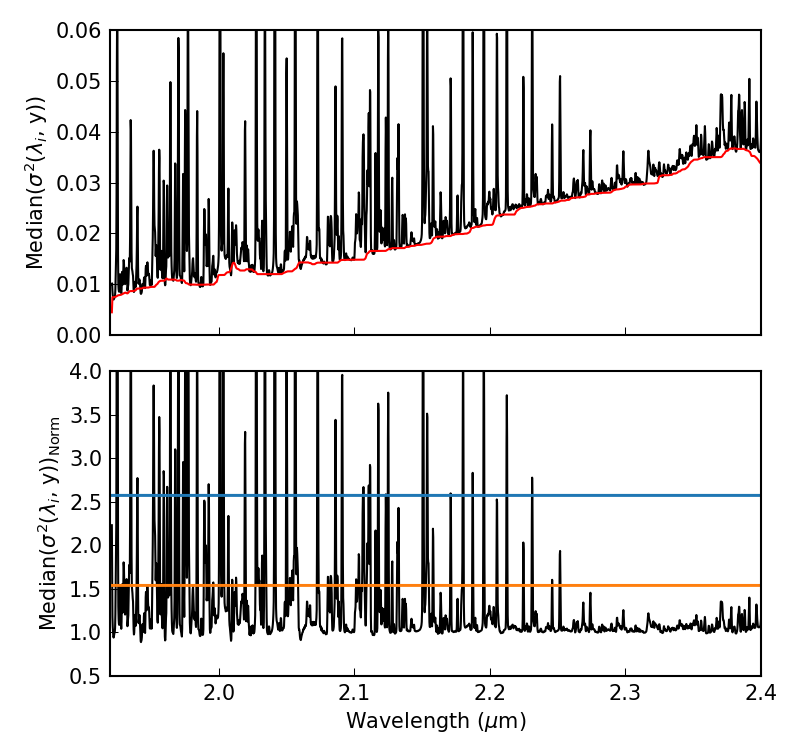}}
	\caption{\textbf{Top:} The median variance per pixel across wavelengths for a sample $K$-band mask (black). The value over which we normalize is shown in red. \textbf{Bottom:} The normalized variance per wavelength (black) as well as the thresholds for weak (orange) and strong (blue) sky line contamination.}
	\label{fig:sl_knorm}
\end{figure}

\begin{figure*}
	\centering{\includegraphics[width=\textwidth,trim=0in 0in 0in 0in, clip=true]{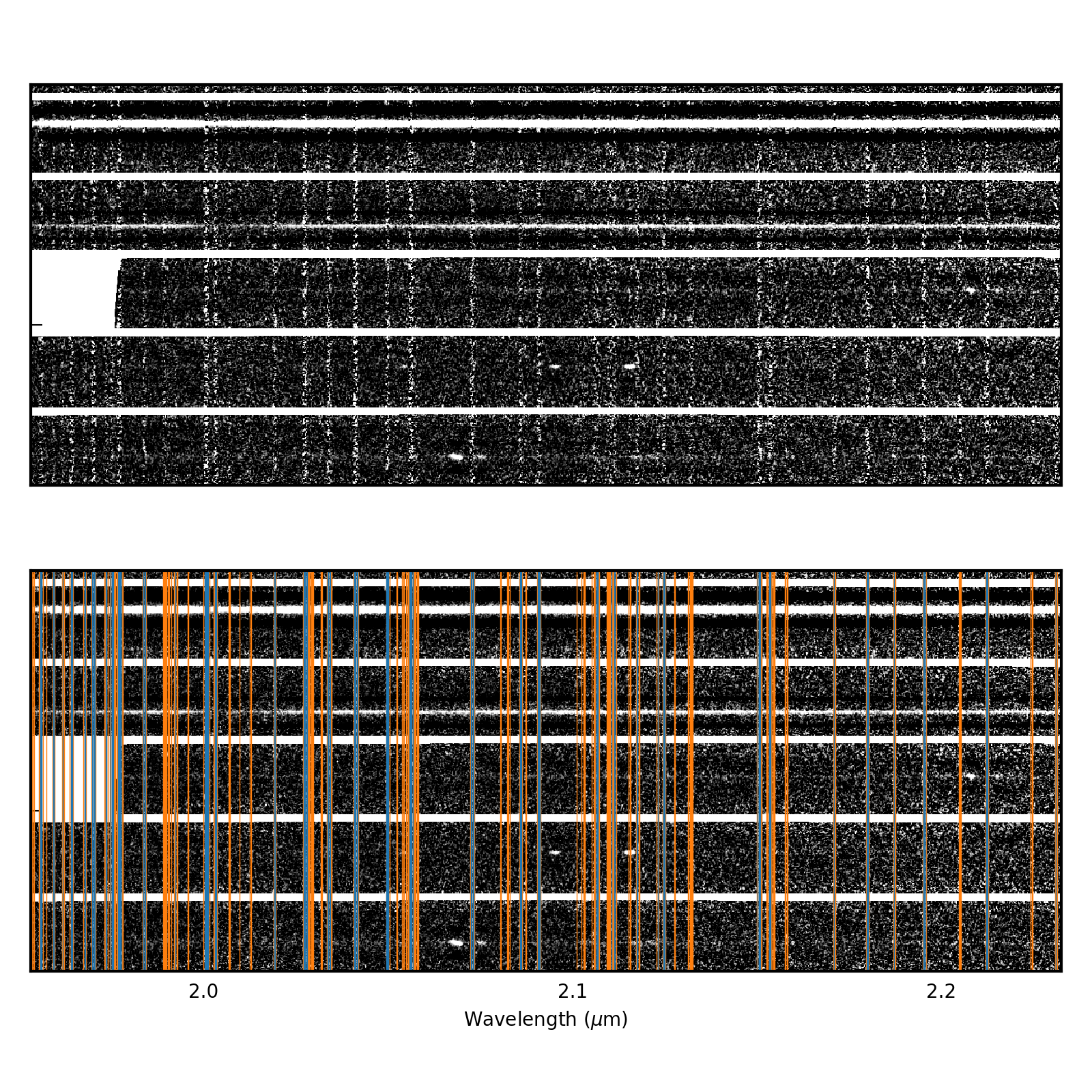}}
	\caption{Same as Figure \ref{fig:sl_h} but for a $K$-band mask, using the normalized variance threshold shown in Figure \ref{fig:sl_knorm}. }
	\label{fig:sl_k}
\end{figure*}

The above method will not work effectively in the $K$-band due to increasing thermal contamination at redder wavelengths.
Applying the same method would result in incorrectly identifying a large percentage of pixels at \mbox{$\lambda>2.2~\mu$m} as sky lines, and failing to identify weak sky lines at bluer wavelengths.
In this case we model the background contribution to variance (\ie\ not due to sky lines) at $\lambda_i$ by calculating a running $P_{10}$($\sigma^2$($\lambda_i\pm0.1$ $\mu$m)) and take the ratio with spectral uncertainty.
We then perform the same calculation as for the $H$-band masks with this `normalized' variance distribution to determine thresholds for sky line identification (see Figure \ref{fig:sl_knorm}).
In the case of both the $H$-band and $K$-band (Figure \ref{fig:sl_k}), these identifications are checked against visual identification and confirmed to be effective at removing sky line contamination.

\clearpage
\section{Telluric Correction Comparison} \label{A:TC}
\setcounter{figure}{0}

The telluric correction used in this work involved modeling the observed spectra of faint stars on the individual MOSFIRE masks.
Slit stars have previously been used successfully for absolute flux calibrations \citep[\eg][]{Kriek2015} and telluric correction \citep[\eg][see Appendix B for details]{Schreiber2018b}.

On many nights we also took spectra of telluric standard stars, all of which were A-type stars, to provide an alternative telluric correction option.
These spectra were processed in the same manner as the science masks (MOSFIRE DRP to obtain 2D spectrum, custom code for 1D extraction).
We subsequently masked and interpolated over hydrogen absorption features present in the 1D spectra of these stars and compared the resultant curve to a blackbody with the temperature of the same spectral subtype.
The normalized ratio of these two yielded a telluric correction factor which was applied to all the spectra on a given mask.

\begin{figure*}
	\centering{\includegraphics[width=\textwidth,trim=0in 0in 0in 0in, clip=true]{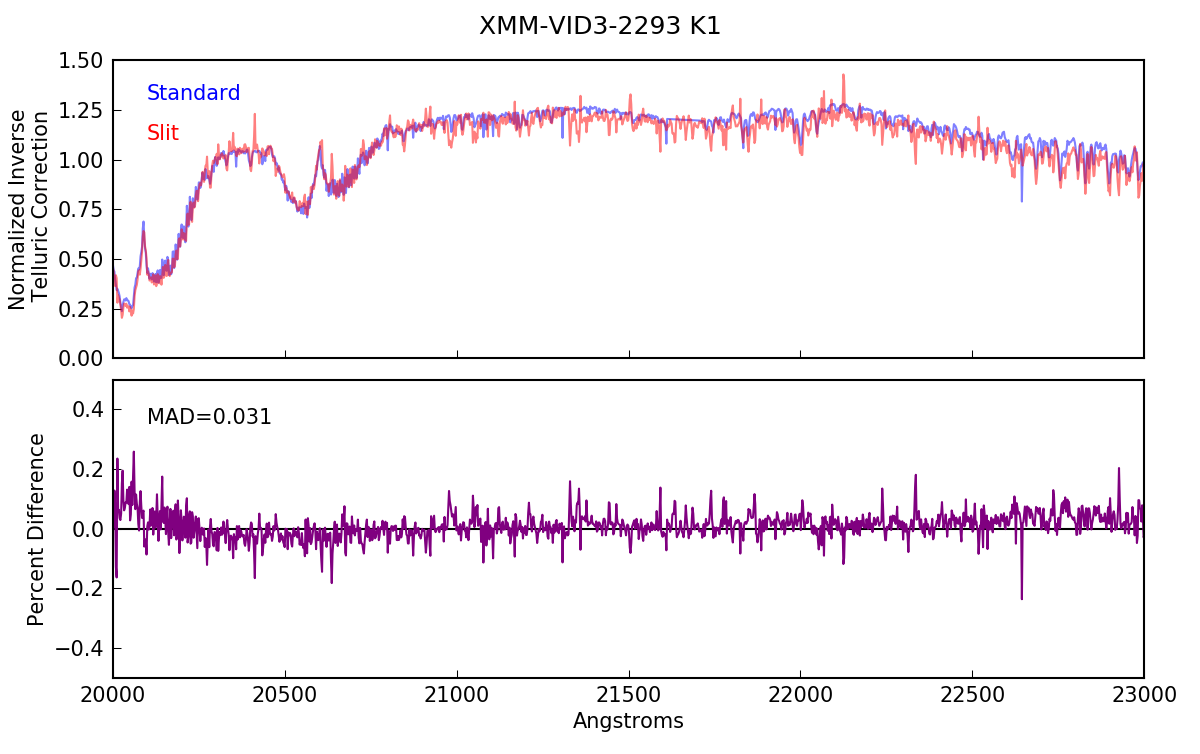}}
	\caption{\textbf{Top:} Comparison of the telluric correction derived using stars on science slits (red) to the telluric corrections derived using a standard star (blue) observed directly after the mask. \textbf{Bottom:} The percent difference between the two corrections as a function of wavelength. The median absolute deviation is given as well.}
	\label{fig:tc_good}
\end{figure*}

\begin{figure*}
	\centering{\includegraphics[width=\textwidth,trim=0in 0in 0in 0in, clip=true]{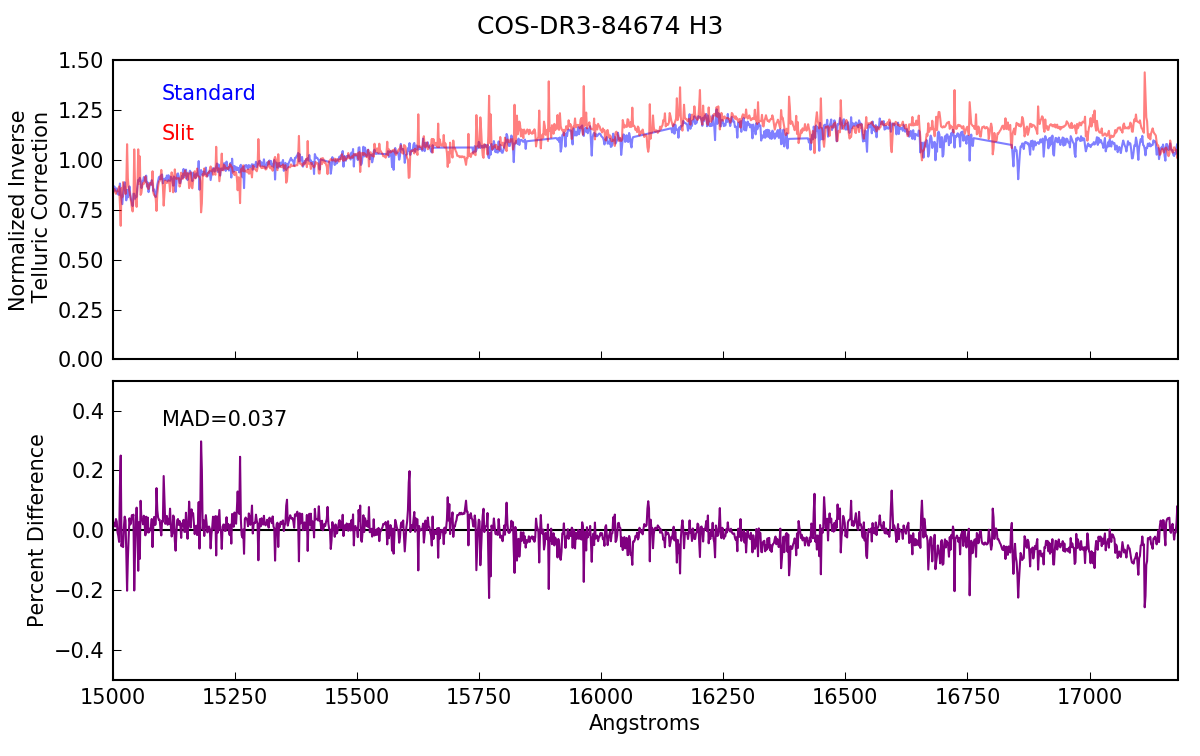}}
	\caption{Same as Figure \ref{fig:tc_good} but for a mask which had two hours between observations of the mask and the standard star.} 
	\label{fig:tc_meh}
\end{figure*}

When compared to the method incorporating the slit stars, this method gave generally similar results.
The standard star correction tends to be smoother, which is reasonable considering that one part of the ratio (the blackbody) is itself a smooth curve, and given the higher SNR of the bright standard star relative to the slit star.
For masks observed directly before or after a standard star observation, the differences in the telluric corrected spectra between the two methods is small (Figure \ref{fig:tc_good}).
However, we find that the corrections tend to be more discrepant as the time between observations of a mask and a standard star increase, almost certainly due to atmospheric changes throughout the night (Figure \ref{fig:tc_meh}).
For determining redshifts and locations of strong emission and absorption features, these differences are inconsequential.

We note that for science goals requiring high SNR spectra, such differences could become relevant.
Also we do not claim that one correction is necessarily better than the other, as this will depend on the targets, observing sequence, and observing conditions.

\clearpage
\section{On the Existence of a Population of Older UMGs} \label{A:DeltaM}
\setcounter{figure}{0}

As we discuss in Sections~\ref{S:ages} and \ref{S:sfh}, spectroscopically confirmed massive galaxies at $z>3$ lacking evidence of star formation appear to be post-starbursts, \ie\ have quenched in the last several hundred Myr.
Given that detection of these objects is very near to the limit of what is possible with current instrumentation, we must consider the fact that this is the result of an observational bias.
We explore this by calculating how much fainter an older, passive UMG would be at these redshifts.

To accomplish this we construct SEDs at various times along several SFHs (using BC03 SPS models) and compare their luminosities in the \Ks-band, assuming a redshift of $z=3.3$, the median of the sample.
We begin with a SFH consisting of 100 Myr of constant (and vigorous) star formation, followed by a quick decline in star formation with $\tau=100$~Myr, shown in Figure \ref{fig:DeltaM}.
The time at which this galaxy quenched is taken to be when star formation drops below 10\% of the SFR during the constant period of star formation, akin to the value calculated for the UMGs.
This SFH is similar to the best-fit SFHs for the UMGs, which have a median starburst period of 122 Myr and steeply declining SFR thereafter.
As expected, we find that galaxies which quenched earlier are significantly redder and fainter, in this case over five times fainter for a galaxy which quenched 1 Gyr before observation. 

\begin{figure*}
	\centerline{\includegraphics[width=\textwidth,trim=0in 0in 0in 0in, clip=true]{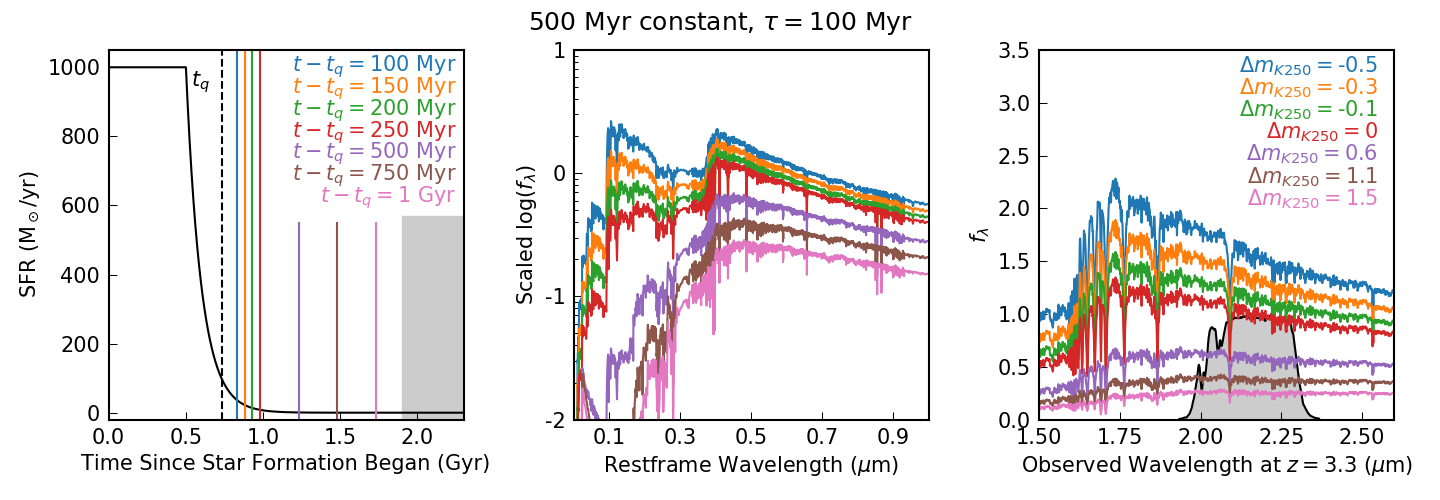}}
	\caption{The same as Figure \ref{fig:DeltaM} but with a SFH of 500 Myr constant followed by an exponential decline of $\tau=100$ Myr. The grayed out region on the left panel indicates an age greater than that of the universe at $z=3.3$.}
	\label{fig:DeltaM_con}
	\end{figure*}

\begin{figure*}
	\centerline{\includegraphics[width=\textwidth,trim=0in 0in 0in 0in, clip=true]{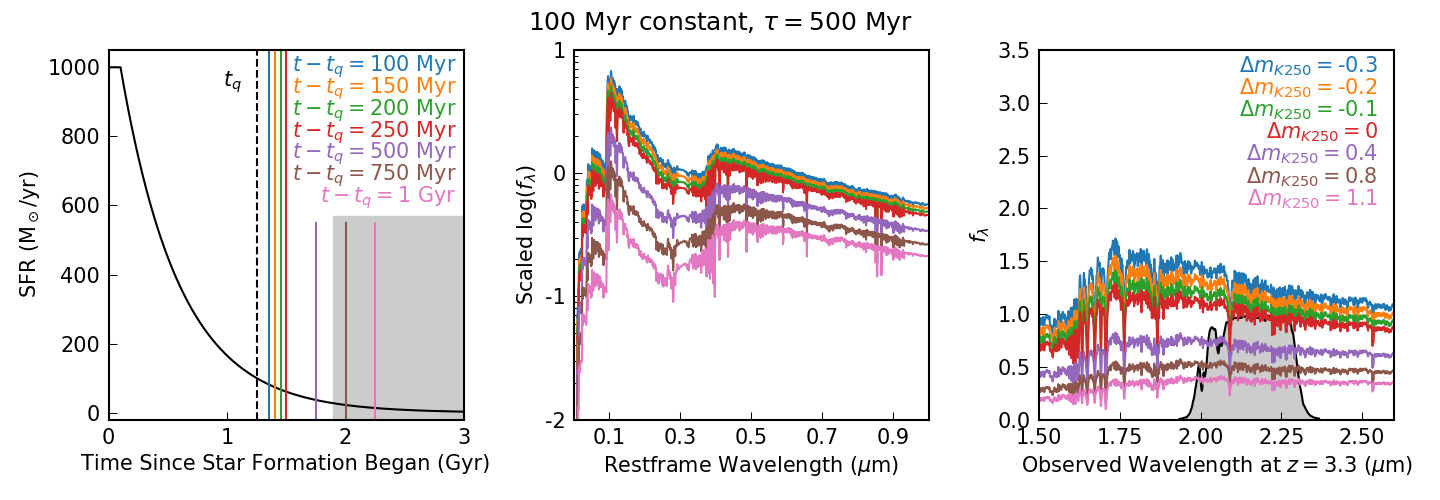}}
	\caption{The same as Figure \ref{fig:DeltaM} but with a SFH of 100 Myr constant followed by an exponential decline of $\tau=500$ Myr. The grayed out region on the left panel indicates an age greater than that of the universe at $z=3.3$. Assuming star-formation began $\sim400$ Myr after the Big Band ($z\sim11$), this extreme example would have dropped below 10\% of its maximum SFR at $z\sim5$.}
	\label{fig:DeltaM_tau}
	\end{figure*}

However, this test is somewhat model dependent.
The difference in magnitude due to time since quenching decreases with an increase in either the duration of the starburst period (Figure \ref{fig:DeltaM_con}) or the $\tau$-parameter (Figure \ref{fig:DeltaM_tau}).
In addition, this limits how long ago the oldest galaxies could have quenched given the young age of the universe (1.9 Gyr at $z=3.3$.)


\end{document}